\documentclass[fleqn,usenatbib]{mnras}

\usepackage{newtxtext}

\usepackage[T1]{fontenc}

\DeclareRobustCommand{\VAN}[3]{#2}
\let\VANthebibliography\thebibliography
\def\thebibliography{\DeclareRobustCommand{\VAN}[3]{##3}\VANthebibliography}

\usepackage{graphicx}	
\usepackage{amsmath}	
\usepackage{amssymb}	
\usepackage{subfigure}
\usepackage{scalerel}

\newcommand\chisqred{$\rm \chi^2_{red.}$}
\newcommand\HII{H\protect\scaleto{$II$}{1.2ex}}
\graphicspath{{figures/}} 

\title[Multiply-lensed SF clumps in a z=1 galaxy]{Multiply lensed star forming clumps in the A521-sys1 galaxy at redshift 1}
\author[M. Messa et al.]{
Matteo Messa,$^{1,2}$\thanks{E-mail: matteo.messa@unige.ch}
Miroslava Dessauges-Zavadsky,$^{1}$
Johan Richard,$^{3}$
Angela Adamo,$^{2}$
David Nagy,$^{1}$
\newauthor
Françoise Combes,$^{4}$
Lucio Mayer$^{5}$
and Harald Ebeling$^{6}$
\\
$^{1}$Observatoire de Genève, Université de Genève, Versoix, Switzerland\\
$^{2}$The Oskar Klein Centre, Department of Astronomy, Stockholm University, AlbaNova, SE-10691 Stockholm, Sweden\\
$^{3}$Univ Lyon, Univ Lyon1, ENS de Lyon, CNRS, Centre de Recherche Astrophysique de Lyon UMR5574, Saint-Genis-Laval, France\\
$^{4}$LERMA, Observatoire de Paris, PSL Research Université, CNRS, Sorbonne Université, UPMC, Paris, France\\
$^{5}$Center for Theoretical Astrophysics and Cosmology, Institute for Computational Science, University of Zurich, Winterthurerstrasse 190,\\
CH-8057 Zürich, Switzerland\\
$^{6}$Institute for Astronomy University of Hawaii, 2680 Woodlawn Drive Honolulu, HI 96822, USA\\
}

\date{Accepted XXX. Received YYY; in original form ZZZ}

\pubyear{2022}

\begin{document}
\label{firstpage}
\pagerange{\pageref{firstpage}--\pageref{lastpage}}
\maketitle

\begin{abstract}
We study the population of star-forming clumps in A521-sys1, a $\rm z=1.04$ system gravitationally lensed by the foreground ($\rm z=0.25$) cluster Abell 0521. The galaxy presents one complete counter--image with a mean magnification of $\rm \mu\sim4$ and a wide arc containing two partial images of A521-sys1 with magnifications reaching $\rm \mu>20$, allowing the investigations of clumps down to scales of $\rm R_{eff}<50$ pc. We identify 18 unique clumps with a total of 45 multiple images. Intrinsic sizes and UV magnitudes reveal clumps with elevated surface brightnesses, comparable to similar systems at redshifts $\rm z\gtrsim1.0$. Such clumps account for $\sim40\%$ of the galaxy UV luminosity, implying that a significant fraction of the recent star-formation activity is taking place there.
Clump masses range from $\rm 10^6\ M_\odot$ to $\rm 10^9\ M_\odot$ and sizes from tens to hundreds of parsec, resulting in mass surface densities from $10$ to $\rm 10^3\ M_\odot\ pc^{-2}$, with a median of $\rm \sim10^2\ M_\odot\ pc^{-2}$. These properties suggest that we detect star formation taking place across a wide range of scale, from cluster aggregates to giant star-forming complexes. 
We find ages of less than $100$ Myr, consistent with clumps being observed close to their natal region.
The lack of galactocentric trends with mass, mass density, or age and the lack of old migrated clumps can be explained either by dissolution of clumps after few $\sim100$ Myr or by stellar evolution making them fall below the detectability limits of our data.
\end{abstract}




\section{Introduction}
The study of galaxies at Cosmic Noon (redshift $\rm z\sim1-3$) reveals morphologies dominated by clumpy structures, particularly at rest-frame ultraviolet (UV) wavelengths \citep[e.g.][]{cowie1995,vandenbergh1996}. 
Clumps have typical sizes of $\lesssim1$ kpc \citep[e.g.][]{elmegreen2007,forsterschreiber2011b}, typical stellar masses of $\rm M_* \sim10^7-10^9\ M_\odot$ \citep[e.g.][]{forsterschreiber2011a,guo2012,soto2017}  and typical star--formation rates (SFRs) from $\rm 0.1-10\ M_\odot/yr$ \citep[e.g.][]{guo2012,soto2017}.
The presence of UV clumps is closely related to gas properties observed in those galaxies, characterized by higher gas fractions \citep{daddi2010,tacconi2010,tacconi2013,genzel2015} and velocity dispersions \citep{elmegreen2005,forsterschreiber2006} than local main sequence (MS) star-forming galaxies; yet, overall they show rotation features indicating the presence of disk structure \citep{forsterschreiber2006,genzel2006,shapiro2008,wisnioski2018}. 
The commonly accepted interpretation of these findings is that clumps result from \textit{in--situ} gas collapse due to gravitational instabilities in the disc, which can fragment at much larger scales at high redshift than in local MS galaxies because of the gas-rich, turbulent composition of these objects \citep[e.g][]{elmegreen2009,immeli2004a,tamburello2015,renaud2021}. This interpretation is supported by recent observations of dense giant molecular cloud complexes from CO data in galaxies at $\rm z\sim1$ \citep{dessauges2019}, as well as by simulations of turbulent high-redshift galaxies \citep[e.g.][]{vandonkelaar2021arxiv} and by observations in nearby analogs \citep[e.g.][]{fisher2017a,fisher2017b,messa2019}.

An additional confirmation of the link between clumps and their host galaxies is given by the evolution of the clump densities with redshift (clumps are denser at higher redshifts), tracing the evolution of star formation (SF) with cosmological time \citep{livermore2015}. We note though that the interpretation of the underlying observations is complicated by the difference in surface-brightness completeness limits \citep{ma2018} and the different resolution achievable at different redshifts and at different gravitational lensing magnifications. 

In addition, high-redshift clumps may affect the process of galaxy assembly; hydro-dynamical and cosmological simulations have suggested that, if clumps are able to survive as bound systems for hundreds of Myr, dynamical friction could cause them to migrate toward the centre of the galaxy \citep{bournaud2014,mandelker2014,mandelker2017}. Such spiralling inward would generate torque that, in turn, funnels inward large amounts of gas, which, along with clump merging, could contribute to the formation of the thick galactic disk and to the bulge growth \citep{noguchi1999,immeli2004b,carollo2007,genzel2008,elmegreen2008,dekel2009,bournaud2007,bournaud2009,bournaud2011,gabor2013}. However, not all simulations predict clumps surviving for long time-scales \citep{oklopcic2017}.
Observations of individual galaxies seem to support this scenario \citep[e.g.][]{guo2012,adamo2013,cava2018}, but the large uncertainties on age determinations and the lack of larger statistical samples prevent us from assessing if, and in what conditions, clumps could survive long enough to migrate from their natal region.

High-redshift clumps contribute by a large fraction to the emission in the rest-frame UV \citep{elmegreen2005b} and in nebular lines (e.g., Balmer transitions, \citealp{livermore2012,mieda2016,zanella2019}) of their host galaxies, suggesting that they trace giant star-forming regions and that those regions constitute the bulk of their host galaxy's recent star-formation activity.
Due to their elevated specific star-formation rate ($\rm sSFR = SFR/M_*$), which can exceed  the integrated sSFR of their host galaxies by orders of magnitude, it has been suggested that clumps are starbursting \citep{bournaud2015,zanella2015,zanella2019}.
We expect feedback from star-forming clumps to affect the evolution of galaxies, suppressing the global star formation and leading to the formation of a multiphase interstellar medium (ISM) \citep[e.g][]{hopkins2012,goldbaum2016}. Evidence from local analogs suggests that stellar feedback from clumps could facilitate the escape of UV radiation into the intergalactic medium (e.g., \citealp{bik2015,bik2018,herenz2017} in local galaxies, \citealp{riverathorsen2019} at $\rm z\sim2$); if this process is efficient, clump feedback could even contribute to the reionization of the Universe \citep{bouwens2015}.

Recent studies of lensed high-redshift galaxies \citep[e.g.][]{livermore2012,adamo2013,johnson2017,cava2018,mestric2022} at higher angular resolution offer the possibility to investigate the substructure of clumps \citep{meng2020}. At the highest resolution, potential clusters have been detected on scales of a few parsecs \citep{vanzella2019,vanzella2021b,vanzella2021}.
One of the challenges for the upcoming James Webb Space Telescope (JWST) and adaptive-optic instruments on the European Extremely Large Telescope (E-ELT) will be the  detection of possible high-redshift progenitors of the globular clusters (GCs) observed in the local universe, to help solve the many open questions about their origin \citep[e.g.][for a review]{bastian2018}.

In the context of analyses of clumps on small physical scales, we here present the study of the strongly lensed arc at $\rm z=1.04$ in Abell 0521 (A521); following the nomenclature in \citet{patricio2018} we will refer to the galaxy as A521-sys1 in the rest of the paper. 
With a stellar mass of $\rm M_*=(7.4\pm1.2)\times10^{10}\ M_\odot$ and a SFR of $\rm (26\pm5)\ M_\odot yr^{-1}$ \citep{nagy2021}, A521-sys1 can be considered a typical main-sequence star-forming galaxy at $\rm z\sim1$ \citep[e.g.][]{speagle2014}. 
The kinematic analysis reveals a rotation-dominated galaxy typical of systems at cosmic noon, with a high velocity dispersion \citep{patricio2018,girard2019}.
In addition, both the molecular gas mass surface density, $\rm \Sigma(M_{mol})$, and the SFR surface density, $\rm \Sigma(SFR)$, are elevated by a factor of $\sim10$ compared to local MS galaxies, as expected for high-z gas-rich galaxies. The radial profiles of $\rm \Sigma(M_{mol})$ and $\rm \Sigma(SFR)$ are very shallow \citep{nagy2021}, suggesting an intense star-formation activity throughout the entire galaxy, as also indicated by the presence of UV clumps in various sub--regions of A521-sys1. 
The gravitational lensing produced by the foreground cluster allows the analysis of A521-sys1 clumps down to scales of few tens of parsecs. In addition, the presence of multiple images of A521-sys1 at different magnification factors allows the comparison of the same clumps seen at different resolution, and hence tests of the effect of resolution on the study of clump populations. 
This paper is structured as follows: we present the data and the lensing model in Section~\ref{sec:data}; the analyses, including the model used to fit the clumps, are described in Section~\ref{sec:datanalysis}. The results are collected in Section~\ref{sec:results_phot} (photometric properties of the clumps) and in Section~\ref{sec:results_sed} (physical properties of the clumps), followed by their discussion in Section~\ref{sec:discussion}. An overall summary of the paper is given in Section~\ref{sec:conclusion}. 
Throughout this paper, we adopt a flat $\rm \Lambda$-CDM cosmology with $H_0=68$ km s$^{-1}$ Mpc$^{-1}$ and $\rm \Omega_M = 0.31$ \citep{planck13_cosmo}, and the \citet{kroupa2001} initial mass function.

\section{Data}
\label{sec:data} 
\subsection{Hubble Space Telescope (HST)}
\label{sec:data_hst}
A521-sys1 was observed with WFC3/UVIS in the F390W passband, with WFC3/IR in F105W and F160W (ID: 15435, PI: Chisholm, exposure times: $2470$, $2610$ and $5220$ s, respectively), with ACS/WFC in the F606W and F814W filters (ID: 16670, PI: Ebeling, exposure times $1200$ s). 
Individual flat-fielded and CTE-corrected exposures were aligned and combined in a single image using the \texttt{AstroDrizzle} procedure from the \texttt{DrizzlePac} package \citep{hoffmann2021}; the final images have pixels scales of 0.06 arcsec/pixel. The astrometry was aligned to the Gaia DR2 \citep{gaia2018}.
We model the instrumental point-spread function (PSF) from a stack of isolated bright stars within the field of view of the observations. The stack in each filter is fitted with an analytical function described by a combination of Moffat, Gaussian, and  $\rm 4^{th}$ degree power-law profiles, to mitigate bias introduced by the choice of a specific function. The fit provides a good description of the stacked stars up to a radius of $\sim20$ pixels (corresponding to $1.20''$).
The minimum detectable magnitude limit, $\rm mag_{lim}$, is estimated from the standard deviation $\rm \sigma$ of the background level in the proximity of A521-sys1; we consider the minimum flux of a PSF light profile whose four brightest pixels are above the $\rm 3\sigma$ level, similarly to the procedure applied to extract sources (see Section~\ref{sec:sextraction}); this minimum flux is converted to an AB magnitude for each filter. We point out that these values are representative of the depth of the observations in the proximity of A521-sys1; the clumps within this system are observed above the diffuse galaxy background, and their detection limits are discussed in Section~\ref{sec:completeness}.
The FWHM values of the PSF, exposure times, zeropoints and depth of the exposures are listed in Tab.~\ref{tab:data}. 
\begin{table}
    \centering
    \begin{tabular}{lrrrrr}
    \multicolumn{1}{l}{Filter} & \multicolumn{1}{c}{$\rm \lambda_{rest}$} & \multicolumn{1}{c}{$\rm t_{exp}$}  & \multicolumn{1}{c}{$\rm ZP_{AB}$} & \multicolumn{1}{c}{$\rm mag_{lim}$} & \multicolumn{1}{c}{$\rm PSF_{FWHM}$} \\
    \multicolumn{1}{c}{\ } & \multicolumn{1}{c}{(\AA)}& \multicolumn{1}{c}{(s)} & \multicolumn{1}{c}{(mag)} & \multicolumn{1}{c}{(mag)} & \multicolumn{1}{c}{(arcsec)} \\
        \hline
        WFC3-UVIS-F390W & 1920 & 2470 & 25.4 & 27.6 & 0.097 \\
        ACS-WFC-F814W   & 2900 & 1200 & 26.5 & 27.5 & 0.112 \\
        ACS-WFC-F606W   & 3940 & 1200 & 25.9 & 27.2 & 0.116 \\
        WFC3-IR-F105W   & 5160 & 2610 & 26.3 & 27.0 & 0.220 \\
        WFC3-IR-F160W   & 7520 & 5220 & 26.0 & 26.8 & 0.237 \\
        \hline
    \end{tabular}
    \caption{Rest--frame pivotal wavelengths ($\rm \lambda_{rest}$), exposure times ($\rm t_{exp}$), AB magnitude zeropoints ($\rm ZP_{AB}$), depth of the observations ($\rm mag_{lim}$) and FWHM of the PSF ($\rm PSF_{FWHM}$).}
    \label{tab:data}
\end{table}

\begin{figure*}
    \centering
    \includegraphics[width=0.99\textwidth]{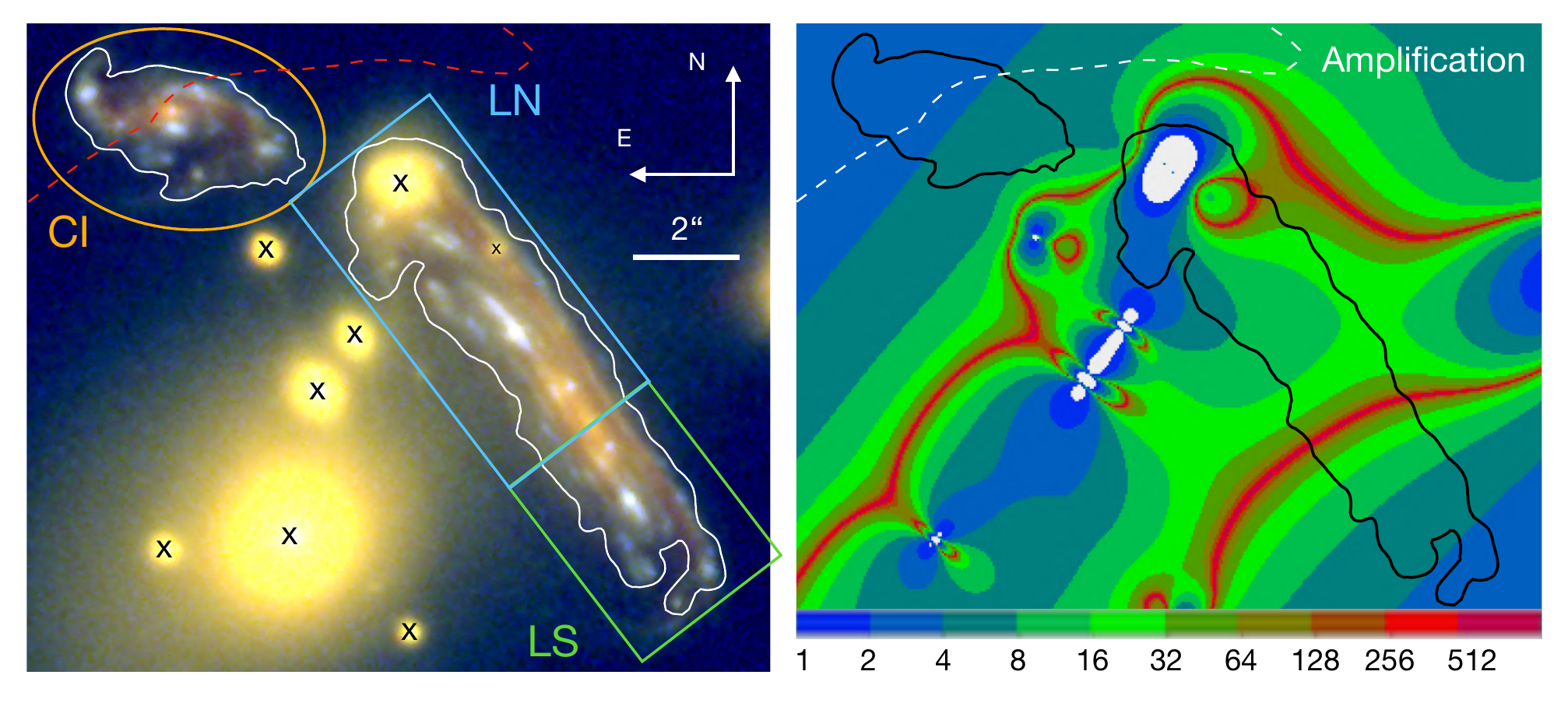}
    \caption{(Left panel): HST observations of A521-sys1 (RGB colors are given by F160W, F105W and F390W, respectively). The division in three sub--regions discussed in the paper (counter--image CI, lensed--north LN and lensed--south LS) is also shown. Foreground cluster members are marked with an `x'. (Right panel): amplification map, showing the magnification factor, $\rm \mu$, at any position of A521-sys1. A white (black) contour enclosing the galaxy is over--plotted to both panels to make the comparison between them easier. The dashed red (white) line delimits the CI region with multiple images.}
    \label{fig:hstdata}
\end{figure*}

A521-sys1 appears as a series of multiple distorted images (Fig.~\ref{fig:hstdata}); in particular, a complete counter--image of A521-sys1 is observed to the north-east of the brightest cluster galaxy (BCG), and two additional, partially lensed images of the galaxy (one mirrored) are observed west and north-west of the BCG. We will refer to these different images of the A521-sys1 galaxy as counter--image (CI), lensed--north (LN) and lensed--south (LS), as showed in the left panel of Fig.~\ref{fig:hstdata}. The division between LN and LS is traced following the critical line, with the help of the lens model described in Section~\ref{sec:lens_model}. 

Black crosses in the left panel of Fig.~\ref{fig:hstdata} mark the position of bright foreground or cluster galaxies in the field of view; the relative contribution from such galaxies to the A521-sys1 photometry increases with the wavelength of the respective observation. 

On the other hand, they would have a strong effect on the analysis of the clumpiness of A521-sys1; for this reason their flux is subtracted in the latter analysis (see Section~\ref{sec:clumpiness} for more details). Single--band observations are shown in Fig.~\ref{fig:390data} for F390W and in Appendix~\ref{sec:app:completetab} for the other filters.

\subsection{Ancillary data}
A521-sys1 was observed with VLT-MUSE as part of the MUSE Guaranteed Time Observations (GTO) Lensing Clusters Programme (ID: 100.A-0249, PI: Richard). Observations and data reductions are presented in \citet{patricio2018}. The PSF of MUSE observation is $0.57''$, almost 5 times larger than the PSF of HST-F390W, the reference filter for our clump extraction and analysis, and therefore MUSE data cannot be used for the study of individual clumps. We use MUSE data to estimate the average extinction in radial regions of the galaxy, using the relative strength of nebular emission lines, as described in Appendix~\ref{sec:app:extinction}. 

ALMA observations of A521-sys1 were acquired during Cycle 4 (ID: 2016.1.00643.S) in band 6, targeting the CO(4-3) emission line, and were presented in \citet{girard2019} and in \citet{nagy2021}, along with their data reduction analysis. 
The high resolution of the ALMA observations (beam size: $0.19''\times0.16''$) allows the study of molecular gas on the same scales as the stellar content; the study of the individual giant molecular clouds (GMCs) is presented in \citet{dessauges2022}.

\subsection{Gravitational lens model}
\label{sec:lens_model}
The gravitational lens model used in this paper to recover the source properties of the individual clumps was constructed using the \textsc{lenstool}\footnote{\hyperlink{https://projets.lam.fr/projects/lenstool/wiki}{https://projets.lam.fr/projects/lenstool/wiki}} software \citep{jullo2007}, and is described in detail in Appendix~\ref{sec:app:lensmodel}.
Its final Root Mean Square (RMS) accuracy in the image plane, based on the positions of 33 multiple images, is $0.08''$ i.e. comparable to the pixel scale of the HST data. 

The amplification map, showing the magnification factor, $\rm \mu$, associated to each position of A521-sys1, is showed in the right panel of Fig.~\ref{fig:hstdata}. The magnification factor in the CI region ranges from $\rm \mu\sim2$ to $\rm \mu\sim6$, with a median of 4 and a shallow spatial gradient across the image. In LN and LS, magnifications are typically higher (median $\rm \mu\sim10$) with sub–regions reaching values $\rm \mu>20$ for the majority of the arc.

\section{Data analysis}
\label{sec:datanalysis}
\subsection{Clump extraction}\label{sec:sextraction}
\begin{figure*}
    \centering
    \includegraphics[width=0.99\textwidth]{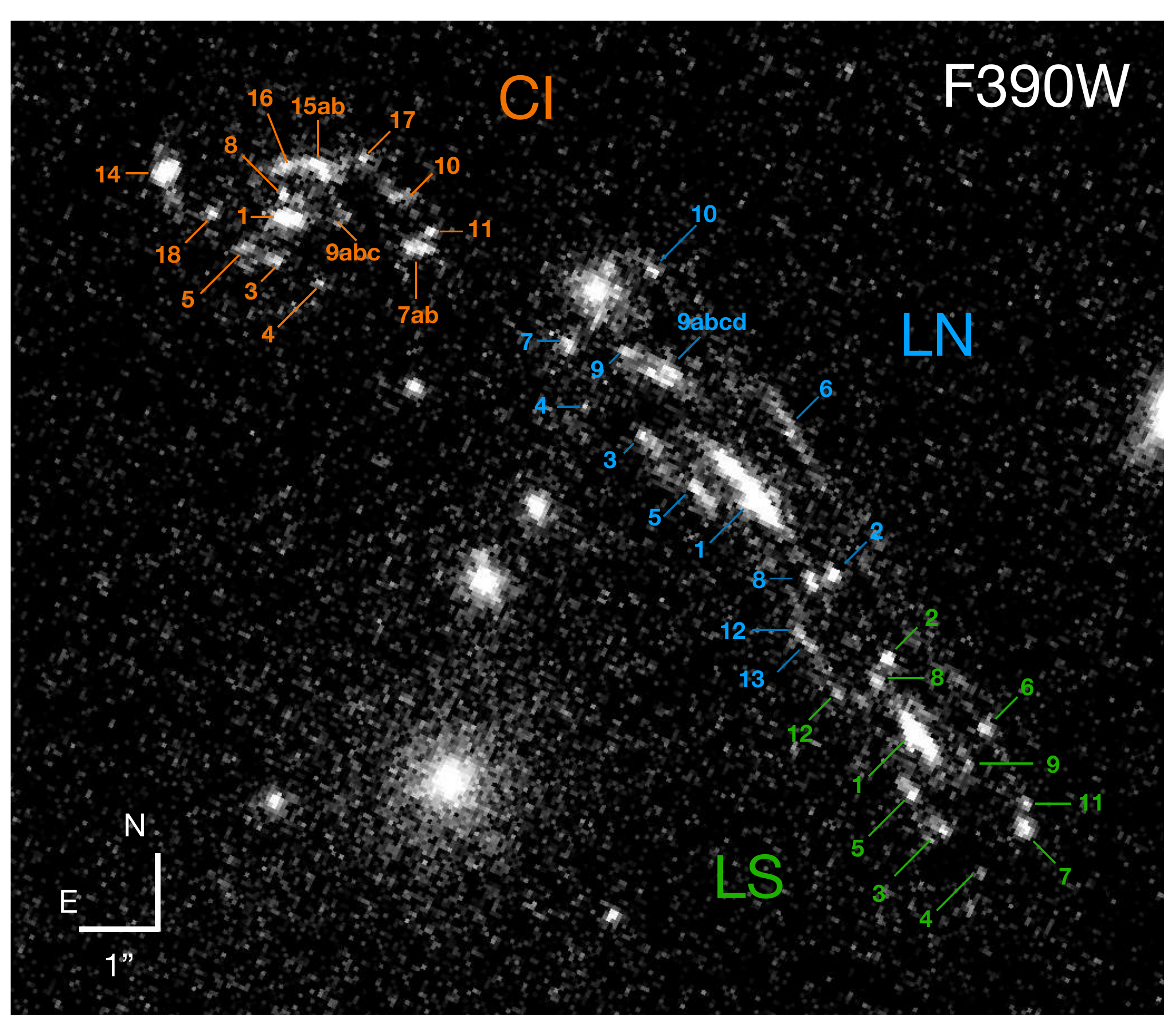}
    \caption{Names and locations of the clumps in A521-sys1 on the F390W data. The coordinates and main properties of the clump sample are given in Tab.~\ref{tab:main_prop}, the complete photometry in all filters is given in Tab.~\ref{tab:photometry} of Appendix~\ref{sec:app:completetab}.}
    \label{fig:390data}
\end{figure*}

We use the F390W observations, corresponding to rest--frame UV, as reference to extract the clump catalog. F390W is the filter where the clumps are more easily detectable; the galaxy looks less clumpy when moving to longer wavelengths, as also quantitatively shown in the clumpiness analysis of Section~\ref{sec:clumpiness}.
We use the \texttt{SExtractor} software \citep{bertin1996} on a portion of the F390W data centred on A521-sys1 to extract sources that have a minimum of 4 pixels with $\rm S/N>3\sigma$ in background-subtracted images. 
The local background is estimated using a convolution grid of 30 pixels ($\rm BACK\_SIZE=30$ in the configuration file); smaller grid would result in considering sources as part of the background, and consequently in removing them. 
Using the galaxy cluster mass model to trace the counter--images of all extracted sources, we notice that one clump (clump `9') is detected in LN but its counter--images in CI and LS are not, the latter being below the detection limits of \texttt{SExtractor}; those were therefore added manually to the catalog.
We also search the images in redder filters looking for red clumps that would have missed in the extraction in F390W; only one such source is found (clump `4'), lying below the detection limit in F390W but bright in all other filters, which is added to the sample. 
Finally, by a visual inspection we verify that none of the UV clump clearly recognizable by eye is missed by our extraction and we remove foreground galaxies from the catalog. 
The final catalog counts 18 unique clumps. Many of those have multiple images; different images of the same clump have been assigned the same ID number, preceded by the sub-region where the image is observed (e.g. `$\rm ci\_1$', `$\rm ln\_1$' and `$\rm ls\_1$' are the same source `1' observed in the counter--image, the lensed-north and the lensed--south regions, respectively). The cross--identification of various images of the same clump  was done with the help of the lens model. 
In addition, some clumps were divided in multiple sub--peaks in the photometric analysis (see Section~\ref{sec:fitmultiple}); each peak was considered as a single entry in the catalog and we add letters to the ID to differentiate the entries (e.g. clumps `ci\_7a' and `ci\_7b' are two peaks of clump `7'). As consequence, the final catalog counts 45 entries, spread across the 3 images of A521-sys1. The position of all clumps on the F390W observations is shown in Fig.~\ref{fig:390data}.

\subsection{Clump modeling}
\label{sec:modelling}
We modelled the clumps on the image plane, deriving their sizes and magnitudes on the observed data, and later convert those to intrinsic values. 
We assume that clumps have intrinsic 2D Gaussian profiles in the source plane and that local lensing transformations still result in Gaussian ellipses in the image plane; in order to describe the observed clump light profile we convolve the 2D Gaussian profiles with the instrumental point spread function i.e. the response of the instrument. Asymmetric gaussian profiles are used to take into account both intrinsic asymmetries in the clump shapes and distortions introduced by the lensing. 

We perform the fits in cutouts of $9\times9$ pixels, centered on each of the clumps. In order to take into account possible background luminosity in the vicinity of the clumps, we add to the clump model a $\rm 1^{st}$ degree polynomial function, described by three parameters ($c_0$, $c_x$ and $c_y$). The choice of a non-uniform background helps avoiding the contamination to the fit from the tails of nearby bright sources.
The `observable' model, $M_f$, to be fitted to the data in filter $f$ can be therefore summarized as:
\begin{multline}
M_f(x,y|x_0,y_0,F,\sigma_x,axr,\theta,c_0,c_x,c_y) = \\ 
F\cdot K_f\ast G_{2D}(x_0,y_0,\sigma_x,axr,\theta)+c_0+c_xx+c_yy,
\end{multline}
where $K_f$ parametrizes the PSF in filter $f$ (as described in Section~\ref{sec:data_hst}) and $F$ parametrizes the observed flux (both the PSF and the gaussian model are normalized). The gaussian model, $G_{2D}$, is parametrized by the minor standard deviation $\sigma_x$, the axis ratio $axr$ defined by $axr\equiv\sigma_y/\sigma_x>1$ and the angle $\theta$, using the \texttt{astropy.modeling} package; by construction we impose that $\sigma_x$ refers to the minimum axis of the 2D gaussian function.
The fit is performed using a least-squared method via the \texttt{python} package \texttt{lmfit} \citep{lmfit}. We calculate and report $\rm1\sigma$ uncertainties derived from the covariance matrix. 

Each clump was fitted separately in each of the filters. Due to the clumps being more easily detectable in F390W, we use the latter as the reference one for determining the clump position and size. As first step, we fit the clumps in F390W  leaving all parameters free. The F390W data, along with clump best--fit models and residuals, are shown in Appendix~\ref{sec:app:completetab}.
For the fit in F606W, F814W, F105W and F160W, we keep the resulting values for the clump centre ($x_0$ and $y_0$) and its size ($\sigma_x$, $axr$, and $\theta$) as fixed parameters, i.e. we fix the gaussian shape and its position, leaving free only the flux (and the background parameters). This choice assumes that the source has intrinsically the same shape and size in all bands.

\subsubsection{Fitting together multiple sources}
\label{sec:fitmultiple}
A variation to the fitting method described above is employed for clumps whose central positions are less than 4 pixels apart. Due to such closeness the fit of each of the sources would be greatly affected by the other one, bringing unreliable results. For this reason we choose to fit nearby clumps in a single fitting run, by using a larger cutout of $11\times11$ px and modelling two separate gaussians within it; this kind of fit applies only to 3 pairs of sources. In naming these cases we use the same numeric ID for the two sources, adding a letter to differentiate them (e.g. clumps `ci\_7a' and `ci\_7b' have been fitted together). In doing so we are therefore considering the two as separate peaks of the same source; this choice is driven solely by the resolution of our data. 
An extreme case is clump `9', that, while in the LS image it appears as a single peak, it can be separated into 4 different sub-peaks (plus a separate image) in LN and into 3 sub-peaks in CI. For the fit of its LN representation we choose to fit at the same time all 4 peaks in a $11\times11$ cutout, imposing circular symmetry for the sources. This last choice is motivated by the too large number of free parameters if asymmetric profiles were considered. The same approach is used to fit the 3 peaks in the CI region.

\subsubsection{Minimum resolvable $\rm \sigma_x$}
\label{sec:minreff}
Our fitting method has an intrinsic resolution limit driven mainly by the instrumental PSF, with a FWHM equal to 1.6, 1.9, 1.9, 3.7 and 4.0 px for F390W, F606W, F814W, F105W and F160W, respectively. 
The convolution of the PSF with very narrow gaussian functions will be indistinguishable from the PSF itself. To test what is the minimum size we can resolve, we simulate clumps with various combinations of $\sigma_x$ and axis ratios, add them on top of the galaxy observations and fit them in the same way we do for the real data. We derive a minimum resolvable size $\rm \sigma_{x,min}=0.4$ px for F390W.  
All the sources whose fit results in $\rm \sigma_{x}<0.4$ px will be considered as upper limits in size, as shown in Fig.~\ref{fig:sizemag_obs}.
More details on the process to derive $\rm \sigma_{x,min}$ are given in Appendix~\ref{sec:app:reffmin}.

\subsubsection{Completeness of the sample}\label{sec:completeness}
We test the magnitude completeness of the clump sample by simulating clumps of various magnitudes, including them at random positions on top of the galaxy, and fitting them in the same way as for the real sources.
We estimate the completeness limit, $\rm lim_{com}$, as the magnitude above which the fit results become unreliable, using simulated sources of different sizes, $\rm \sigma_x=0.4$, $1.0$ and $2.0$ px, corresponding to $0.024"$, $0.06"$ and $0.12"$ respectively.
More details on the completeness test are given in Appendix~\ref{sec:app:completeness}.

The derived values for F390W are compared to the photometry of the actual clump sample in Fig.~\ref{fig:sizemag_obs}; for an easier comparison to clump magnitudes we we corrected $\rm lim_{com}$ values by the Galaxy reddening in the figure. We find a completeness $\rm lim_{com}=27.4$ mag for point--like sources ($\rm \sigma_x\leq0.4$ px), consistent with the faintest unresolved clumps of our sample. This value is only slightly brighter than the minimum detectable magnitude ($\rm mag_{lim}$) discussed in Section~\ref{sec:data_hst}.
The completeness values get brighter for larger sources, namely $\rm lim_{com}=26.7$ mag and $25.2$ mag for sources with $\rm \sigma_x=1.0$ px ($0.06"$) and $2.0$ px ($0.12"$), respectively. These values are still consistent with the faintest clumps we observed at the corresponding sizes and suggest that $\rm lim_{com}$ traces the magnitudes of the sources which are $3\sigma$ above their local background, i.e. the lower limit chosen for extracting the clump catalog (as seen in Section~\ref{sec:sextraction}).

\subsection{Conversion to intrinsic sizes and magnitudes}\label{sec:convert_intrinsic}
The fluxes, F (in $\rm e^-/s$), are converted into observed AB magnitudes by considering the instrumental zeropoints relative to each filter (Tab.~\ref{tab:data}); the reddening introduced by the Milky Way ($0.29$, $0.19$, $0.11$, $0.07$ and $0.04$ magnitudes for F390W, F606W, F814W, F105W and F160W, respectively) is subtracted in each filter. 
The photometry of all A521-sys1 clumps is collected in Appendix~\ref{sec:app:completetab} for all filters.

In order to convert observed magnitudes into absolute ones we subtract the distance modulus ($44.3$ mag) and we add the $k$ correction, a factor $\rm 2.5\log(1+z)$. 
Concerning the clump sizes measured in F390W, we calculate the geometrical mean of the minor and major $\rm \sigma$ derived from the fit, i.e. $\rm \sigma_{xy}\equiv \sqrt{\sigma_x\sigma_y}=\sigma_x\sqrt{axr}$, and we convert it to an effective radius. In the case of the gaussian function, the effective radius is equivalent to the half width at half maximum, $\rm HWHM=FWHM/2$ and therefore $\rm R_{eff,xy}\equiv FWHM/2=\sigma_{xy}\sqrt{2\ln{2}}$. The conversion from pixels to parsec is $\rm 1\ px\equiv 498.5\ pc$, derived considering the angular diameter distance of the galaxy of $1713$ Mpc and the pixel scale of the observations, $0.06$ arcsec/px.

The fitting method and the steps just described return sizes and luminosities as observed in the image plane, i.e. after the effect of the gravitational lensing. In order to recover the intrinsic properties of the clumps, we consider the lensing model, described in detail in Appendix~\ref{sec:app:lensmodel}. First, we focus on the best fit model, resulting in the magnification map shown in Fig.~\ref{fig:hstdata} (right panel);
for each clump we identify the region enclosed within $\rm R_{eff}$ and use the median amplification value of the selection as the face--value considered for de-lensing sizes and luminosities. We use the standard deviation of the values within the selected region as a first estimate of the uncertainty on the magnification, $\rm \delta\mu_1$. 
Second, we consider 500 models from the MCMC chain produced with \textsc{lenstool} (Appendix~\ref{sec:app:lensmodel}). These models sample the posterior distribution of each parameter in the mass model of the cluster. For each of those realisations, we re-measure the median amplification value of each clump and use their standard deviation as a measure of the uncertainties related to the best fit model, $\rm \delta\mu_2$. We have checked that for each clump the magnification of the best fit model is not biased against the median of the distribution of magnifications for the 500 models. 
We account for both the magnification uncertainty related to the clump extension ($\rm \delta\mu_1$) and the one related to the lens model uncertainties ($\rm \delta\mu_2$) by considering their sum root squared, $\rm \delta\mu = \sqrt{\delta\mu_1^2+\delta\mu_2^2}$.

Intrinsic luminosities and sizes are derived by dividing the observed quantities by the magnification value and by its square-root, respectively.
The final uncertainties combine both photometric and magnification uncertainties via the root sum squared. In this way they include possible magnification gradients close to the source positions; regions with higher magnifications also have a steeper $\rm \mu$ gradient, such that the sources within those regions have large uncertainties associated.

\subsection{Broadband SED fitting}\label{sec:BBSED}
We use the broadband photometry to estimate ages and masses of the clumps. The limited number of filters available, covering the rest--frame wavelength range $\sim1700-8500$ \AA, do not allow to fully break the degeneracy between ages and extinctions, nor to constrain the metallicity or the star formation history of the clumps. In order to mitigate the effect of degeneracies, we limit the number of free--parameters making some \textit{a--priori} assumptions. In detail, we use the Yggdrasil stellar population synthesis code \citep{zackrisson2011};
Yggdrasil models are based on Starburst99 Padova-AGB tracks \citep{leitherer1999,vazquez2005} with a universal \citet{kroupa2001} initial mass function (IMF) in the interval $\rm 0.1-100\ M_\odot$. Starburst99 tracks are processed through \texttt{Cloudy} software \citep{ferland2013} to obtain the evolution of the nebular continuum and line emission, produced by the ionized gas surrounding the clumps. Yggdrasil adopts a spherical gas distribution around the emitting source, with hydrogen number density $\rm n_H = 10^2\ cm^{-3}$ and gas filling factor (describing the porosity of the gas) $\rm f_{fill} = 0.01$, typical of \HII\ regions \citep{kewley2002}, and assumes that the gas and the stars form from material of the same metallicity. We choose the models with a gas covering fraction $\rm f_{cov} = 0.5$, i.e. only $50\%$ of the Lyman continuum photons produced by the central source ionize the gas, but we point out that our fit results are basically not affected by the choice of $\rm f_{cov}$. 

As fiducial model we consider the stellar tracks obtained assuming a continuum star formation for 10 Myr (C10), a Milky Way extinction law \citep{cardelli1989} and Solar metallicity ($\rm Z=0.02$ as suggested by the analysis in \citealp{patricio2018}).
The C10 assumption is motivated by most of the clumps in the sample having physical sizes of $\sim100$ pc. For star--forming regions at larger scales we can expect more complex star formation histories (SFHs), in particular prolonged star--formation events; the opposite is true at smaller scales, for stellar clusters and small clumps (few tens of parsecs), where the hypothesis of instantaneous burst (`single stellar population' model, or SSP) is usually assumed.
Our clump sample contains sources with a wide range of physical scales (Section~\ref{sec:sizelum}); for this reason, in addition to the fiducial model, we consider a SSP model and a model assuming a continuum star formation for 100 Myr (C100). The comparison between these two `extreme' assumptions will give the magnitude of the effect of the SFH on the derived properties. 

To test the effects of the choice of the extinction curve, we consider a fourth model with the starburst curve \citep{calzetti2000} instead of the MW one.
Due to the uncertainties associated to the study of stellar metallicity in A521-sys1 in \citet{patricio2018}, we consider a further model, assuming sub--Solar metallicity ($\rm Z=0.008$).
All the models used in the SED-fitting are summarized in Tab.~\ref{tab:SED_models}.

Considering the assumptions described above, we are left with 3 free parameters in our fits, age, mass and extinction, parametrised by the color excess $\rm E(B-V)$.
The photometric data of our catalog are fitted to the spectra from the models considered using a minimum-$\chi^2$ technique. Only sources with magnitude uncertainties below $0.6$ mag in more than 3 filters have been fitted. 
We report in Section~\ref{sec:results_sed} the face--values relative to the minimum reduced $\rm \chi^2$ ($\rm \chi^2_{red.,min}$) for each clump, and we assign to it an uncertainty given by the range in properties spanned by the results satisfying the condition $\rm \chi^2_{red.}\le 1.07$ (consistent with $\rm 1\sigma$ uncertainties for fits with two degrees of freedom). In cases where the minimum \chisqred\ is above that threshold, we retained within the uncertainty range the values within $10\%$ of $\rm \chi^2_{red.,min}$.
The differences in derived properties for each clump given by the choice of the different models of Tab.~\ref{tab:SED_models} are considered and discussed in Section~\ref{sec:results_sed}. 

\begin{table}
    \centering
    \begin{tabular}{lllll}
       Model & SFH & Ext. curve  & Z  \\
       \hline
       C10 (reference) & Const. SFR (10 Myr) & MW  & 0.020 \\
       SSP & Single burst & MW & 0.020 \\
       C100 & Const. SFR (100 Myr) & MW & 0.020 \\
       C10-SB & Const. SFR (10 Myr) & Starburst  & 0.020 \\
       C10-008 & Const. SFR (10 Myr) & MW  & 0.008 \\
       \hline
       \end{tabular}
    \caption{Models and relative assumptions used in the broad--band SED-fitting process. 
    In all cases spectra from the Yggdrasil stellar population synthesis code \citep{zackrisson2011} (based on Starburst99 Padova-AGB tracks), with \citet{kroupa2001} IMF, are considered.}
    \label{tab:SED_models}
\end{table}

\subsection{Alternative clump selection and photometry}\label{sec:alternative}
Literature studies offer a variety of methods for extracting clump samples and analyzing them. 
To test the reliability of our extraction and photometric analysis we consider an alternative method: we draw elliptical regions that best follow $3\sigma$ contours above the level of the galaxy background to define the clump extent and measure the flux of the clumps within those regions.
Such method is used in the analysis on GMC complexes from CO data \citep[e.g][]{dessauges2019,dessauges2022} but has also been applied to the study of stellar clumps \citep[e.g.][]{cava2018}.  
More details on the source extraction, size and photometry measurements with this alternative method are given in Appendix~\ref{sec:app:alternative}, while the derived properties and their differences to the ones of the reference method are discussed in Section~\ref{sec:alternative_results}.

\section{Photometric Results}
\label{sec:results_phot}
\begin{table*}
    \centering
    \begin{tabular}{lccrrrrrrrr}
        \multicolumn{1}{c}{ID} & \multicolumn{1}{c}{RA} & \multicolumn{1}{c}{Dec} & \multicolumn{1}{c}{$\mu$} & \multicolumn{1}{c}{$\rm R_{eff}$} & \multicolumn{1}{c}{$\rm Mag_{UV}$} & \multicolumn{1}{c}{Age} & \multicolumn{1}{c}{$\rm log(M)$} & \multicolumn{1}{c}{E(B-V)} & \multicolumn{1}{c}{$\rm log\langle\Sigma_M\rangle$} & \multicolumn{1}{c}{$\rm T_{cr}$} \\ 
        \ & \multicolumn{1}{c}{[hh:mm:ss]} & \multicolumn{1}{c}{[hh:mm:ss]} & \ & \multicolumn{1}{c}{[pc]} & \multicolumn{1}{c}{[AB]} & \multicolumn{1}{c}{[Myr]} & \multicolumn{1}{c}{[$\rm M_\odot$]} & \multicolumn{1}{c}{[mag]} & \multicolumn{1}{c}{[$\rm M_\odot pc^{-2}$]} & \multicolumn{1}{c}{[Myr]}  \\ 
        \multicolumn{1}{c}{(0)} & \multicolumn{1}{c}{(1)} & \multicolumn{1}{c}{(2)} & \multicolumn{1}{c}{(3)} & \multicolumn{1}{c}{(4)} & \multicolumn{1}{c}{(5)} & \multicolumn{1}{c}{(6)} & \multicolumn{1}{c}{(7)} & \multicolumn{1}{c}{(8)} & \multicolumn{1}{c}{(9)} & \multicolumn{1}{c}{(10)} \\ 
\hline
\hline
ci\_1	& 4:54:07.0521	& -10:13:16.964	& $3.7^{\pm 0.2}$	& <$138.0^{\pm 3.7}$	& $-17.6^{\pm 0.1}$ & $4^{+2}_{-3}$  	& $7.38^{+0.11}_{-0.06}$     	& $0.22^{+0.01}_{-0.04}$    	& >$2.3^{+0.11}_{-0.06}$    	& <$1.7^{+0.1}_{-0.3}$    	 \\[1mm]
ci\_3	& 4:54:07.0607	& -10:13:17.565	& $3.9^{\pm 0.2}$	& $314.8^{\pm 89.0}$	& $-16.3^{\pm 0.2}$ & $30^{+10}_{-0}$  	& $7.89^{+0.05}_{-0.02}$     	& $0.18^{+0.01}_{-0.03}$    	& $2.1^{+0.25}_{-0.25}$    	& $3.2^{+1.4}_{-1.4}$    	 \\[1mm]
ci\_4	& 4:54:07.0179	& -10:13:17.879	& $4.8^{\pm 0.3}$	& $132.5^{\pm 115.8}$	& $-15.0^{\pm 0.2}$ & $11^{+2}_{-3}$  	& $7.64^{+0.08}_{-0.06}$     	& $0.53^{+0.07}_{-0.09}$    	& $2.6^{+0.76}_{-0.76}$    	& $1.2^{+1.5}_{-1.5}$    	 \\[1mm]
ci\_5	& 4:54:07.0897	& -10:13:17.389	& $3.5^{\pm 0.2}$	& <$237.5^{\pm 60.9}$	& $-15.7^{\pm 0.2}$ & $50^{+0}_{-0}$  	& $7.28^{+0.00}_{-0.00}$     	& $0.00^{+0.00}_{-0.00}$    	& >$1.74^{+0.22}_{-0.22}$    	& <$4.2^{+1.6}_{-1.6}$    	 \\[1mm]
ci\_7a	& 4:54:06.9343	& -10:13:17.386	& $6.0^{\pm 0.5}$	& $196.4^{\pm 68.9}$	& $-15.5^{\pm 0.2}$ & $50^{+50}_{-49}$  	& $7.94^{+0.13}_{-0.50}$     	& $0.19^{+0.41}_{-0.13}$    	& $2.55^{+0.33}_{-0.58}$    	& $1.5^{+0.9}_{-0.8}$    	 \\[1mm]
ci\_7b	& 4:54:06.9206	& -10:13:17.390	& $6.3^{\pm 0.5}$	& $298.0^{\pm 99.3}$	& $-16.1^{\pm 0.2}$ & $11^{+69}_{-10}$  	& $7.28^{+0.54}_{-0.09}$     	& $0.31^{+0.16}_{-0.31}$    	& $1.53^{+0.61}_{-0.3}$    	& $6.0^{+3.0}_{-8.0}$    	 \\[1mm]
ci\_8	& 4:54:07.0529	& -10:13:16.650	& $3.5^{\pm 0.2}$	& <$138.0^{\pm 57.7}$	& $-16.1^{\pm 0.1}$ & $20^{+0}_{-5}$  	& $8.29^{+0.07}_{-0.28}$     	& $0.47^{+0.06}_{-0.04}$    	& >$3.21^{+0.37}_{-0.46}$    	& <$0.6^{+0.4}_{-0.4}$    	 \\[1mm]
ci\_9a	& 4:54:07.0006	& -10:13:16.819	& $4.1^{\pm 0.2}$	& <$115.7^{\pm 3.3}$	& $-14.5^{\pm 0.3}$ & $15^{+5}_{-1}$  	& $6.91^{+0.34}_{-0.13}$     	& $0.41^{+0.08}_{-0.09}$    	& >$1.98^{+0.34}_{-0.13}$    	& <$2.2^{+0.3}_{-1.3}$    	 \\[1mm]
ci\_9b	& 4:54:06.9922	& -10:13:16.951	& $4.3^{\pm 0.3}$	& $148.9^{\pm 41.8}$	& $-15.0^{\pm 0.3}$ & $50^{+0}_{-10}$  	& $6.89^{+0.04}_{-0.02}$     	& $0.00^{+0.05}_{-0.00}$    	& $1.75^{+0.25}_{-0.24}$    	& $3.3^{+1.4}_{-1.4}$    	 \\[1mm]
ci\_9c	& 4:54:07.0007	& -10:13:17.050	& $4.3^{\pm 0.3}$	& <$113.3^{\pm 3.4}$	& $-14.7^{\pm 0.3}$ & $60^{+40}_{-10}$  	& $7.06^{+0.20}_{-0.05}$     	& $0.00^{+0.04}_{-0.00}$    	& >$2.15^{+0.2}_{-0.06}$    	& <$1.8^{+0.1}_{-0.5}$    	 \\[1mm]
ci\_10	& 4:54:06.9492	& -10:13:16.684	& $4.7^{\pm 0.3}$	& $163.2^{\pm 124.4}$	& $-15.0^{\pm 0.4}$ & $14^{+26}_{-5}$  	& $7.36^{+0.54}_{-0.08}$     	& $0.40^{+0.17}_{-0.15}$    	& $2.14^{+0.86}_{-0.67}$    	& $2.2^{+2.5}_{-3.7}$    	 \\[1mm]
ci\_11	& 4:54:06.9141	& -10:13:17.163	& $5.9^{\pm 0.4}$	& $111.4^{\pm 26.8}$	& $-15.2^{\pm 0.2}$ & $12^{+18}_{-11}$  	& $6.84^{+0.47}_{-0.08}$     	& $0.26^{+0.14}_{-0.18}$    	& $1.95^{+0.52}_{-0.22}$    	& $2.3^{+0.8}_{-2.4}$    	 \\[1mm]
ci\_14	& 4:54:07.1624	& -10:13:16.335	& $2.7^{\pm 0.1}$	& $448.6^{\pm 46.3}$	& $-18.1^{\pm 0.1}$ & $40^{+0}_{-27}$  	& $8.61^{+0.04}_{-0.47}$     	& $0.13^{+0.15}_{-0.02}$    	& $2.5^{+0.1}_{-0.48}$    	& $2.4^{+0.9}_{-0.4}$    	 \\[1mm]
ci\_15a	& 4:54:07.0211	& -10:13:16.236	& $3.6^{\pm 0.2}$	& <$278.1^{\pm 7.3}$	& $-17.1^{\pm 0.1}$ & $5^{+2}_{-1}$  	& $7.76^{+0.02}_{-0.08}$     	& $0.40^{+0.01}_{-0.05}$    	& >$2.08^{+0.03}_{-0.09}$    	& <$3.1^{+0.3}_{-0.1}$    	 \\[1mm]
ci\_15b	& 4:54:07.0140	& -10:13:16.392	& $3.7^{\pm 0.2}$	& $137.3^{\pm 3.6}$	& $-15.6^{\pm 0.1}$ & $20^{+0}_{-0}$  	& $7.73^{+0.02}_{-0.04}$     	& $0.31^{+0.01}_{-0.02}$    	& $2.66^{+0.03}_{-0.04}$    	& $1.1^{+0.1}_{-0.1}$    	 \\[1mm]
ci\_16	& 4:54:07.0497	& -10:13:16.259	& $3.4^{\pm 0.2}$	& $577.3^{\pm 115.7}$	& $-17.3^{\pm 0.2}$ & $60^{+0}_{-0}$  	& $8.14^{+0.02}_{-0.00}$     	& $0.00^{+0.01}_{-0.00}$    	& $1.82^{+0.18}_{-0.17}$    	& $6.0^{+1.8}_{-1.8}$    	 \\[1mm]
ci\_17	& 4:54:06.9778	& -10:13:16.144	& $3.9^{\pm 0.2}$	& <$126.3^{\pm 81.2}$	& $-15.3^{\pm 0.2}$ & $4^{+2}_{-1}$  	& $6.76^{+0.08}_{-0.09}$     	& $0.27^{+0.04}_{-0.06}$    	& >$1.76^{+0.56}_{-0.57}$    	& <$3.0^{+2.9}_{-2.9}$    	 \\[1mm]
ci\_18	& 4:54:07.1194	& -10:13:16.912	& $3.1^{\pm 0.1}$	& $178.2^{\pm 70.9}$	& $-16.1^{\pm 0.1}$ & $20^{+0}_{-8}$  	& $7.76^{+0.02}_{-0.27}$     	& $0.24^{+0.12}_{-0.01}$    	& $2.46^{+0.35}_{-0.44}$    	& $1.6^{+1.0}_{-0.9}$    	 \\[1mm]
ln\_1	& 4:54:06.6065	& -10:13:20.897	& $11.0^{\pm 0.8}$	& <$80.1^{\pm 2.8}$	& $-17.6^{\pm 0.1}$ & $11^{+1}_{-2}$  	& $7.12^{+0.06}_{-0.06}$     	& $0.07^{+0.04}_{-0.05}$    	& >$2.52^{+0.07}_{-0.07}$    	& <$1.0^{+0.1}_{-0.1}$    	 \\[1mm]
ln\_2	& 4:54:06.5362	& -10:13:21.911	& $21.8^{\pm 1.6}$	& <$50.3^{\pm 9.9}$	& $-15.2^{\pm 0.1}$ & $11^{+1}_{-1}$  	& $6.58^{+0.04}_{-0.03}$     	& $0.19^{+0.03}_{-0.04}$    	& >$2.38^{+0.18}_{-0.17}$    	& <$0.9^{+0.3}_{-0.3}$    	 \\[1mm]
ln\_3	& 4:54:06.7141	& -10:13:20.003	& $6.4^{\pm 0.6}$	& $214.1^{\pm 72.9}$	& $-15.6^{\pm 0.3}$ & $7^{+93}_{-6}$  	& $7.48^{+0.54}_{-0.11}$     	& $0.47^{+0.13}_{-0.42}$    	& $2.02^{+0.61}_{-0.32}$    	& $2.9^{+1.5}_{-3.8}$    	 \\[1mm]
ln\_4	& 4:54:06.7692	& -10:13:19.588	& $3.4^{\pm 0.4}$	& <$140.2^{\pm 61.4}$	& $-15.0^{\pm 0.3}$ & $10^{+30}_{-9}$  	& $7.61^{+0.45}_{-0.12}$     	& $0.55^{+0.16}_{-0.28}$    	& >$2.52^{+0.59}_{-0.4}$    	& <$1.3^{+0.9}_{-1.5}$    	 \\[1mm]
ln\_5	& 4:54:06.6649	& -10:13:20.718	& $8.0^{\pm 0.7}$	& $170.2^{\pm 35.1}$	& $-15.3^{\pm 0.2}$ & $40^{+20}_{-32}$  	& $7.15^{+0.09}_{-0.53}$     	& $0.03^{+0.27}_{-0.03}$    	& $1.89^{+0.2}_{-0.56}$    	& $3.0^{+1.4}_{-1.0}$    	 \\[1mm]
ln\_6	& 4:54:06.5781	& -10:13:19.957	& $16.5^{\pm 1.0}$	& <$57.9^{\pm 36.6}$	& $-13.7^{\pm 0.3}$ & $4^{+1}_{-1}$  	& $6.40^{+0.06}_{-0.06}$     	& $0.37^{+0.03}_{-0.04}$    	& >$2.07^{+0.55}_{-0.55}$    	& <$1.4^{+1.3}_{-1.3}$    	 \\[1mm]
ln\_7	& 4:54:06.7850	& -10:13:18.739	& $1.5^{\pm 0.2}$	& $484.2^{\pm 112.4}$	& $-17.2^{\pm 0.2}$ & $1^{+99}_{-0}$  	& $8.07^{+0.40}_{-0.31}$     	& $0.46^{+0.05}_{-0.46}$    	& $1.91^{+0.44}_{-0.37}$    	& $5.0^{+2.1}_{-4.1}$    	 \\[1mm]
ln\_8	& 4:54:06.5573	& -10:13:22.002	& $20.0^{\pm 1.7}$	& <$98.1^{\pm 16.2}$	& $-14.5^{\pm 0.2}$ & $15^{+15}_{-4}$  	& $7.02^{+0.47}_{-0.07}$     	& $0.31^{+0.13}_{-0.10}$    	& >$2.24^{+0.49}_{-0.16}$    	& <$1.5^{+0.4}_{-1.5}$    	 \\[1mm]
ln\_9	& 4:54:06.7297	& -10:13:18.834	& $15.8^{\pm 7.8}$	& $115.8^{\pm 62.5}$	& $-14.4^{\pm 0.6}$ & $90^{+110}_{-89}$  	& $7.17^{+0.40}_{-0.83}$     	& $0.01^{+0.56}_{-0.01}$    	& $2.25^{+0.62}_{-0.95}$    	& $1.6^{+1.5}_{-1.8}$    	 \\[1mm]
ln\_9a	& 4:54:06.6850	& -10:13:19.162	& $119.4^{\pm 76.4}$	& $25.3^{\pm 9.5}$	& $-12.1^{\pm 0.7}$ & $---$			& $---$				& $---$				& $---$				& $---$ \\[1mm]
ln\_9b	& 4:54:06.6938	& -10:13:19.247	& $40.8^{\pm 11.4}$	& $42.8^{\pm 11.2}$	& $-13.2^{\pm 0.4}$ & $7^{+93}_{-6}$  	& $6.39^{+0.66}_{-0.21}$     	& $0.43^{+0.17}_{-0.43}$    	& $2.33^{+0.7}_{-0.31}$    	& $0.9^{+0.4}_{-1.7}$    	 \\[1mm]
ln\_9c	& 4:54:06.7074	& -10:13:19.115	& $106.9^{\pm 44.8}$	& $39.1^{\pm 11.9}$	& $-12.2^{\pm 0.5}$ & $50^{+252}_{-49}$  	& $6.82^{+0.44}_{-0.73}$     	& $0.26^{+0.50}_{-0.26}$    	& $2.84^{+0.52}_{-0.78}$    	& $0.5^{+0.3}_{-0.5}$    	 \\[1mm]
ln\_9d	& 4:54:06.6937	& -10:13:19.053	& $642.4^{+1338.6}_{-641.4}$	& $14.1^{+15.0}_{-14.1}$	& $-10.1^{+2.3}_{-1.1}$ & $---$			& $---$				& $---$				& $---$				& $---$ \\[1mm]
ln\_10	& 4:54:06.7057	& -10:13:17.705	& $2.7^{\pm 0.5}$	& <$237.1^{\pm 54.2}$	& $-16.3^{\pm 0.3}$ & $1^{+89}_{-0}$  	& $7.49^{+0.44}_{-0.33}$     	& $0.39^{+0.07}_{-0.39}$    	& >$1.95^{+0.48}_{-0.38}$    	& <$3.3^{+1.4}_{-3.1}$    	 \\[1mm]
ln\_12	& 4:54:06.5671	& -10:13:22.744	& $50.2^{\pm 11.4}$	& <$74.2^{\pm 15.1}$	& $-13.3^{\pm 0.3}$ & $3^{+7}_{-2}$  	& $6.07^{+0.24}_{-0.16}$     	& $0.34^{+0.07}_{-0.10}$    	& >$1.53^{+0.3}_{-0.24}$    	& <$3.0^{+1.0}_{-1.4}$    	 \\[1mm]
ln\_13	& 4:54:06.5553	& -10:13:22.927	& $225.6^{\pm 78.5}$	& $19.3^{+27.6}_{-19.3}$	& $-10.6^{\pm 0.5}$ & $---$			& $---$				& $---$				& $---$				& $---$ \\[1mm]
ls\_1	& 4:54:06.4604	& -10:13:24.085	& $7.8^{\pm 0.5}$	& <$84.1^{\pm 2.9}$	& $-17.3^{\pm 0.1}$ & $5^{+1}_{-2}$  	& $7.18^{+0.04}_{-0.03}$     	& $0.19^{+0.03}_{-0.02}$    	& >$2.53^{+0.05}_{-0.04}$    	& <$1.0^{+0.1}_{-0.1}$    	 \\[1mm]
ls\_2	& 4:54:06.4853	& -10:13:23.066	& $25.9^{\pm 2.5}$	& <$46.1^{\pm 4.1}$	& $-14.6^{\pm 0.1}$ & $3^{+1}_{-1}$  	& $6.61^{+0.07}_{-0.04}$     	& $0.34^{+0.02}_{-0.02}$    	& >$2.48^{+0.11}_{-0.09}$    	& <$0.8^{+0.1}_{-0.1}$    	 \\[1mm]
ls\_3	& 4:54:06.4322	& -10:13:25.466	& $4.3^{\pm 0.3}$	& <$142.2^{\pm 68.6}$	& $-15.5^{\pm 0.2}$ & $12^{+1}_{-0}$  	& $7.39^{+0.02}_{-0.04}$     	& $0.38^{+0.01}_{-0.06}$    	& >$2.29^{+0.42}_{-0.42}$    	& <$1.7^{+1.3}_{-1.3}$    	 \\[1mm]
ls\_4	& 4:54:06.3976	& -10:13:26.049	& $3.4^{\pm 0.2}$	& <$165.8^{\pm 54.3}$	& $-15.2^{\pm 0.2}$ & $8^{+32}_{-7}$  	& $7.71^{+0.41}_{-0.07}$     	& $0.58^{+0.14}_{-0.30}$    	& >$2.48^{+0.5}_{-0.29}$    	& <$1.5^{+0.7}_{-1.4}$    	 \\[1mm]
ls\_5	& 4:54:06.4618	& -10:13:24.964	& $5.5^{\pm 0.4}$	& <$142.0^{\pm 50.2}$	& $-15.4^{\pm 0.2}$ & $40^{+20}_{-32}$  	& $7.20^{+0.08}_{-0.54}$     	& $0.03^{+0.26}_{-0.03}$    	& >$2.1^{+0.32}_{-0.62}$    	& <$2.1^{+1.4}_{-1.2}$    	 \\[1mm]
ls\_6	& 4:54:06.3934	& -10:13:24.044	& $5.6^{\pm 0.4}$	& $276.8^{\pm 57.9}$	& $-16.0^{\pm 0.2}$ & $40^{+50}_{-33}$  	& $7.71^{+0.13}_{-0.51}$     	& $0.11^{+0.29}_{-0.10}$    	& $2.03^{+0.22}_{-0.54}$    	& $3.3^{+1.5}_{-1.2}$    	 \\[1mm]
ls\_7	& 4:54:06.3565	& -10:13:25.426	& $3.4^{\pm 0.2}$	& $404.7^{\pm 59.2}$	& $-17.0^{\pm 0.2}$ & $50^{+50}_{-10}$  	& $8.28^{+0.14}_{-0.05}$     	& $0.13^{+0.05}_{-0.12}$    	& $2.27^{+0.19}_{-0.14}$    	& $3.0^{+0.7}_{-0.9}$    	 \\[1mm]
ls\_8	& 4:54:06.4956	& -10:13:23.390	& $18.1^{\pm 1.9}$	& $115.6^{\pm 35.0}$	& $-14.3^{\pm 0.2}$ & $12^{+1}_{-2}$  	& $7.12^{+0.03}_{-0.04}$     	& $0.45^{+0.05}_{-0.05}$    	& $2.19^{+0.27}_{-0.27}$    	& $1.7^{+0.8}_{-0.8}$    	 \\[1mm]
ls\_9	& 4:54:06.4098	& -10:13:24.546	& $5.0^{\pm 0.3}$	& <$206.4^{\pm 78.0}$	& $-15.2^{\pm 0.3}$ & $5^{+2}_{-1}$  	& $6.98^{+0.03}_{-0.10}$     	& $0.38^{+0.03}_{-0.06}$    	& >$1.55^{+0.33}_{-0.34}$    	& <$4.9^{+2.8}_{-2.8}$    	 \\[1mm]
ls\_11	& 4:54:06.3552	& -10:13:25.084	& $3.6^{\pm 0.2}$	& <$168.8^{\pm 31.7}$	& $-15.7^{\pm 0.1}$ & $13^{+1}_{-1}$  	& $7.20^{+0.02}_{-0.02}$     	& $0.28^{+0.04}_{-0.04}$    	& >$1.94^{+0.16}_{-0.16}$    	& <$2.8^{+0.8}_{-0.8}$    	 \\[1mm]
ls\_12	& 4:54:06.5318	& -10:13:23.573	& $22.5^{\pm 2.4}$	& <$80.8^{\pm 26.1}$	& $-13.6^{\pm 0.3}$ & $3^{+3}_{-2}$  	& $5.91^{+0.14}_{-0.12}$     	& $0.26^{+0.03}_{-0.05}$    	& >$1.3^{+0.31}_{-0.3}$    	& <$4.1^{+2.0}_{-2.1}$    	 \\[1mm]
\hline
\end{tabular}
    \caption{Main intrinsic properties of the clumps in A521-sys1 and relative uncertainties: (1)-(2) RA and Dec coordinates; (3)-(5) magnification factors, effective radii and absolute UV magnitudes (from F390W), derived as described in \S~\ref{sec:modelling} and \S~\ref{sec:convert_intrinsic} and presented in \S~\ref{sec:sizelum}; (6)-(8) ages, masses and color excesses, for the reference SSP model (Tab.~\ref{tab:SED_models}), derived as described in \S~\ref{sec:BBSED} and presented in \S~\ref{sec:results_sed}; (9) mass surface densities, defined as $\rm \langle\Sigma_M\rangle = M/(2\pi R_{eff}^2)$ and discussed in \S~\ref{sec:masses}; (10) crossing times, defined as $\rm T_{cr}\equiv 10 \sqrt{{R^{3}_{eff}/GM}}$ and discussed in \S~\ref{sec:ages}. Upper and lower limits are indicated by `$<$' and `$>$', respectively. }
    \label{tab:main_prop}
\end{table*}

\subsection{UV sizes and magnitudes of the clumps}
\label{sec:sizelum}
\begin{figure}
    \centering
    \includegraphics[width=\columnwidth]{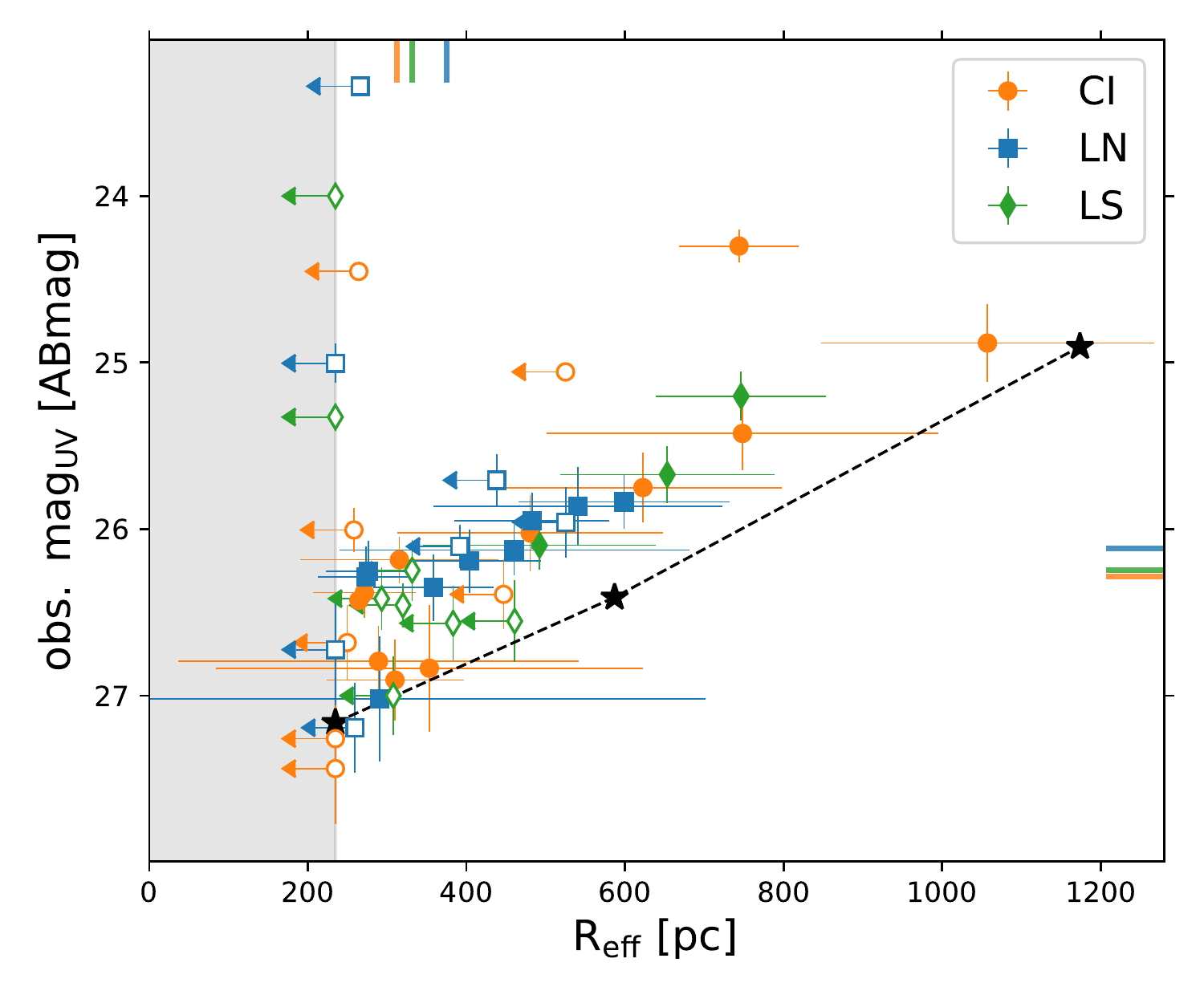}
    \caption{Apparent F390W magnitudes and sizes of the clumps (color--coded by the region where they belong) as they appear in the image--frame, i.e. before taking into account the de--lensing. The black stars joint by the dashed line are the completeness limits ($\rm lim_{com}$) discussed in Section~\ref{sec:completeness} and Appendix~\ref{sec:app:completeness}. The solid line at the top and on the right side of the panel indicate median values for size and magnitudes, respectively. Size upper limits (defined as $\rm \sigma_x<0.4$ px, see Section~\ref{sec:minreff}) are shown as empty markers. The grey area is below the size resolution limit ($<235$ pc).}
    \label{fig:sizemag_obs}
\end{figure}

We show the distribution of observed sizes and F390W magnitudes of the clumps in Fig.~\ref{fig:sizemag_obs}. Magnitudes have been considered after correcting for Galactic reddening. We plot apparent sizes, i.e. not corrected for the effect of magnification. The observed magnitudes ranges mostly between 27 and 25 mag (AB system), while sizes are mainly clustered below  600 pc. The minimum size, $235$ pc, is set by the choice of $\rm \sigma_{x,min}=0.4$ px described in Section~\ref{sec:minreff} and Appendix~\ref{sec:app:reffmin}. 
Many of the clumps observed have upper limits in size, i.e. they show a light profile consistent with the instrumental PSF, at least on their minor axis.
We do not observe systematic differences for clumps in different counter--images of the galaxy as can be verified comparing the median sizes and magnitudes reported at the top and on the right side of Fig.~\ref{fig:sizemag_obs}. In the same figure we report the completeness limits, $\rm lim_{com}$, derived in Appendix~\ref{sec:app:completeness} and discussed in Section~\ref{sec:completeness}, as black stars connected by a dashed line; all sources are above the $\rm lim_{com}$ value or consistent with it.

\begin{figure}
\centering
\subfigure{\includegraphics[width=0.50\textwidth]{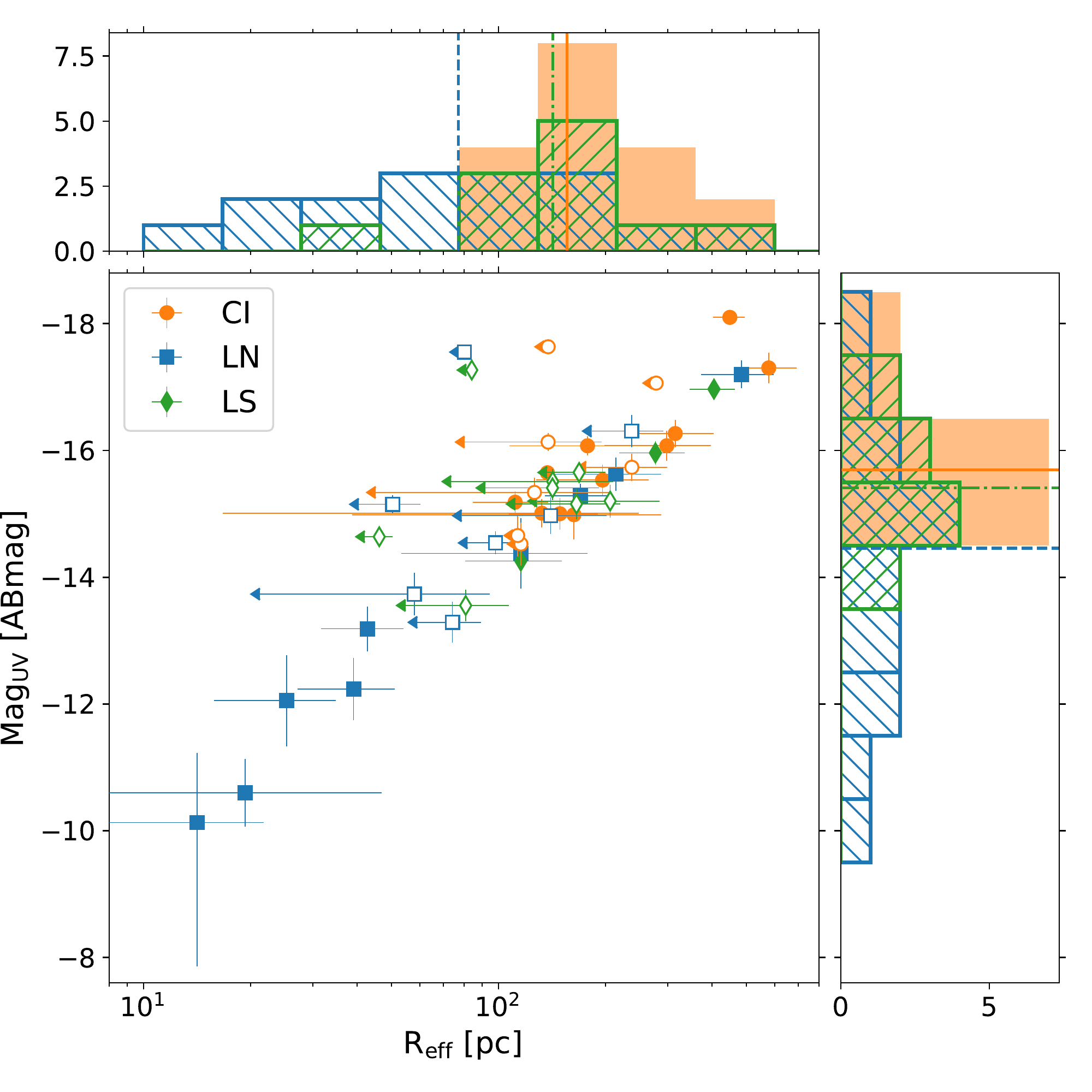}}
\subfigure{\includegraphics[width=0.50\textwidth]{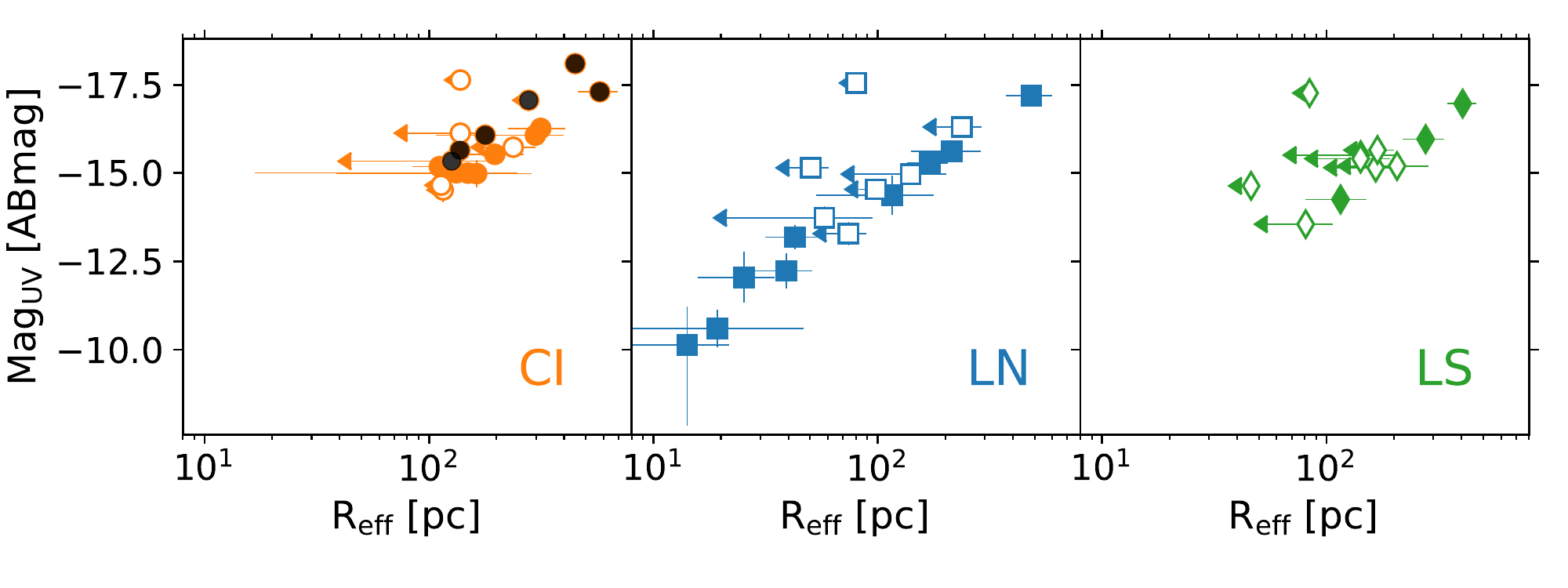}}
    \caption{Clumps' de--lensed sizes and absolute F390W magnitudes, color--coded by the sub--region where the clumps are observed (CI, LN and LS). Size upper limits are shown as empty markers and the length of their arrows reflects the size uncertainty (coming from the uncertainty in the magnification). The top and right histograms show the distributions of sizes and magnitudes in each of the sub--regions, with solid, dashed and dash--dotted lines giving the median values for CI, LN and LS, respectively. 
    The bottom panels show separately the sizes and magnitudes of sources in each of the sub-regions. The black sources in the CI panel (bottom--left) are clumps without a counterpart in either LN or LS.}
    \label{fig:sizemag_dl}
\end{figure}

Absolute UV magnitudes and clump sizes after correcting for the de-lensing are shown in Fig.~\ref{fig:sizemag_dl}. The values shown are the intrinsic sizes and luminosities of the clumps, also reported in Tab.~\ref{tab:main_prop}. De--lensing reveals a wide range of intrinsic properties spanning $\sim$~$8$ magnitudes and sizes between $\sim10$ and $\sim600$ pc. 
This suggests that we are observing a wide variety of clumps, from large star-forming regions on scales of hundreds of parsecs to almost star clusters. 
The distribution of sizes and magnitudes are summarized in histograms in Fig.~\ref{fig:sizemag_dl}; while clumps in the CI and LS regions have similar distribution of properties, clumps in the LN region are on average smaller and less bright, as suggested by the median values, $\rm med(R_{eff})=77$, $142$ and $156$ pc and $\rm med(Mag_{UV})=-14.5$, $-15.4$ and $-15.7$ mag for LN, LS and CI, respectively. Such difference is driven by the large amplification factors reached in some sub-regions of the LN image and, is specifically due to few sources in the LN that, thanks to such amplification, can be resolved in their sub--components; four of those sources are the peaks of the same clump `9', already described in Section~\ref{sec:fitmultiple}. We remind that many size measurements return only upper limits, affecting the distributions and median values just discussed. Nevertheless, the differences found between median values in CI, LN and LS remain even when removing clumps with size upper--limits.
Some of the brightest and largest sources in the CI are outside the region that produces multiple images (see Fig.~\ref{fig:hstdata}) and therefore do not have a counterpart either in LN or in LS (black circles in the bottom panel of Fig.~\ref{fig:sizemag_dl}). Neglecting clumps without multiple images would produce a minimal effect on the median values discussed above. 
Despite differences in median magnitude and sizes, clumps appear to share similar surface brightnesses between the three sub-regions, consistent with the conservation of surface brightness by gravitational lensing.

\subsection{Clumpiness}
\label{sec:clumpiness}
We measure the clumpiness of A521-sys1 in its three sub--regions for each filter; we consider clumpiness as the fraction of the galaxy luminosity coming from clumps, with respect to the total luminosity of the galaxy. This definition was already used in literature \citep[e.g.][]{messa2019} and in high redshift galaxies has been used also as a proxy for the cluster formation efficiency \citep{vanzella2021b}. 
To avoid contamination from nearby cluster members, we subtract them out of the observations using the \texttt{Ellipse} class in the \texttt{photutils} \texttt{python} library, providing the tools for an elliptical isophote analysis (following the methods described by \citealp{jedrzejewski1987}). Such subtraction was not needed in the F390W filter; at the redshift of A521-sys1 this filter corresponds to rest--frame FUV regime and therefore we do not expect significant contamination, as confirmed by visual inspection. 
The orange ellipse and blue and green boxes in Fig.~\ref{fig:hstdata} (left panel) mark the regions of the galaxy included in the extraction of the total flux of the system. These contours are driven by ensuring that all the extracted clumps lie within the area and are the same for all filters. 
We check that increasing the area covered by these regions we would add $<5\%$ of the galaxy flux, while including mostly local background emission. 
In order to exclude the contribution of local background from the measure of the galaxy flux we perform aperture photometry in the aforementioned elliptical and rectangular regions, employing an annular sky region with a width of $0.3"$ (5 px) around each of the three apertures.
A foreground galaxy is located on top of the northern part of the LN image. Despite the subtraction of the galaxy some residuals remains and for this reason a small circular region covering the galaxy is excluded from the flux measurement. Since we are interested in measuring the source-plane flux of the galaxy, the nearby region within the close critical line (in red in the magnification map of the right panel of Fig.~\ref{fig:hstdata}), corresponding to the position of the clumps ln\_9a,b,c,d, is also excluded, as it represent a further multiple image of a fraction of the A521-sys1 galaxy.

\begin{figure}
    \centering
    \includegraphics[width=\columnwidth]{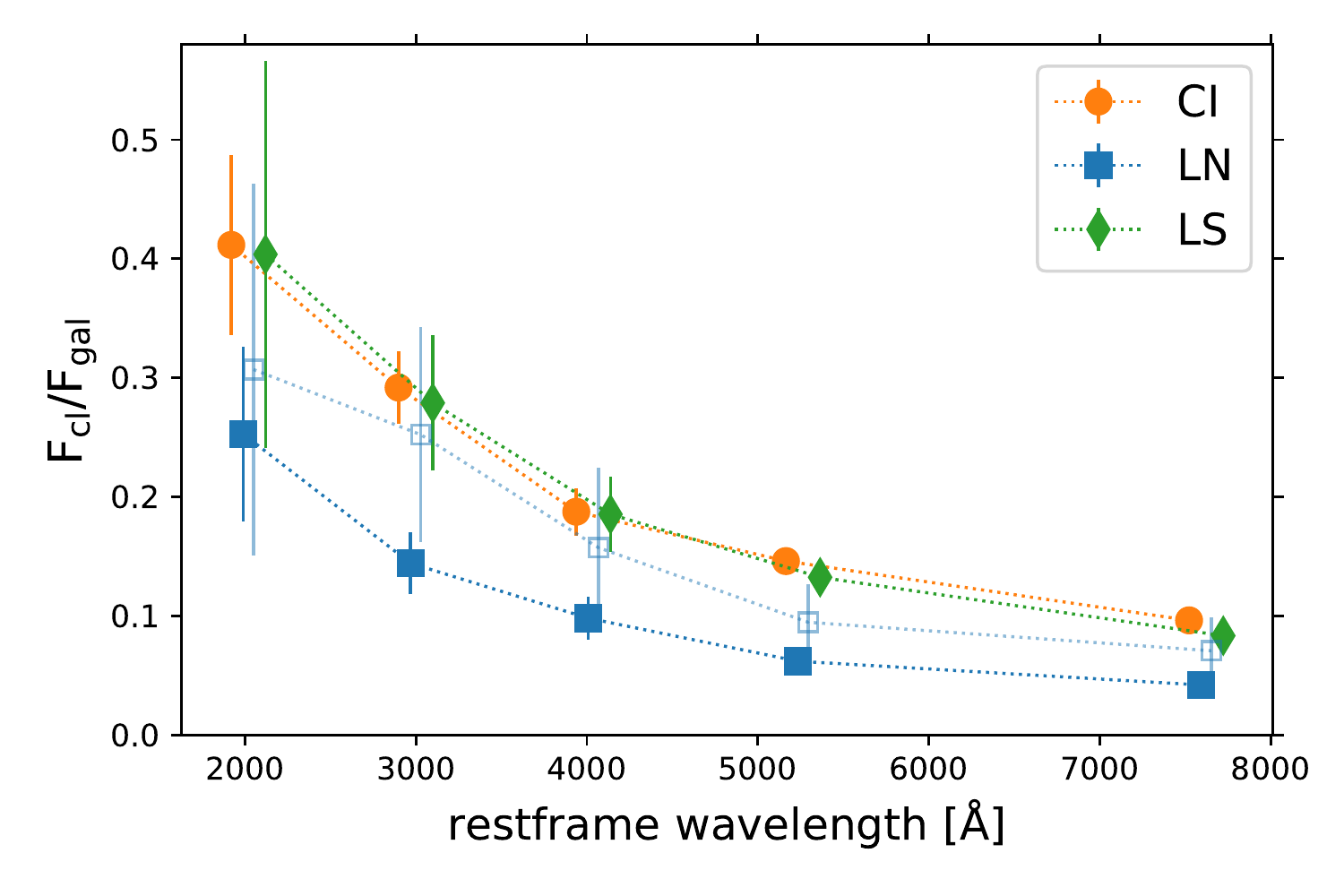}
    \caption{Clumpiness, measured as the ratio of the galaxy luminosity coming from clumps, in function of the rest--frame wavelengths for filters F390W, F606W, F814W, F105W and F160W from left to right, for the three images of A521-sys1 (orange circles for CI, blue squares for LN and green diamonds for LS). The clumpiness is measured using de-lensed galaxy and clump fluxes and therefore represent the source-plane value. The empty blue squares represent an alternative measure carried out excluding the northern part of the LN sub-region, possibly contaminated by the residual of a bright foreground galaxy. A small shift to the values on the x-axis have been applied for clarity of the plot, even though the same wavelengths are observed in CI, LN and LS.}
    \label{fig:clumpiness}
\end{figure}

The source-plane flux of each of the sub-regions is calculated by dividing the observed flux by its magnification, on a pixel-by-pixel basis.
The de-lensed flux of clumps is calculated by dividing the clump photometry by the amplification factor assigned to it, as already described in Section~\ref{sec:convert_intrinsic}. The ratios of these two measurements, for each filter and in each sub-region, give the clumpiness values, reported in Fig.~\ref{fig:clumpiness}.

The main trend observed is that clumpiness is high in the UV and decreases when moving to longer wavelength. This trend confirms what can be noticed from the single-band observations collected in Appendix~\ref{sec:app:completetab}, i.e. that the galaxy has a less clumpy appearance at redder wavelengths. The clumpiness in F390W, tracing rest-frame UV wavelengths ($\sim1900$ \AA) and therefore the massive stars from recent star-formation, suggests that a considerable fraction ($20\%-50\%$) of recent star formation is taking place in the observed clumps. Redder wavelengths trace older population of stars distributed along the entire galaxy. 
The clumpiness measurement for the LN sub-region is lower than the ones for CI and LS, though $2\sigma$ consistent in the bluest band. We attribute this difference mainly to the presence of residuals from the foreground galaxy in the north part of LN. This is confirmed by a second measure of the clumpiness in LN, done by excluding the northern part of the sub-region (the one encompassing the clumps ln\_4, ln\_7, ln\_9 and ln\_10); this further measure is plotted as empty blue markers in Fig.~\ref{fig:clumpiness}.
A second cause to this difference could be the lower average physical resolution reached in CI and LS, compared to LN, as literature studies have shown how low clump resolutions lead to over-estimate their contribution to the galaxy luminosity \citep{tamburello2017,messa2019}.  

\subsection{Color--color diagrams}
\label{sec:colorcolor}
\begin{figure}
    \centering
    \includegraphics[width=\columnwidth]{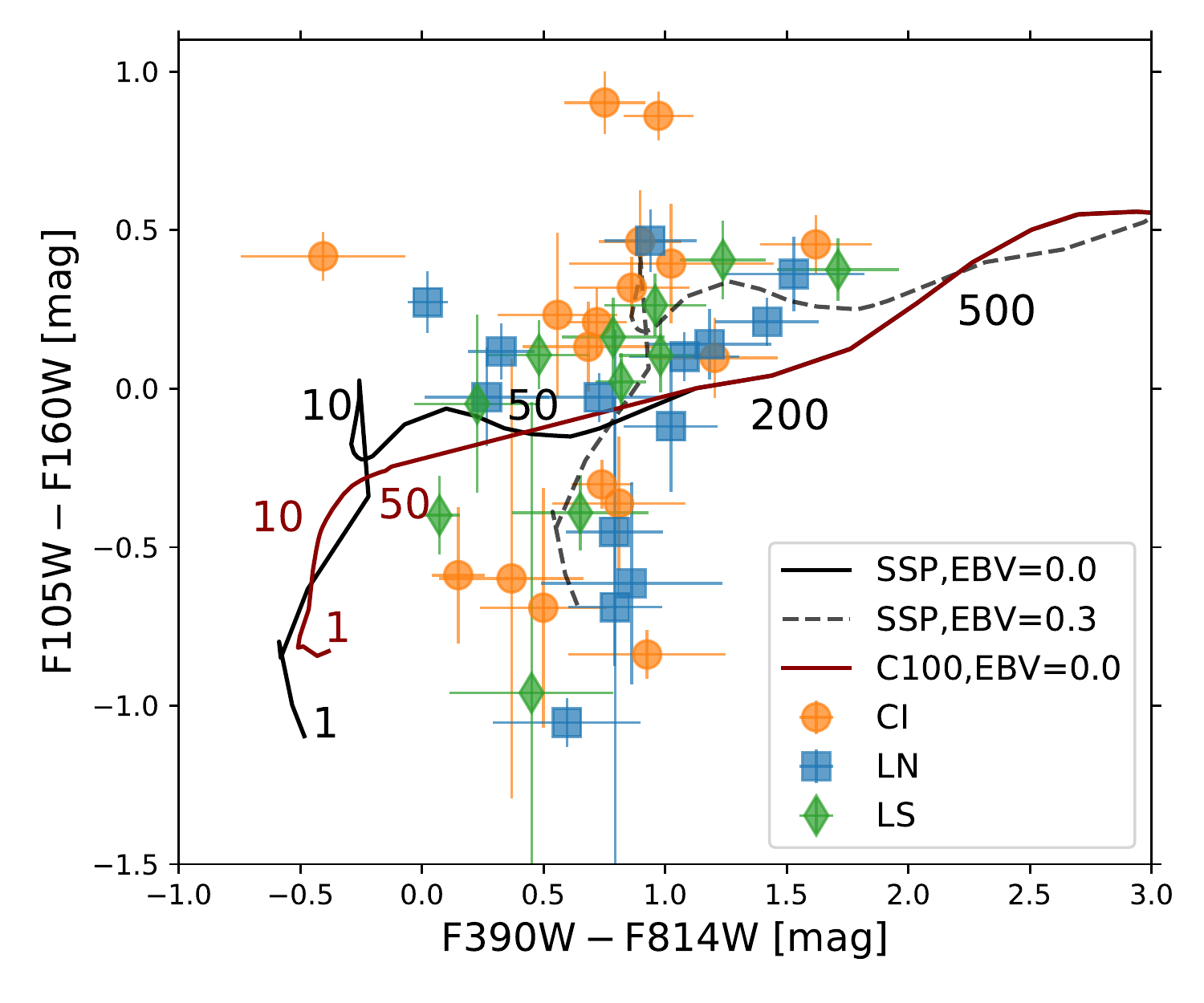}
    \caption{Color--color diagram of the clumps, with $UV$--$B$ and $V$--$I$ colors on $x$ and $y-$axis, respectively.  Over--imposed are the stellar track from the SSP and C100 models used for the SED fitting, as black and dark-red solid lines, respectively. The colors at the ages of 1, 10, 50, 200 and 500 Myr, are marked. The colors at 200 and 500 Myr are the same for the two models. The black dashed line show the SSP track at an extinction of $\rm E(B-V)=0.3$ mag (assuming Milky Way curve).}
    \label{fig:color_color}
\end{figure}

Color--color diagrams provide an intuitive way of estimating the age range covered by the clumps in our sample. In particular we focus in Fig.~\ref{fig:color_color} on the colors given by the filters F390W--F814W (on the x-axis) and F105W--F160W (on the y-axis); because of the rest-frame wavelengths probed by these filters ($\sim2000$, $\sim4000$, $\sim5300$ and $\sim7700$ \AA) we call these colors $UV-B$ (x-axis) and $V-I$ (y-axis), although no conversion to the Johnson filter system is applied. 
We over-plot on such a diagram the stellar evolution tracks used for the broadband SED fitting (described in Section~\ref{sec:BBSED}), and in particular the SSP and C100 tracks, i.e. the two extreme cases of SFH considered. We notice that they show similar behaviours, with the $UV-B$ color remaining almost constant for ages $1$ to $10$ Myr and then changing by $\sim3$ magnitudes for ages $10$ to $500$ Myr; the opposite is true for the $V-I$ color, that changes by $1$ mag in the first 10 Myr and then remains almost constant for the rest of the stellar evolution. Extinction moves the curve towards redder color and therefore specifically towards the top-right of the diagram in Fig.~\ref{fig:color_color}. 
The colors of our clump sample are scattered by $\sim1.5$ mag on both x and y axes. They all fall in the age range $\sim10-200$ Myr, if the no--extinction tracks are considered. However, while their scatter in the UV--B color can be due to a spread in ages in the range $10-200$ Myr, the large spread in $V-I$ suggests the presence of some extinction and of younger ages ($1-10$ Myr). In particular, data--points seem to be well aligned along the track with an extinction of $\rm E(B-V)=0.3$ mag.

\section{Results of Broadband-SED Fitting}
\label{sec:results_sed}
\begin{figure*}
\centering
\includegraphics[width=\textwidth]{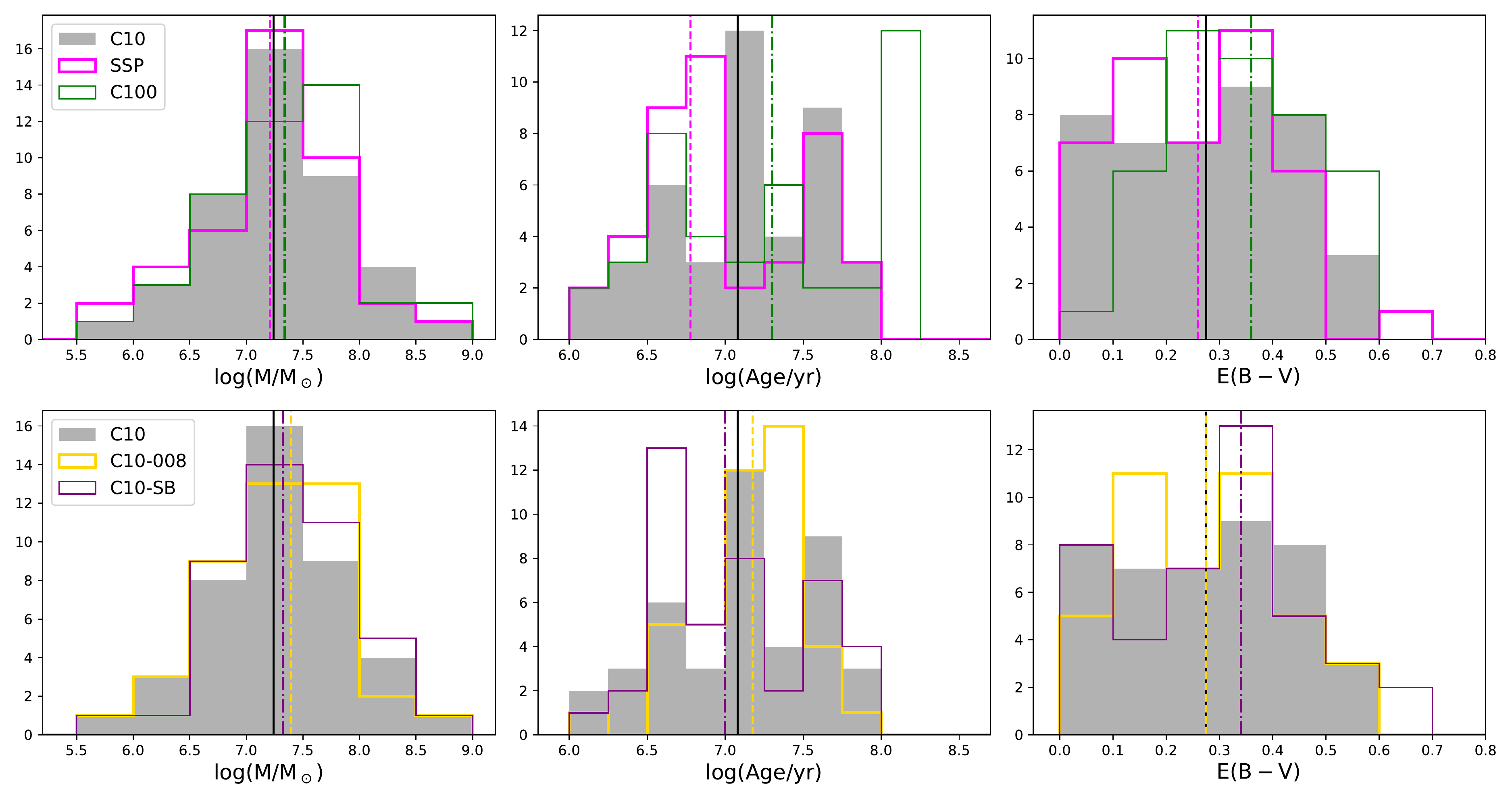}
\caption{Distributions of masses (left panels), ages (central panels) and color-excesses (right panels) for all the SED models listed in Tab.~\ref{tab:SED_models}. Vertical lines give the median value for each of the distributions. The 100 Myr continuum SF model (C100) gives on average the oldest ages, highest masses and highest extinctions; The instantaneous burst (SSP) gives the youngest ages but masses and extinctions similar to the reference C10 model. The assumption of either a \citealt{calzetti2000} extinction curve (C10-SB) or a lower metallicity model (C10-008) has on average a small effect on the derived properties.
The clump masses are much less sensitive than ages to the model assumption and remain overall stable within $\sim0.2$ dex.}
\label{fig:property_comparison_values}
\end{figure*}

Individual values for the derived masses, ages and extinctions in the case of our reference (SSP) model, are collected in Tab.~\ref{tab:main_prop}; their distributions are shown in Fig.~\ref{fig:property_comparison_values}.
Three clumps have detections in less than 4 filters and therefore were not fitted.
Masses range mainly between $\rm 10^6$ and $\rm 10^8\ M_\odot$, but extends up to $\rm \sim10^9\ M_\odot$; ages are distributed between $1$ and $100$ Myr, with the majority of clumps resulting younger than 20 Myr. 
Extinctions range between $\rm E(B-V)=0.0$ mag and $\rm E(B-V)=0.6$ mag, with a peak around $\rm E(B-V)\sim0.3$ mag. 

As discussed in Section~\ref{sec:BBSED}, the limited number of filters available implies taking assumptions on the models to be adopted. We show in Fig.~\ref{fig:property_comparison_values} the distribution of derived properties using the combination of assumptions listed in Tab.~\ref{tab:SED_models}, to help unveiling possible biases associated to the choice of stellar models.

The assumption of longer star formation histories (C100) produce older derived ages, on average (as already pointed out in the literature, e.g. \citealp{adamo2013}), and the opposite is true for instantaneous burst of star formation (SSP); ages derived using our reference model, C10, are on average in-between (top panel of Fig.~\ref{fig:property_comparison_values}). We point out that the difference in median ages for those three models is only $\sim10$ Myr; the main difference is the presence of a considerable fraction of sources (almost one third of the sample) with ages $\gtrsim100$ Myr in the case of C100.
The C100 model also produces on average larger masses (by only $\sim0.10$ dex) and higher extinctions (by $\sim0.1$ mag).
Smaller difference are observed if either a lower metallicity (C10-008) or a difference extinction curve (C10-SB) are assumed (bottom panel of Fig.~\ref{fig:property_comparison_values}).
Overall, we notice that the distribution of ages is the one most affected by the model assumptions, while the distribution of derived masses is similar in all cases. 
We point out that the lowest median \chisqred\ value is found considering the reference C10 model is considered. We find 4 sources of the sample (ci\_8, ci\_9a, ci\_15b, ln\_1) whose SED fit with the SSP model gives a much lower \chisqred\ than with our reference one; the difference in derived properties with the two models is however negligible.

The distributions just discussed only show the best fit values and are associated in some cases to large uncertainties.
The uncertainties within the reference model range to $\sim0.5$ dex, $\sim1.0$ dex and $\sim0.3$ mag for log(M), log(Age) and E(B-V), respectively, but their distributions are mainly distributed around zero. 
The difference in derived properties caused by the choice of different models are mostly consistent with the intrinsic uncertainty within the single model. 

\subsection{Masses and Densities}
\label{sec:masses}
\begin{figure*}
\centering
\subfigure{\includegraphics[width=0.48\textwidth]{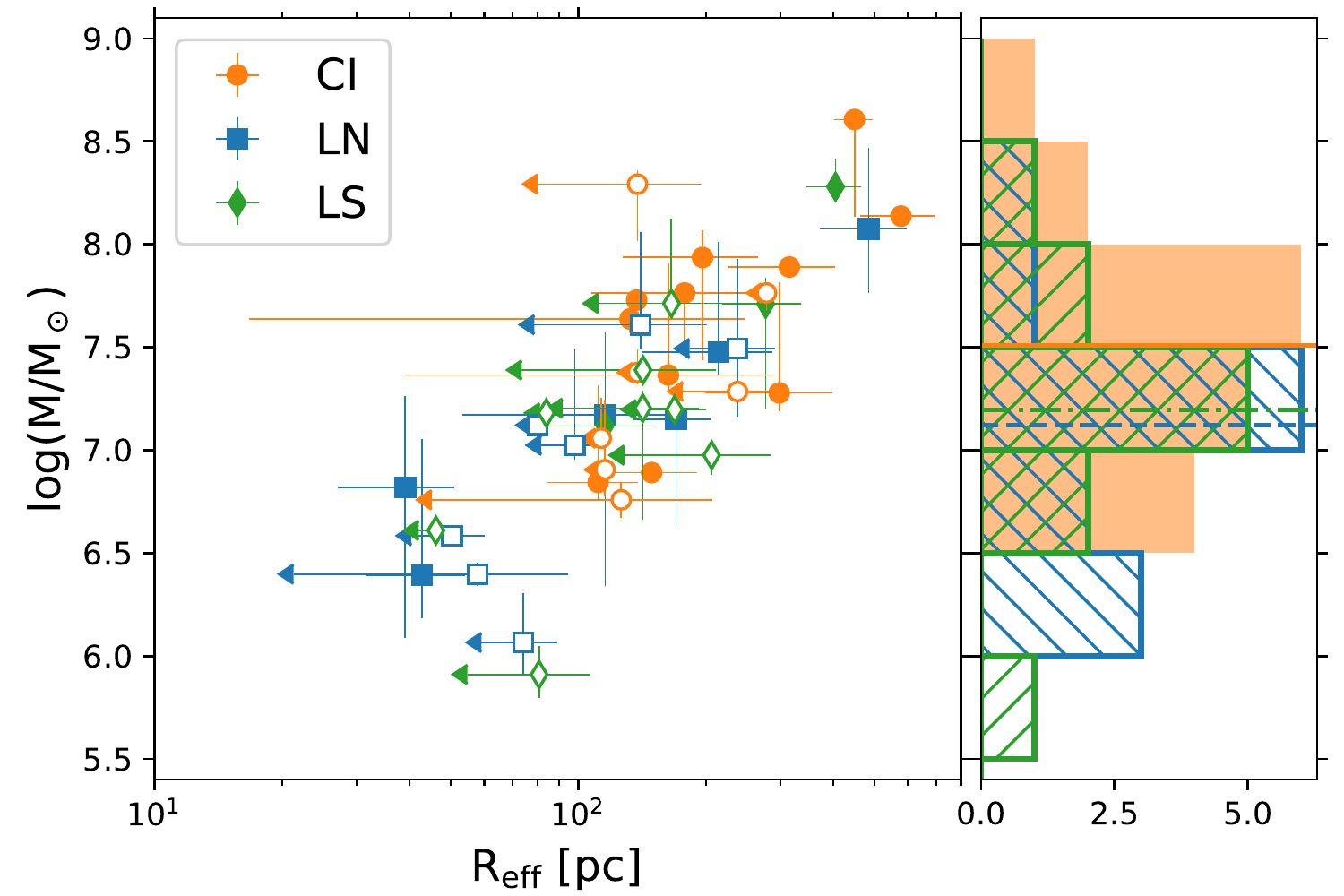}}
\subfigure{\includegraphics[width=0.48\textwidth]{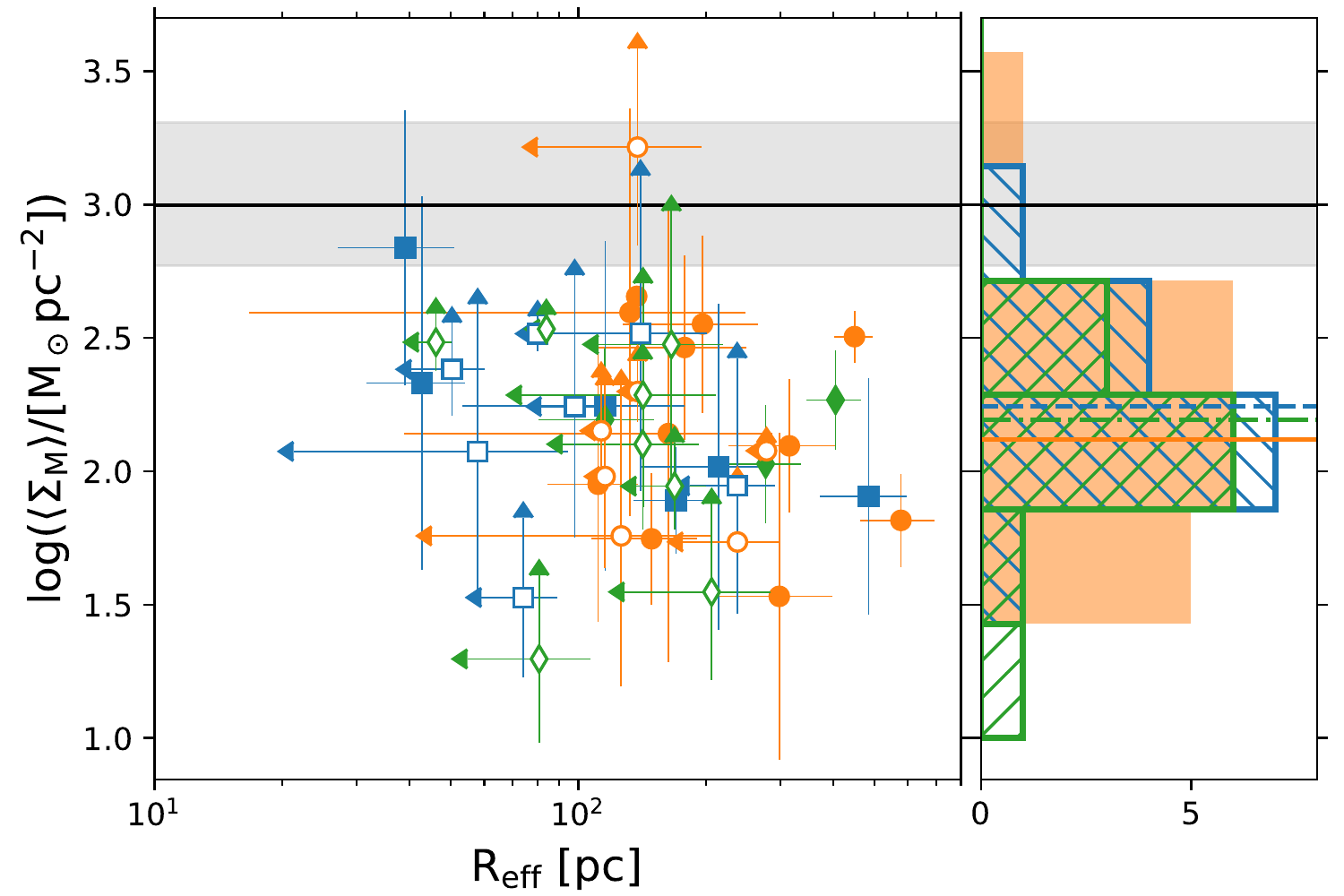}}
\caption{Size versus $\rm log(M/M_\odot)$ (left panel) and mass surface density (right panel). Sources are color--coded according to the sub--region where the clumps are observed. Size upper limits are shown as empty markers and the length of their arrows reflects the size uncertainty. Empty markers in the right plot are upper limits on the size and, as consequence, lower limits on the density. Each panel has a box showing the distribution of $\rm log(M/M_\odot)$ and mass density in each of the subregions of A521-sys1. The black horizontal line (and the grey shaded area) in the right panel represent the typical surface density of a nearby massive cluster (and the uncertainty associated), with $\rm M=10^5\ M_\odot$ and $\rm R_{eff}=4$ pc \citep{brown2021}. Three clumps do not have derived masses due to the lack of enough filter detections (see Section~\ref{sec:results_sed} and Tab.~\ref{tab:main_prop}) and therefore are not shown in the plots.}
\label{fig:mass_size_density}
\end{figure*}

We compare the derived masses to the sizes of the clumps in Fig.~\ref{fig:mass_size_density} (left panel). 
As pointed out in the previous paragraph, the range of masses spans more than two orders of magnitude; this range is similar in all three images of A521-sys1 and difference in the median mass is $\sim0.4$ dex between clumps in the LN field (less massive) and the ones in CI. 
We observe quite large scatters in mass ($\rm \gtrsim0.5$ dex) at any given clump size but also a robust correlation between mass and size (Spearman's coefficient: 0.78, p-value: $10^{-9}$), probably driven by incompleteness effects, as low--mass large clumps will fall below our detection limits. 
By combining masses and sizes we study the clump average mass density. We choose to focus on the surface densities instead of the volume ones because in many cases we are dealing with star-forming regions of hundreds of parsecs in size and we do not know their 3D intrinsic shape, therefore we cannot assume spherical symmetries.  
We define $\rm \langle\Sigma_M\rangle = M/(2\pi R_{eff}^2)$\footnote{The factor 2 at denominator is driven by $\rm R_{eff}$ being defined as the radius enclosing half of the source mass.} and plot the derived values in Fig.~\ref{fig:mass_size_density} (right panel). They span $\sim2$ orders of magnitude, in the range $\rm 10-1000\ M_\odot/pc^2$. We observe only a weak anti-correlation between clump size and surface density (Spearman's $\rho_s=-0.3$, \textit{p-val}: $0.06$). There is not a significant density difference for clumps in different fields, with a $0.12$ dex difference between LN (denser clumps) and CI.
For comparison, a typical low-redshift young massive star cluster of $\rm 10^5\ M_{\odot}$ has a median size of $4$ pc \citep{brown2021} and therefore a typical surface density of $\rm 10^3\ M_\odot/pc^2$; this value, shown as a black solid line on the right panel of Fig.~\ref{fig:mass_size_density} is almost one order of magnitude larger than the median values found for our sample, but we remind that a good fraction of our measurements are upper limits in size and therefore lower limit in terms of mass density. Two clumps have $\rm \langle\Sigma_M\rangle$ values comparable to the one of local massive clusters, namely one of the sub-peaks of clump ln\_9 and ci\_8. The latter displays a large mass density despite being observed at scales $>10$ times larger in size than local massive clusters and is discussed in more detail in Section~\ref{sec:radial_trends}.  

\subsection{Age distributions}
\label{sec:ages}
Fig.~\ref{fig:property_comparison_values} suggests that the bulk of clumps in A521-sys1 has ages close to $\sim10$ Myr, with few possibly as old as $\sim100$ Myr. This picture does not drastically change when considering age uncertainties and other stellar models; we observe that all clumps have derived ages $<200$ Myr, and the majority of them $<100$ Myr. The derived age distribution is therefore consistent with clumps being clearly detected in F390W, covering rest-frame $~2000$ \AA\ UV emission, associated to young stars. 
Taking $100$ Myr as an upper limit on the age of the clumps (as suggested by our reference C10 model), we estimate SFRs of individual clumps; the derived values span the range $\rm 0.008-4\ M_\odot/yr$, consistent with the range covered by UV magnitudes, if those are converted to SFR values using the factor from \citet{kennicutt2012} (see also Section~\ref{sec:literature_clumps} and Fig.~~\ref{fig:sizemag_literature}).
Summing the contributions from all clumps we obtain $12.4$, $2.9$ and $\rm 3.9\ M_\odot/yr$ in CI, LN and LS, respectively. Compared to the total SFR of the galaxy, $\rm \sim 16\ M_\odot/yr$ \citep{nagy2021}\footnote{The original value $\rm SFR=26\ M_\odot/yr$ reported in \citet{nagy2021} was derived assuming a \citet{salpeter1955} IMF and is here converted to match the assumption of \citet{kroupa2001} IMF used to derive clump masses.}, clumps appear to represent a good fraction of the galaxy current SFR, as already suggested by the clumpiness analysis in Section~\ref{sec:completeness}. 
We remind that the clump SFR values just derived are based over an age range of $100$ Myr and therefore constitute lower limits; larger values (by a factor $\sim10$) would result from taking the best-fit individual clump ages, suggesting an increase in the very recent SF activity of A521-sys1. 

Clump ages can be compared to their crossing time, which in terms of empirical parameters can be found as:
\begin{equation}
\rm T_{cr}\equiv 10 \left(\frac{R^{3}_{eff}}{GM}\right)^{1/2}
\end{equation}
Their ratio, named dynamical age $\rm \Pi\equiv Age/T_{cr}$ \citep[e.g.][]{gieles2011}, is used to distinguish bound ($\rm \Pi>1$) and unbound ($\rm \Pi<1$) agglomerates \citep[e.g.][for star clusters in local galaxies]{ryon2015,ryon2017,krumholz2019}.
Clumps in A521-sys1 have crossing times in the range $\rm T_{cr}=0.5-6.0$ Myr. 
Considering the best-fit age values we derive dynamical ages $\Pi>1$ for most of the sample ($\sim90\%$), suggesting that many clumps may be gravitationally stable against expansion. This result is discussed in light of the apparent lack of old clumps in Section~\ref{sec:radial_trends}.
Similar fractions are found if either the SSP or the C100 models are assumed.

\subsection{Extinctions}
\label{sec:extinctions}
As a sanity check for the extinction values obtained, we leverage archival VLT-MUSE observations of A521 to derive extinction values in annular sub--regions of the galaxy, using the Balmer decrement, i.e. the observed ratio of $\rm H\gamma$ and $\rm H\delta$ emission lines (technical details of this analysis are given in Appendix~\ref{sec:app:extinction}); the depth of the VLT-MUSE data prevents us from constraining with high precision the extinction map of A521-sys1 but the analysis suggests $E(B-V)$ values below $\sim0.7$ mag, confirming the range of extinctions found via the SED fitting process.

\begin{figure*}
\centering
\includegraphics[width=\textwidth]{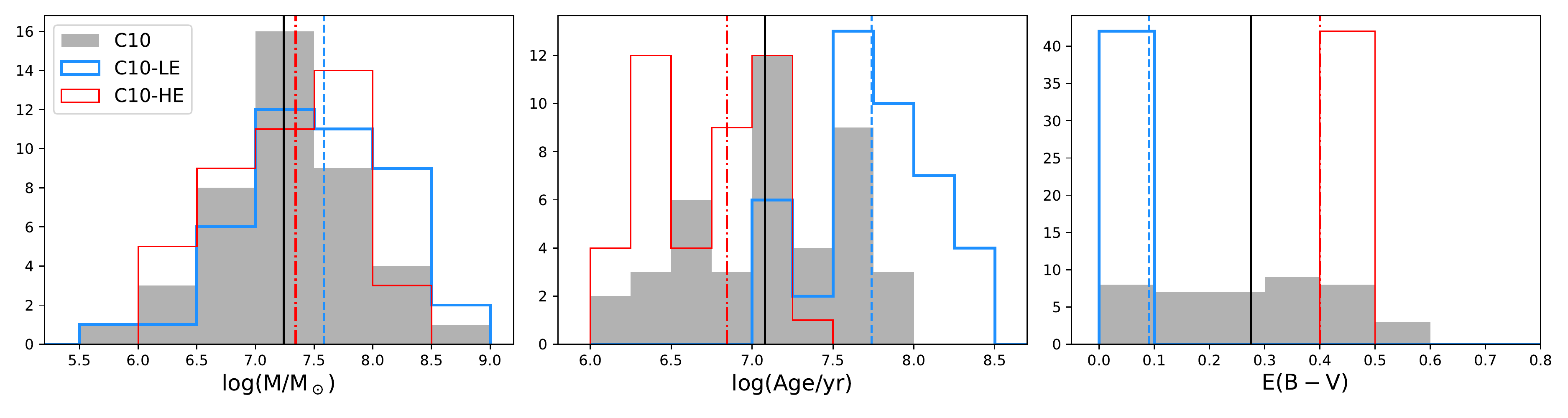}
\caption{Distributions of masses (left panels), ages (central panels) and color-excesses (right panels) for the low--extinction (C10-LE) and high--extinction (C10-HE) models, compared to the reference C10 model. Vertical lines give the median value for each of the distributions. }
\label{fig:extinction_comparison_values}
\end{figure*}

We perform an additional test to estimate the impact of assuming \textit{a-priori} an extinction value on the ages and masses derived via broadband SED fit; this test is motivated by the lack of HST multi-band detections affecting the study of high-z clumps (due to rest-frame optical-UV emission falling beyond the observable wavelength range), implying taking further assumptions on the clump models.
We consider two models, taking the same main assumptions of the reference C10 model but limiting the range of extinction values allowed by the fit:
\begin{itemize}
    \item C10-LE: the low extinction model, allowing extinctions only in the range $\rm E(B-V)<0.1$ mag;
    \item C10-HE: the high extinction model, allowing extinctions only in the range $\rm 0.4<E(B-V)<0.5$ mag. 
\end{itemize}
The results of these two models are shown in Fig.~\ref{fig:extinction_comparison_values}; as could be expected, lower (higher) extinctions force the fit to find older (younger) age values. In the case of our sample the low--extinction model is the one performing worst, with the age distribution shifted by $\sim0.75$ dex; we point out again that masses are less affected by the choice of model and in the low--extinction model are shifted to larger values by $0.3$ dex only. 

\section{Discussion}
\label{sec:discussion}

\subsection{UV size-magnitude comparison to z=0-3 literature samples}
\label{sec:literature_clumps}
\begin{figure}
    \centering
    \includegraphics[width=\columnwidth]{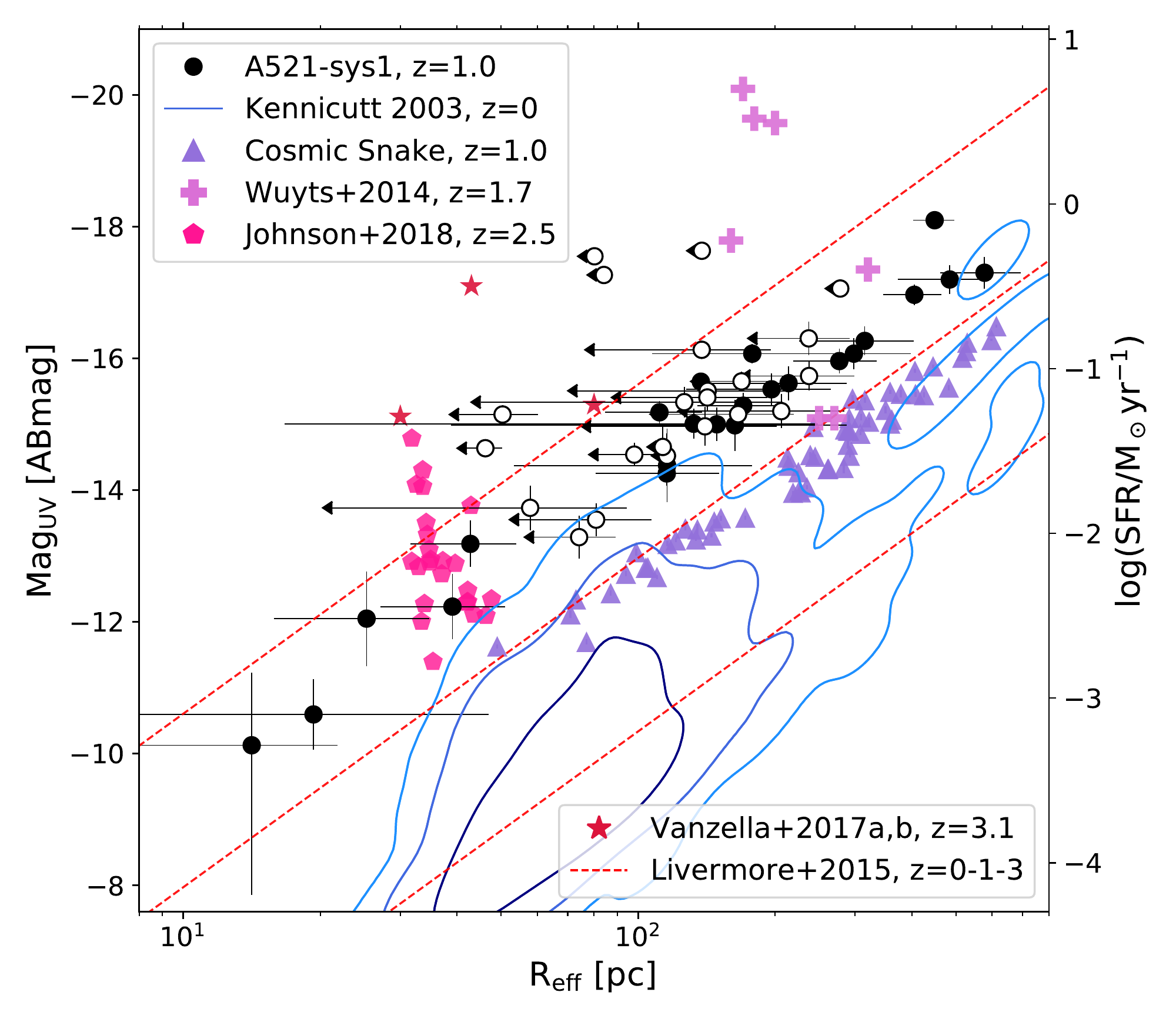}
    \caption{Intrinsic sizes and UV magnitudes of the clumps in A521-sys1 (black circles, empty markers used for size upper limits) compared to literature samples: SINGS \citep{kennicutt2003} at z=0, shown as blue contours enclosing $68\%$, $95\%$ and $99.7\%$ of the sample; Cosmic Snake \citep{cava2018} at z=1.0 as purple triangles; \citet{wuyts2014} sample at z=1.7 as pink plus ($\rm +$) symbols; \citet{johnson2017} sample at z=2.5 as fuchsia pentagons; \citet{vanzella2017a,vanzella2017b} sources at z=3.1-3.2 as red stars. Lines of median surface brightness at redshifts 0, 1 and 3 as derived by \citet{livermore2015} are shown as red dotted lines.}
    \label{fig:sizemag_literature}
\end{figure}

We compare the intrinsic sizes and luminosities of clumps in A521-sys1, presented in Section~\ref{sec:sizelum}, to other samples available in the literature in Fig.~\ref{fig:sizemag_literature}. Although clump masses and ages are derived for A521-sys1 clumps, we remind that it is worth discussing UV magnitudes as tracers of the recent SFR and mass of the clumps for two main reasons; first, they are widely available for many systems both at low and high redshift (while mass estimates are much less common) and, second, they avoid comparing physical quantities typically derived using different assumptions among different samples. 

In the same figure we show the sizes and luminosities of \HII\ regions in local ($z=0$) main-sequence (MS) galaxies from the SINGS sample \citep{kennicutt2003}. The SFR values of the SINGS sample have been converted to UV magnitudes using the conversion factor in \citet{kennicutt2012}. We observe that clumps in A521-sys1 are brighter than the ones in \citep{kennicutt2003} when sources at similar scales are compared, suggesting that star--forming regions in A521-sys1 are denser than local \HII\ regions.
Similar sizes and magnitudes are measured in clumps in the redshift range $\rm z=1-3$; we show in Fig.~\ref{fig:sizemag_literature} the clumps samples of the Cosmic Snake (z=1.0, \citealp{cava2018}), \citet{wuyts2014} (z=1.7), \citet{johnson2017} (z=2.5) and three highly magnified clumps from \citet{vanzella2017a,vanzella2017b} ($\rm z\sim3.1$). 
Studies of clumps at $z\geqslant1$ suggest an evolution of the clumps' average density with redshift \citep[e.g][]{livermore2015}. We plot the average surface brightness at $z=0$, $1$ and $3$ derived by \citet{livermore2015} using clumps from samples of SINGS, WiggleZ \citep{wisnioski2012}, SHiZELS \citep{swinbank2012}, and the lensed arcs from \citet{jones2010}, \citet{swinbank2007,swinbank2009} and \citet{livermore2012}; our sample of clumps in A521-sys1 lies, similarly to the other samples just presented, in the range of expected densities for redshifts $z=1-3$. 
The main possible cause of clumps' density redshift evolution is the effect of galactic environment within galaxies \citep[e.g][]{livermore2015}, at higher redshift characterized by higher gas turbulence and higher hydrostatic pressure at the disk midplane, fragmenting as denser clouds \citep{dessauges2019,dessauges2022}. Detection limit differences could also partly explain the trends as, typically, galaxies at higher redshifts have worse detection limits. 

Supporting the hypothesis of the (internal) galactic environmental effect, studies of nearby samples of high-z analogs, e.g. GOALS LIRGs \citep{armus2009,larson2020}, DYNAMO gas-rich galaxies \citep{green2014,fisher2017a} and LARS starbursts \citep{ostlin2014,messa2019}, find clumps with surface densities comparable to the ones observed at redshift 1 and above. We point out that such galaxies sit above the MS for local galaxies (while instead the SINGS sample contain typical MS galaxies at z=0) but are consistent with MS galaxies at $z\gtrsim1$.

\subsection{Properties derived via the alternative photometry method}
\label{sec:alternative_results}
\begin{figure}
    \centering
    \includegraphics[width=0.49\textwidth]{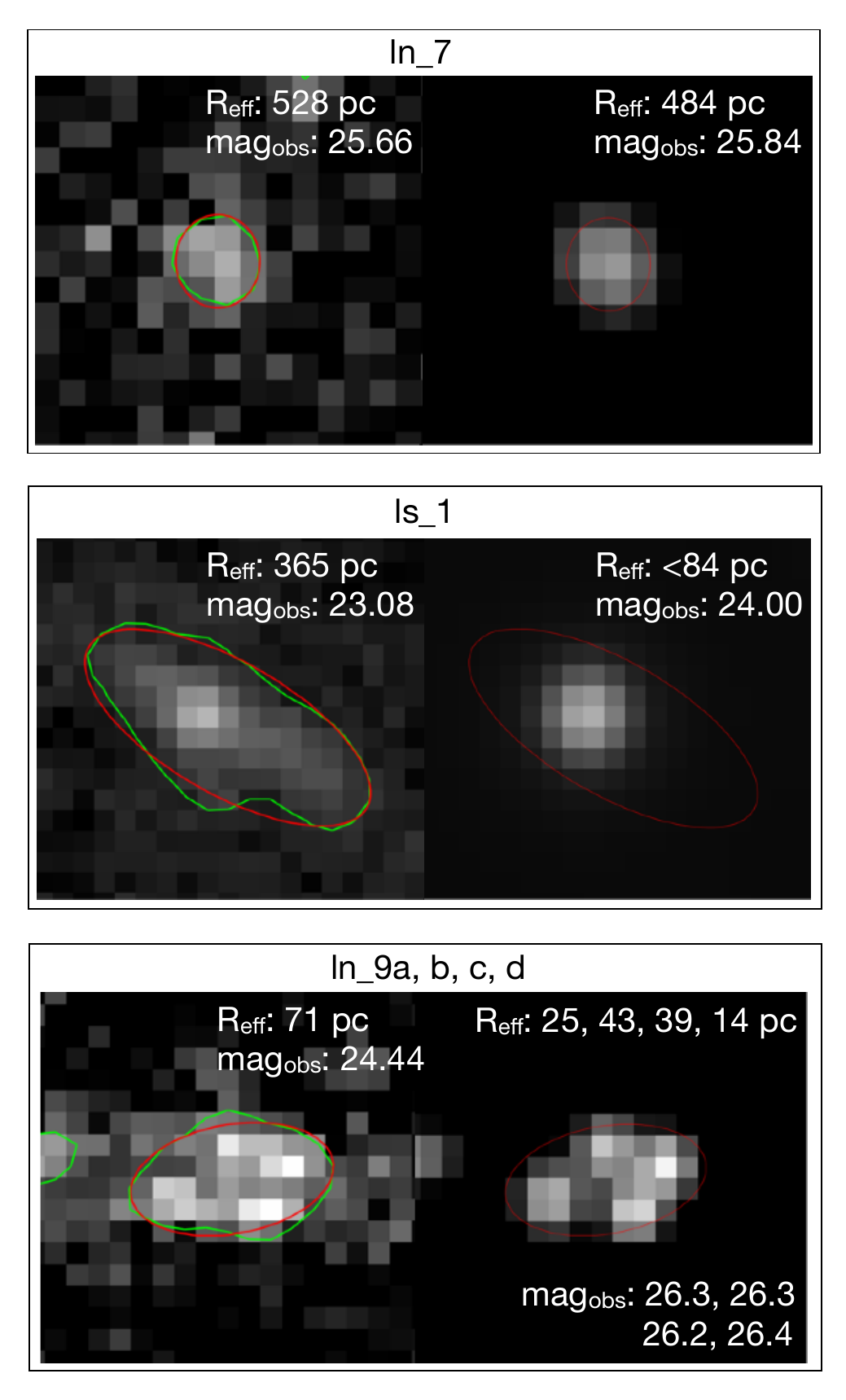}
    \caption{Examples of the extraction via $3\sigma$ contours vs the best-fit of the 2D model. For each panel, the figure on the left shows the data with $3\sigma$ contours in green and the ellipse used for the aperture photometry and for estimating the size in red, over--plotted. The figure on the right shows the best-fit model according to the reference 2D--fit photometry. Intrinsic sizes ($\rm R_{eff}$) and observed magnitudes derived via the two methods are reported.}
    \label{fig:contours}
\end{figure}
We compare the results presented in Section~\ref{sec:results_phot} and \ref{sec:results_sed} to the ones obtained with the alternative extraction and photometry method introduced in Section~\ref{sec:alternative}. Overall, the alternative method miss to extract 5 sources (2 in CI, 1 in LN and 2 in LS). We checked that for bright isolated sources (e.g. top panel of Fig.~\ref{fig:contours}) we get similar results with the two methods (radii are different by less than a factor 1.5, magnitude differences are $<0.3$ mag). 
Large differences are observed for clumps consisting of a bright narrow peak and a diffuse tail (e.g. middle panel of Fig.~\ref{fig:contours}). The 2D fit of the reference method recover only the bright peak, i.e. the densest core of the star-forming region, while the $3\sigma$ contour also include the diffuse tail. This is the case for 6 clumps (ci\_1, ln\_1, ls\_1, ln\_3, ln\_5 and ls\_5); the derived sizes can differ up to factors 4, and magnitudes up to $\sim1$ mag. These differences, in turn, convert into mass values larger by $\sim1$ order of magnitude and mass surface densities lower by $\sim0.4$ dex, for sources ci\_1, ln\_1 and ls\_1, in the case of the alternative photometry. We deduce that, in the cases just mentioned,we are studying large star-forming regions via the alternative method, while the standard method focus on their dense cores.

Another class of sources where we see differences between the two methods are clumps fitted by multiple peaks in the 2D fit but falling within the same $3\sigma$ profile and therefore considered as a single source in the alternative photometry. This is the case for 3 clumps (the groups ln\_9a,b,c,d, see bottom panel of Fig.~\ref{fig:contours},  ci\_7a,b and ci\_17a,b).

Despite the differences just mentioned, the overall distribution of clump sizes and F390W magnitudes are similar in the two analysis; the alternative method recovers, as median values, brighter (by $\sim 0.5$ mag) and larger (by less than a factor $1.5$) clumps, but the median surface brightness of the clumps is the same with both methods. Similarly, the median mass recovered with the alternative method is larger by $0.2$ dex, but its surface density is smaller (by $0.2$ dex) with respect to the median values from the reference method. Age and extinction distributions are similar in the two cases. 
We conclude that the methodology for extracting and analyzing clumps can have a strong effect especially when studying non--Gaussian or multiple--peaked systems; on the other hand the average differences between considering $3\sigma$ contours or 2D Gaussian fits in our sample are negligible. 

\subsection{Lensing effect on derived properties}
\label{sec:lensing_effects}
Studying the same clumps imaged in the three regions introduced in Section~\ref{sec:data_hst} allows us to understand the effects of gravitational lensing on clump samples overall and on single sources. 
Clumps that appear similar, in terms of size and magnitude, on the image plane, i.e. in terms of observed properties (Fig.~\ref{fig:sizemag_obs}), show intrinsic properties that differ on average by a factor $\sim2$ in size and by $\sim1$ mag if clumps in CI and in LN are compared. Despite these differences the surface brightness values observed are similar in all sub--regions, as consequence of its conservation through gravitational lensing.
The mass values resulting from the SED fitting, confirm the photometric results, as clumps in the CI region appear more massive by $0.5$ dex compared to the ones in LN, but median surface densities are similar in all sub--regions. 
Overall we are able to observe, on average, smaller, less massive clumps, in regions with larger magnification, but the distribution of such properties are not drastically different in the three sub--regions.
The clumpiness estimates are also similar (Fig.~\ref{fig:clumpiness}) and the slightly lower values retrieved in LN can be mainly attributed the the presence of a bright foreground galaxy, difficult to subtract completely from the data (Section~\ref{sec:clumpiness}). 

Moving from the overall distributions to one--to--one analysis of individual clumps as observed in CI, LN and LS, we find that clumps with magnification differences smaller than a factor $\sim2$ between one image and another, e.g. source 4 (ci\_4, ln\_4 and ls\_4 have $\mu=4.8$, $3.4$ and $3.4$ respectively),  display similar photometric and physical properties, consistent within uncertainties. 
On the other hand, larger differences can be observed when clumps are greatly magnified in some sub--regions, as for clump 1, with an amplification $\mu=11$ in the LN image (ln\_1) but $\mu=3.7$ in the CI (ci\_1); in the latter case the derived mass value is larger by 0.25 dex but with a lower limit on the mass density which is 0.25 dex smaller than the one derived for ln\_1.
A similar case is clump $9$ (bottom--right panel of Fig.~\ref{fig:contours}), which in the LS region (magnification $\mu=5$) appears like a single-peaked source, with an estimated size upper limit $\rm R_{eff}<200$ pc , but with the large magnification of the LN region ($\mu\gtrsim50$) can be separated into 4 narrow peaks, with physical scales between 15 and 50 pc. Individual sub--peaks have smaller derived sizes and masses than the single source ls\_9, but their derived mass surface densities are larger, suggesting that at smaller physical scales we are able to observe denser cores of clumps (Fig.~\ref{fig:mass_size_density}); such trend is confirmed by simulations of resolution effects on derived clump properties \citep{meng2020}. 

One extreme case is clump 8, being magnified by $\mu=20$ in the LN and LS images, compared to $\mu=3.5$ in the CI; in case of ci\_8 we derive a mass of $\rm \log(M/M_\odot)=8.3$, more than one order of magnitude larger than for ln\_8 and ls\_8 ($\rm \log(M/M_\odot)=7.0$ and $7.1$); also its mass surface density is one order of magnitude larger than what is found for ln\_8 and ls\_8.
We attribute such large values of mass and density to the position of ci\_8, being consistent with the bulge of the galaxy and with a massive cloud of molecular gas, as found by the analysis of \citet{dessauges2022}. Its derived age, 20 Myr, seems to suggest that some star formation is still going on even there. The image of clump 8 on the lensed arc is heavily distorted and magnified, therefore what we observe as ln\_8 and ls\_8 could be a dense star--forming core within source 8 itself.

\subsection{Galactocentric trends}
\label{sec:radial_trends}
\begin{figure*}
\centering
\subfigure{\includegraphics[width=0.49\textwidth]{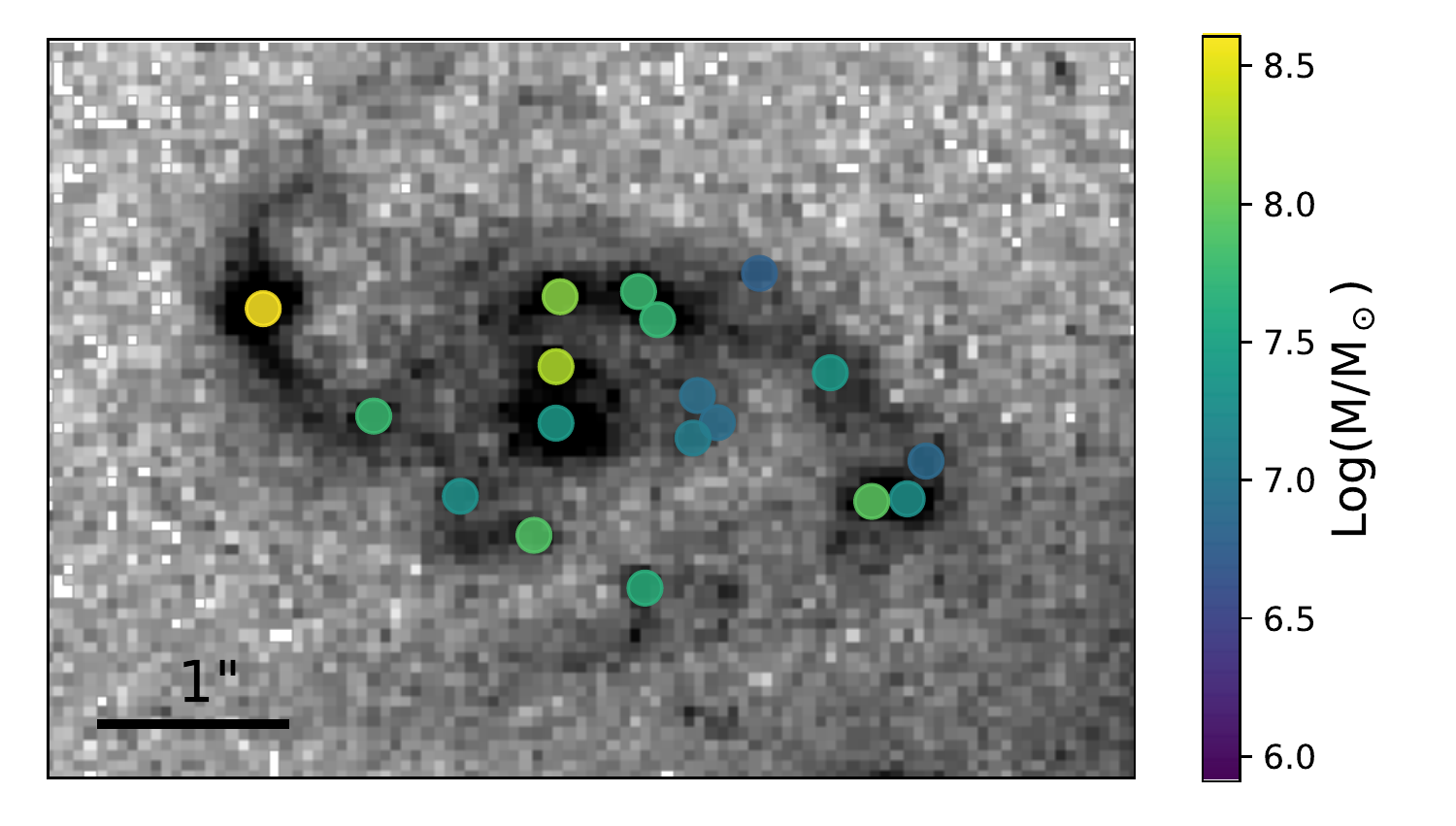}}
\subfigure{\includegraphics[width=0.49\textwidth]{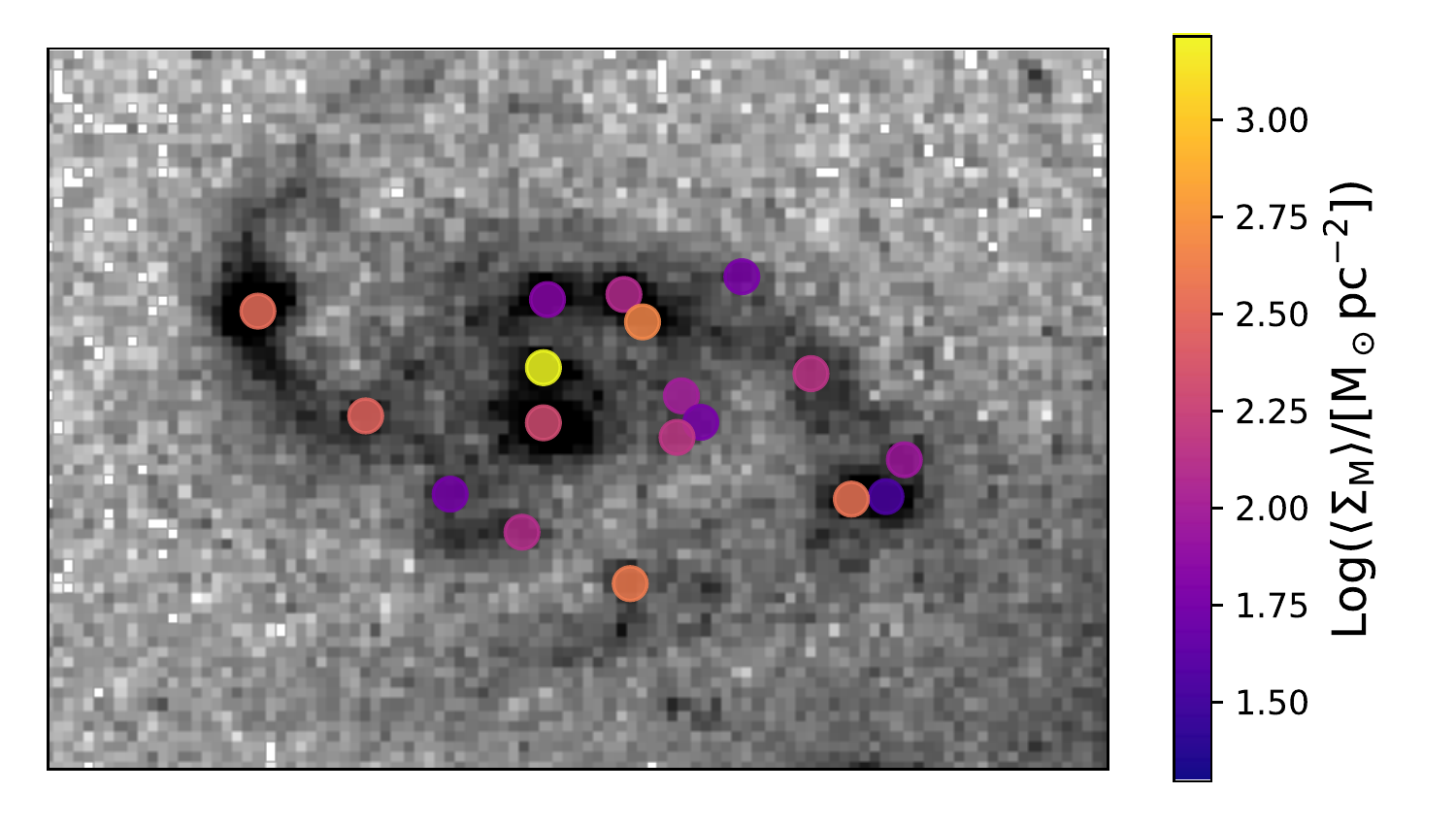}}
\subfigure{\includegraphics[width=0.49\textwidth]{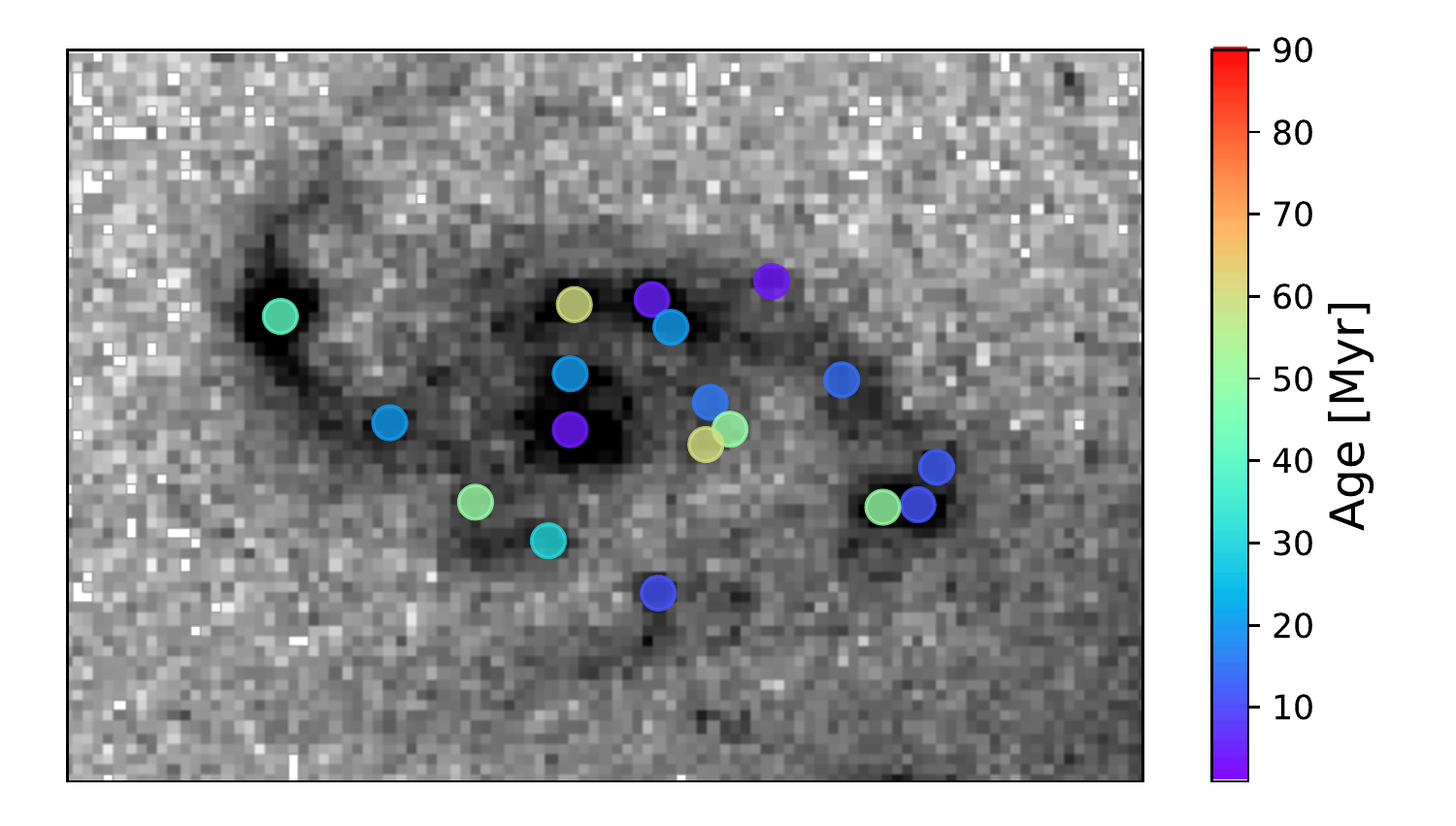}}
\subfigure{\includegraphics[width=0.49\textwidth]{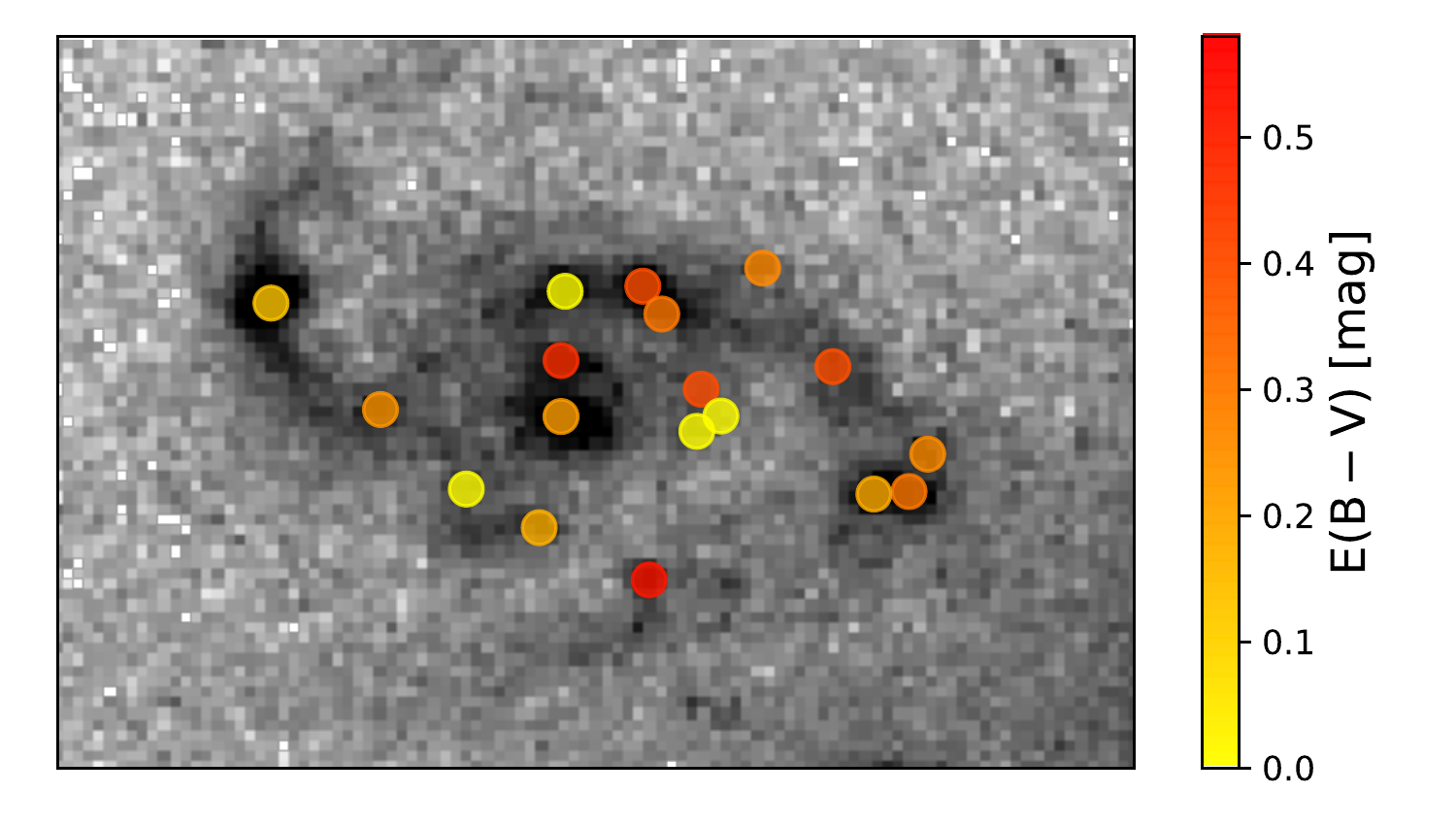}}
    \caption{F814W observations of the CI region, with the position of detected clumps, color--coded by their derived mass (top-left), mass surface density (top-right), age (bottom-left) and extinction (bottom-right). The Spearman's correlation test do not recover significant any correlations between the galactocentric distance of the clumps and their properties shown here.}
    \label{fig:f814_vs_properties}
\end{figure*}

Focusing on the CI, where the entire galaxy can be studied with an almost uniform magnification, we test for possible radial trends of A521-sys1 clumps' properties. In Fig.~\ref{fig:f814_vs_properties} we plot the positions of clumps in the CI, color-coded by their derived properties, on the F814W observations.
Radial trends in clumps' ages and masses can be used to test their survival and evolution within the host galaxies and, as consequence, to test formation models of galaxies and their bulges. The presence of older and more massive clumps near the centre of the galaxy has been interpreted as a sign of the more massive clumps being able to survive bound for hundred of Myr, migrating toward the centre of the galaxy, and there merging to form the galactic bulge, as suggested by simulations by e.g. \citealp{bournaud2007,krumholz2010}, while other simulations argue that such migrating clumps would have marginal effect on bulge growth \citep[e.g.][]{tamburello2015}. 
Running Spearman's correlation test we do not find any statistically significant correlation between the clump physical properties plotted in Fig.~\ref{fig:f814_vs_properties} and the galactocentric radius. 
We observe massive clumps all over the spiral arms, with the most massive one being at $\rm \sim7.5$ kpc from the centre (ci\_14). In the same way, we observe dense clumps both very close to the centre and further away, along the spiral arms (e.g. ci\_4). 
In particular, we observe two massive clumps close to the centre of the galaxy, namely ci\_1 and ci\_8 (the latter sitting at the coordinates of the bulge, \citealp{nagy2021}); their young ages (4 and 20 Myr, respectively) suggest that star formation is taking place also at the centre of the galaxy. At the same time, the large mass, $\rm log(M/M_\odot)=8.3$, and density, $\rm \langle\Sigma_M\rangle> 10^3\ M_\odot pc^{-2}$ of clump ci\_8, may suggest that we are looking at the formation of a proto-bulge.

Fig.~\ref{fig:f814_vs_properties} suggests the presence of an age and extinction asymmetry between the two spiral arms, with the western arm being younger and more extincted than the eastern one. The difference is small (on average $\sim20$ Myr in age, and $0.1$ mag in color excess) but consistent across the stellar models tested. Asymmetries are very common in late-type galaxies but the uncertainties associated to the derived ages prevent us to drive robust conclusions for A521-sys1.  

Another useful metric to test the possible migration of clumps is the dynamic time of the galaxy, defined as the ratio between the rotation velocity and the radius; when compared to the age of the clumps it probes whether a clump is still close to the natal region, age$\rm \lesssim t_{dyn}$, or it had survived enough $\rm t_{dyn}$ to have possibly migrated, age$\rm \gtrsim10\times t_{dyn}$ \citep[e.g.][]{forsterschreiber2011b,adamo2013}.
Considering the rotation curve of A521-sys1 \citep[][from MUSE data]{patricio2018} we derive a $\rm t_{dyn}$ varying from $\sim10$ Myr near the centre to $\sim100$ Myr at $6$ kpc; these values are consistent with the ages spanned by the clumps, indicating that they observed close to their natal region. In addition, the clumpiness analysis (Section~\ref{sec:clumpiness}) show that clumps are not dominating the light at wavelengths longer than (rest--frame) $\rm \gtrsim3000$ \AA, suggesting that clumps are not surviving as bound structures for time--scales longer than 100 Myr.

The lack of old and migrating clumps seems in contrast with the large dynamical ages retrieved (Section~\ref{sec:ages}), suggesting that clumps should be gravitationally stable against expansions. One possible cause of this inconsistency could be that the dynamical age is not a suitable metric for the gravitational stability of clumps, at scales $>10$ pc; dynamical ages were introduced to study the stability of stellar clusters on scales of few pc and assuming virial equilibrium \citep{gieles2011}. On the other hand, stellar evolution changes the clump colors to redder values such that a $500$ Myr old clump with $\rm M=2\cdot10^7\ M_\odot$ (the median value for our sample, found in Section~\ref{sec:results_sed}) would have, at the distance of A521-sys1 an observed magnitude of $29.64$ mag in F814W (and fainter magnitudes in bluer filters); while the depth of the observations in F814W reaches $27.5$ mag (Tab.~\ref{tab:data}), the completeness within A521-sys is shallower by $>0.5$ mag and therefore we would expect to observe such old clumps only in case of large magnifications, $\mu\gtrsim10$, thus only in limited regions. Moving to the NIR filters (F105W and F160W) would result in brighter observed magnitudes, but, at the cost of worse spatial resolution and worse completeness, leading similarly to low chances of observing old clumps in A521-sys1. 
%
%

\section{Conclusions}
\label{sec:conclusion}
We analyzed the clump population of the gravitationally-lensed galaxy A521-sys1, a $\rm z=1.04$ galaxy with properties typical of main sequence systems at similar redshift, i.e. elevated star formation ($\rm SFR=16\pm5\ M_\odot yr^{-1}$) and gas-rich, rotation-dominated disk with high velocity dispersion \citep{patricio2018,girard2019,nagy2021}. A521-sys1 is characterized by a clumpy morphology in the NUV band, observed with HST WFC3-F390W; we use this as the reference filter for extracting the clump catalog and study the sizes and rest-frame UV photometry. Four additional HST filters, F606W, F814W from ACS and F105W, F160W from WFC3/IR, are used to characterize ages and masses of the clumps via broad-band SED fitting.

The appearance of A521-sys is heavily affected by gravitational lensing, producing multiple images of the same system and allowing the study of clumps seen at different intrinsic scales, in the range $10-600$ pc. Roughly half of the galaxy is stretched into a wide arc, with magnification, $\rm \mu$, reaching factors 10 and above; the arc is made by two mirrored images, which we call lensed-north (LN) and lensed-south (LS). The entire system is observable via a counter-image (CI) with a mean magnification $\rm \mu\sim4$.
A gravitational lens model is constructed for the entire A521 galaxy cluster \citep{richard2010} and is later fine-tuned to constrain with better precision the area enclosing the A521-sys1 images, giving a final positional accuracy of $\rm 0.08''$, comparable to the pixel scale of the HST observations.

We derive the following results via photometric and broad-band SED analyses:
\begin{itemize}
    \item we extract a sample of 18 unique clumps; many of those are imaged multiple times and some are resolved into sub-clumps when observed at high magnifications. As consequence, the final sample counts 45 entries;
    \item the intrinsic clump sizes range from $\sim10$ to $\sim600$ pc, suggesting that we are observing systems that span from almost single clusters to large star-forming regions. Scales below $\sim50$ pc are resolved only in the LN region, hosting small areas close to the critical lines with extreme magnifications ($\rm \mu>20$). Half of the recovered values are upper limits, suggesting that in many cases clumps are more compact that what we are able to resolve; 
    \item the interval of absolute UV clump magnitudes is comparable to the ones of other literature clump samples at similar redshift and at similar physical scales. We confirm that the surface brightnesses of clumps in $\rm z\gtrsim1$ galaxies are much larger than the corresponding star-forming regions in local galaxies. On the other hand, the completeness analysis reveals that, given the depth of our observations, we would not be able to observe clumps with lower surface brightness;
    \item the galaxy appears less clumpy in redder bands; this is quantitatively confirmed by the clumpiness analysis, measuring what fraction of the galaxy luminosity is produced by clumps. The clumpiness is high (around $40\%$) in rest-frame NUV, suggesting that a large fraction of the recent star formation is taking place in the clumps we observe, and decreases moving to V and I bands, where the old stellar population of the galaxy dominates the emission;
    \item the derived clump masses range from $\rm 10^{5.9}\ M_\odot$ to $\rm 10^{8.6}\ M_\odot$, confirming that we are studying both cluster or cluster aggregations and large star-forming regions. The overall mass distribution and its median value ($\rm \sim2\cdot10^{7}\ M_\odot$), do not change considerably if either a 10 Myr continuum star formation models (C10, used as reference), a single stellar population model (SSP) or a 100 Myr continuum star formation model are considered; the same is true when testing different extinction models (\citealt{cardelli1989} and \citealt{calzetti2000}) and different metallicities. 
    
    The clump sample has a median mass surface density of $\rm \sim10^2\ M_\odot\ pc^{-2}$ but few clumps reach densities typical of the most massive compact ($<5$ pc) stellar clusters observed in local galaxies ($\rm \sim10^3\ M_\odot\ pc^{-2}$). No statistically significant galactocentric trend is observed with either mass or mass density. Dense and massive clumps are observed both close to the galactic bulge and along the outskirts of the spiral arms; 
    \item the majority of derived ages are $<100$ Myr, with many clumps having a best-fit age close to $10$ Myr. Clumps of such young ages are consistent with being observed close to their natal region, making impossible the study of possible clump migration. The study of the dynamical age, defined by the comparison between clump ages and their density, suggests that most of the clumps may be gravitationally stable against expansion; 
    \item clump extinctions are distributed in the range $\rm E(B-V)=0.0-0.6$ mag, consistent with the analysis of the Balmer decrement derived from VLT-MUSE observations. Testing the SED fitting with extinction fixed in narrow intervals reveals that inaccurate assumptions (e.g. $\rm E(B-V)\sim0.0$ mag for the entire sample) would result in biasing the derived ages by roughly a factor 10, while having a much smaller impact on the masses; 
    \item the lack of galactocentric trends for any of the physical properties available and the lack of old migrated clumps can be explained either by dissolution of clumps after few $\sim100$ Myr or by stellar evolution making them fall below the detectability limits of our data.
    \item when comparing the properties observed in different galaxy images (CI, LN and LS), clumps appear on average smaller and less bright (and less massive) in LN, suggesting that in regions with large magnifications we are able to observe the cores of the $>100$ pc star-forming regions seen with no or little magnification. Surface brightnesses and mass surface densities are overall very similar in all sub-regions. 
\end{itemize}

\section*{Acknowledgements}

We thank the anonymous referee for the useful comments that helped improving the quality of the paper.
This research made use of Photutils, an Astropy package for detection and photometry of astronomical sources \citep{bradley2020}. 
M.M. acknowledges the support of the Swedish Research Council, Vetenskapsrådet (internationell postdok grant 2019-00502). 
\section*{Data Availability}
The HST data underlying this article are accessible from the Hubble Legacy Archive (HLA) at \url{https://hla.stsci.edu/} or through the MAST portal at \url{https://mast.stsci.edu/portal/Mashup/Clients/Mast/Portal.html} (proposal IDs 15435 and 16670). The derived data generated in this research will be shared on reasonable request to the corresponding author.



\bibliographystyle{mnras}
\bibliography{references} 

\begin{thebibliography}{}
\makeatletter
\relax
\def\mn@urlcharsother{\let\do\@makeother \do\$\do\&\do\#\do\^\do\_\do\%\do\~}
\def\mn@doi{\begingroup\mn@urlcharsother \@ifnextchar [ {\mn@doi@}
  {\mn@doi@[]}}
\def\mn@doi@[#1]#2{\def\@tempa{#1}\ifx\@tempa\@empty \href
  {http://dx.doi.org/#2} {doi:#2}\else \href {http://dx.doi.org/#2} {#1}\fi
  \endgroup}
\def\mn@eprint#1#2{\mn@eprint@#1:#2::\@nil}
\def\mn@eprint@arXiv#1{\href {http://arxiv.org/abs/#1} {{\tt arXiv:#1}}}
\def\mn@eprint@dblp#1{\href {http://dblp.uni-trier.de/rec/bibtex/#1.xml}
  {dblp:#1}}
\def\mn@eprint@#1:#2:#3:#4\@nil{\def\@tempa {#1}\def\@tempb {#2}\def\@tempc
  {#3}\ifx \@tempc \@empty \let \@tempc \@tempb \let \@tempb \@tempa \fi \ifx
  \@tempb \@empty \def\@tempb {arXiv}\fi \@ifundefined
  {mn@eprint@\@tempb}{\@tempb:\@tempc}{\expandafter \expandafter \csname
  mn@eprint@\@tempb\endcsname \expandafter{\@tempc}}}

\bibitem[\protect\citeauthoryear{{Adamo}, {{\"O}stlin}, {Bastian},
  {Zackrisson}, {Livermore}  \& {Guaita}}{{Adamo} et~al.}{2013}]{adamo2013}
{Adamo} A.,  {{\"O}stlin} G.,  {Bastian} N.,  {Zackrisson} E.,  {Livermore}
  R.~C.,   {Guaita} L.,  2013, \mn@doi [\apj] {10.1088/0004-637X/766/2/105},
  \href {https://ui.adsabs.harvard.edu/abs/2013ApJ...766..105A} {766, 105}

\bibitem[\protect\citeauthoryear{{Armus} et~al.,}{{Armus}
  et~al.}{2009}]{armus2009}
{Armus} L.,  et~al., 2009, \mn@doi [\pasp] {10.1086/600092}, \href
  {https://ui.adsabs.harvard.edu/abs/2009PASP..121..559A} {121, 559}

\bibitem[\protect\citeauthoryear{{Bastian} \& {Lardo}}{{Bastian} \&
  {Lardo}}{2018}]{bastian2018}
{Bastian} N.,  {Lardo} C.,  2018, \mn@doi [\araa]
  {10.1146/annurev-astro-081817-051839}, \href
  {https://ui.adsabs.harvard.edu/abs/2018ARA&A..56...83B} {56, 83}

\bibitem[\protect\citeauthoryear{{Bertin} \& {Arnouts}}{{Bertin} \&
  {Arnouts}}{1996}]{bertin1996}
{Bertin} E.,  {Arnouts} S.,  1996, \mn@doi [\aaps] {10.1051/aas:1996164}, \href
  {https://ui.adsabs.harvard.edu/abs/1996A&AS..117..393B} {117, 393}

\bibitem[\protect\citeauthoryear{{Bik}, {{\"O}stlin}, {Hayes}, {Adamo},
  {Melinder}  \& {Amram}}{{Bik} et~al.}{2015}]{bik2015}
{Bik} A.,  {{\"O}stlin} G.,  {Hayes} M.,  {Adamo} A.,  {Melinder} J.,   {Amram}
  P.,  2015, \mn@doi [\aap] {10.1051/0004-6361/201525850}, \href
  {http://adsabs.harvard.edu/abs/2015A\%26A...576L..13B} {576, L13}

\bibitem[\protect\citeauthoryear{{Bik}, {{\"O}stlin}, {Menacho}, {Adamo},
  {Hayes}, {Herenz}  \& {Melinder}}{{Bik} et~al.}{2018}]{bik2018}
{Bik} A.,  {{\"O}stlin} G.,  {Menacho} V.,  {Adamo} A.,  {Hayes} M.,  {Herenz}
  E.~C.,   {Melinder} J.,  2018, \mn@doi [\aap] {10.1051/0004-6361/201833916},
  \href {https://ui.adsabs.harvard.edu/\#abs/2018A&A...619A.131B} {619, A131}

\bibitem[\protect\citeauthoryear{{Bournaud}, {Elmegreen}  \&
  {Elmegreen}}{{Bournaud} et~al.}{2007}]{bournaud2007}
{Bournaud} F.,  {Elmegreen} B.~G.,   {Elmegreen} D.~M.,  2007, \mn@doi [\apj]
  {10.1086/522077}, \href
  {https://ui.adsabs.harvard.edu/abs/2007ApJ...670..237B} {670, 237}

\bibitem[\protect\citeauthoryear{{Bournaud}, {Elmegreen}  \&
  {Martig}}{{Bournaud} et~al.}{2009}]{bournaud2009}
{Bournaud} F.,  {Elmegreen} B.~G.,   {Martig} M.,  2009, \mn@doi [\apjl]
  {10.1088/0004-637X/707/1/L1}, \href
  {https://ui.adsabs.harvard.edu/abs/2009ApJ...707L...1B} {707, L1}

\bibitem[\protect\citeauthoryear{{Bournaud}, {Dekel}, {Teyssier}, {Cacciato},
  {Daddi}, {Juneau}  \& {Shankar}}{{Bournaud} et~al.}{2011}]{bournaud2011}
{Bournaud} F.,  {Dekel} A.,  {Teyssier} R.,  {Cacciato} M.,  {Daddi} E.,
  {Juneau} S.,   {Shankar} F.,  2011, \mn@doi [\apjl]
  {10.1088/2041-8205/741/2/L33}, \href
  {https://ui.adsabs.harvard.edu/abs/2011ApJ...741L..33B} {741, L33}

\bibitem[\protect\citeauthoryear{{Bournaud} et~al.,}{{Bournaud}
  et~al.}{2014}]{bournaud2014}
{Bournaud} F.,  et~al., 2014, \mn@doi [\apj] {10.1088/0004-637X/780/1/57},
  \href {https://ui.adsabs.harvard.edu/abs/2014ApJ...780...57B} {780, 57}

\bibitem[\protect\citeauthoryear{{Bournaud}, {Daddi}, {Wei{\ss}}, {Renaud},
  {Mastropietro}  \& {Teyssier}}{{Bournaud} et~al.}{2015}]{bournaud2015}
{Bournaud} F.,  {Daddi} E.,  {Wei{\ss}} A.,  {Renaud} F.,  {Mastropietro} C.,
  {Teyssier} R.,  2015, \mn@doi [\aap] {10.1051/0004-6361/201425078}, \href
  {https://ui.adsabs.harvard.edu/abs/2015A&A...575A..56B} {575, A56}

\bibitem[\protect\citeauthoryear{{Bouwens}, {Illingworth}, {Oesch}, {Caruana},
  {Holwerda}, {Smit}  \& {Wilkins}}{{Bouwens} et~al.}{2015}]{bouwens2015}
{Bouwens} R.~J.,  {Illingworth} G.~D.,  {Oesch} P.~A.,  {Caruana} J.,
  {Holwerda} B.,  {Smit} R.,   {Wilkins} S.,  2015, \mn@doi [\apj]
  {10.1088/0004-637X/811/2/140}, \href
  {http://adsabs.harvard.edu/abs/2015ApJ...811..140B} {811, 140}

\bibitem[\protect\citeauthoryear{Bradley et~al.,}{Bradley
  et~al.}{2020}]{bradley2020}
Bradley L.,  et~al., 2020, astropy/photutils: 1.0.0,
  \mn@doi{10.5281/zenodo.4044744}, \url
  {https://doi.org/10.5281/zenodo.4044744}

\bibitem[\protect\citeauthoryear{{Brinchmann}, {Charlot}, {White}, {Tremonti},
  {Kauffmann}, {Heckman}  \& {Brinkmann}}{{Brinchmann}
  et~al.}{2004}]{brinchmann2004}
{Brinchmann} J.,  {Charlot} S.,  {White} S.~D.~M.,  {Tremonti} C.,  {Kauffmann}
  G.,  {Heckman} T.,   {Brinkmann} J.,  2004, \mn@doi [\mnras]
  {10.1111/j.1365-2966.2004.07881.x}, \href
  {https://ui.adsabs.harvard.edu/abs/2004MNRAS.351.1151B} {351, 1151}

\bibitem[\protect\citeauthoryear{{Brown} \& {Gnedin}}{{Brown} \&
  {Gnedin}}{2021}]{brown2021}
{Brown} G.,  {Gnedin} O.~Y.,  2021, arXiv e-prints, \href
  {https://ui.adsabs.harvard.edu/abs/2021arXiv210612420B} {p. arXiv:2106.12420}

\bibitem[\protect\citeauthoryear{{Calzetti}, {Armus}, {Bohlin}, {Kinney},
  {Koornneef}  \& {Storchi-Bergmann}}{{Calzetti} et~al.}{2000}]{calzetti2000}
{Calzetti} D.,  {Armus} L.,  {Bohlin} R.~C.,  {Kinney} A.~L.,  {Koornneef} J.,
   {Storchi-Bergmann} T.,  2000, \mn@doi [\apj] {10.1086/308692}, \href
  {https://ui.adsabs.harvard.edu/abs/2000ApJ...533..682C} {533, 682}

\bibitem[\protect\citeauthoryear{{Cappellari}}{{Cappellari}}{2017}]{cappellari2017}
{Cappellari} M.,  2017, \mn@doi [MNRAS] {10.1093/mnras/stw3020}, 466, 798

\bibitem[\protect\citeauthoryear{{Cardelli}, {Clayton}  \& {Mathis}}{{Cardelli}
  et~al.}{1989}]{cardelli1989}
{Cardelli} J.~A.,  {Clayton} G.~C.,   {Mathis} J.~S.,  1989, \mn@doi [\apj]
  {10.1086/167900}, \href
  {https://ui.adsabs.harvard.edu/abs/1989ApJ...345..245C} {345, 245}

\bibitem[\protect\citeauthoryear{{Carollo}, {Scarlata}, {Stiavelli}, {Wyse}  \&
  {Mayer}}{{Carollo} et~al.}{2007}]{carollo2007}
{Carollo} C.~M.,  {Scarlata} C.,  {Stiavelli} M.,  {Wyse} R.~F.~G.,   {Mayer}
  L.,  2007, \mn@doi [\apj] {10.1086/511125}, \href
  {https://ui.adsabs.harvard.edu/abs/2007ApJ...658..960C} {658, 960}

\bibitem[\protect\citeauthoryear{{Cava}, {Schaerer}, {Richard},
  {P{\'e}rez-Gonz{\'a}lez}, {Dessauges-Zavadsky}, {Mayer}  \&
  {Tamburello}}{{Cava} et~al.}{2018}]{cava2018}
{Cava} A.,  {Schaerer} D.,  {Richard} J.,  {P{\'e}rez-Gonz{\'a}lez} P.~G.,
  {Dessauges-Zavadsky} M.,  {Mayer} L.,   {Tamburello} V.,  2018, \mn@doi
  [Nature Astronomy] {10.1038/s41550-017-0295-x}, \href
  {http://adsabs.harvard.edu/abs/2018NatAs...2...76C} {2, 76}

\bibitem[\protect\citeauthoryear{{Cowie}, {Hu}  \& {Songaila}}{{Cowie}
  et~al.}{1995}]{cowie1995}
{Cowie} L.~L.,  {Hu} E.~M.,   {Songaila} A.,  1995, \mn@doi [\aj]
  {10.1086/117631}, \href {http://adsabs.harvard.edu/abs/1995AJ....110.1576C}
  {110, 1576}

\bibitem[\protect\citeauthoryear{{Daddi} et~al.,}{{Daddi}
  et~al.}{2010}]{daddi2010}
{Daddi} E.,  et~al., 2010, \mn@doi [\apj] {10.1088/0004-637X/713/1/686}, \href
  {https://ui.adsabs.harvard.edu/abs/2010ApJ...713..686D} {713, 686}

\bibitem[\protect\citeauthoryear{{Dekel}, {Sari}  \& {Ceverino}}{{Dekel}
  et~al.}{2009}]{dekel2009}
{Dekel} A.,  {Sari} R.,   {Ceverino} D.,  2009, \mn@doi [\apj]
  {10.1088/0004-637X/703/1/785}, \href
  {https://ui.adsabs.harvard.edu/#abs/2009ApJ...703..785D} {703, 785}

\bibitem[\protect\citeauthoryear{{Dessauges-Zavadsky}
  et~al.,}{{Dessauges-Zavadsky} et~al.}{2019}]{dessauges2019}
{Dessauges-Zavadsky} M.,  et~al., 2019, \mn@doi [Nature Astronomy]
  {10.1038/s41550-019-0874-0}, \href
  {https://ui.adsabs.harvard.edu/abs/2019NatAs...3.1115D} {3, 1115}

\bibitem[\protect\citeauthoryear{{Dessauges-Zavadsky}, {Richard}, {Messa}  \&
  {et al.}}{{Dessauges-Zavadsky} et~al.}{2022}]{dessauges2022}
{Dessauges-Zavadsky} M.,  {Richard} J.,  {Messa} M.,   {et al.} 2022, submitted
  to \mnras

\bibitem[\protect\citeauthoryear{{Dopita} \& {Sutherland}}{{Dopita} \&
  {Sutherland}}{2003}]{dopita2003}
{Dopita} M.~A.,  {Sutherland} R.~S.,  2003, {Astrophysics of the diffuse
  universe}

\bibitem[\protect\citeauthoryear{{Elmegreen} \& {Elmegreen}}{{Elmegreen} \&
  {Elmegreen}}{2005}]{elmegreen2005}
{Elmegreen} B.~G.,  {Elmegreen} D.~M.,  2005, \mn@doi [\apj] {10.1086/430514},
  \href {https://ui.adsabs.harvard.edu/abs/2005ApJ...627..632E} {627, 632}

\bibitem[\protect\citeauthoryear{{Elmegreen}, {Elmegreen}, {Rubin}  \&
  {Schaffer}}{{Elmegreen} et~al.}{2005}]{elmegreen2005b}
{Elmegreen} D.~M.,  {Elmegreen} B.~G.,  {Rubin} D.~S.,   {Schaffer} M.~A.,
  2005, \mn@doi [\apj] {10.1086/432502}, \href
  {http://adsabs.harvard.edu/abs/2005ApJ...631...85E} {631, 85}

\bibitem[\protect\citeauthoryear{{Elmegreen}, {Elmegreen}, {Ravindranath}  \&
  {Coe}}{{Elmegreen} et~al.}{2007}]{elmegreen2007}
{Elmegreen} D.~M.,  {Elmegreen} B.~G.,  {Ravindranath} S.,   {Coe} D.~A.,
  2007, \mn@doi [\apj] {10.1086/511667}, \href
  {https://ui.adsabs.harvard.edu/abs/2007ApJ...658..763E} {658, 763}

\bibitem[\protect\citeauthoryear{{Elmegreen}, {Bournaud}  \&
  {Elmegreen}}{{Elmegreen} et~al.}{2008}]{elmegreen2008}
{Elmegreen} B.~G.,  {Bournaud} F.,   {Elmegreen} D.~M.,  2008, \mn@doi [\apj]
  {10.1086/592190}, \href
  {https://ui.adsabs.harvard.edu/abs/2008ApJ...688...67E} {688, 67}

\bibitem[\protect\citeauthoryear{{Elmegreen}, {Elmegreen}, {Fernandez}  \&
  {Lemonias}}{{Elmegreen} et~al.}{2009}]{elmegreen2009}
{Elmegreen} B.~G.,  {Elmegreen} D.~M.,  {Fernandez} M.~X.,   {Lemonias} J.~J.,
  2009, \mn@doi [\apj] {10.1088/0004-637X/692/1/12}, \href
  {https://ui.adsabs.harvard.edu/abs/2009ApJ...692...12E} {692, 12}

\bibitem[\protect\citeauthoryear{{Ferland} et~al.,}{{Ferland}
  et~al.}{2013}]{ferland2013}
{Ferland} G.~J.,  et~al., 2013, \rmxaa, \href
  {https://ui.adsabs.harvard.edu/abs/2013RMxAA..49..137F} {49, 137}

\bibitem[\protect\citeauthoryear{{Fisher} et~al.,}{{Fisher}
  et~al.}{2017a}]{fisher2017a}
{Fisher} D.~B.,  et~al., 2017a, \mn@doi [\mnras] {10.1093/mnras/stw2281}, \href
  {https://ui.adsabs.harvard.edu/abs/2017MNRAS.464..491F} {464, 491}

\bibitem[\protect\citeauthoryear{{Fisher} et~al.,}{{Fisher}
  et~al.}{2017b}]{fisher2017b}
{Fisher} D.~B.,  et~al., 2017b, \mn@doi [\apjl] {10.3847/2041-8213/aa6478},
  \href {https://ui.adsabs.harvard.edu/abs/2017ApJ...839L...5F} {839, L5}

\bibitem[\protect\citeauthoryear{{F{\"o}rster Schreiber} et~al.,}{{F{\"o}rster
  Schreiber} et~al.}{2006}]{forsterschreiber2006}
{F{\"o}rster Schreiber} N.~M.,  et~al., 2006, \mn@doi [\apj] {10.1086/504403},
  \href {https://ui.adsabs.harvard.edu/abs/2006ApJ...645.1062F} {645, 1062}

\bibitem[\protect\citeauthoryear{{F{\"o}rster Schreiber}, {Shapley}, {Erb},
  {Genzel}, {Steidel}, {Bouch{\'e}}, {Cresci}  \& {Davies}}{{F{\"o}rster
  Schreiber} et~al.}{2011a}]{forsterschreiber2011a}
{F{\"o}rster Schreiber} N.~M.,  {Shapley} A.~E.,  {Erb} D.~K.,  {Genzel} R.,
  {Steidel} C.~C.,  {Bouch{\'e}} N.,  {Cresci} G.,   {Davies} R.,  2011a,
  \mn@doi [\apj] {10.1088/0004-637X/731/1/65}, \href
  {https://ui.adsabs.harvard.edu/abs/2011ApJ...731...65F} {731, 65}

\bibitem[\protect\citeauthoryear{{F{\"o}rster Schreiber} et~al.,}{{F{\"o}rster
  Schreiber} et~al.}{2011b}]{forsterschreiber2011b}
{F{\"o}rster Schreiber} N.~M.,  et~al., 2011b, \mn@doi [\apj]
  {10.1088/0004-637X/739/1/45}, \href
  {https://ui.adsabs.harvard.edu/abs/2011ApJ...739...45F} {739, 45}

\bibitem[\protect\citeauthoryear{{Gabor} \& {Bournaud}}{{Gabor} \&
  {Bournaud}}{2013}]{gabor2013}
{Gabor} J.~M.,  {Bournaud} F.,  2013, \mn@doi [\mnras] {10.1093/mnras/stt1046},
  \href {https://ui.adsabs.harvard.edu/abs/2013MNRAS.434..606G} {434, 606}

\bibitem[\protect\citeauthoryear{{Gaia Collaboration} et~al.,}{{Gaia
  Collaboration} et~al.}{2018}]{gaia2018}
{Gaia Collaboration} et~al., 2018, \mn@doi [\aap]
  {10.1051/0004-6361/201833051}, \href
  {https://ui.adsabs.harvard.edu/abs/2018A&A...616A...1G} {616, A1}

\bibitem[\protect\citeauthoryear{{Genzel} et~al.,}{{Genzel}
  et~al.}{2006}]{genzel2006}
{Genzel} R.,  et~al., 2006, \mn@doi [\nat] {10.1038/nature05052}, \href
  {https://ui.adsabs.harvard.edu/abs/2006Natur.442..786G} {442, 786}

\bibitem[\protect\citeauthoryear{{Genzel} et~al.,}{{Genzel}
  et~al.}{2008}]{genzel2008}
{Genzel} R.,  et~al., 2008, \mn@doi [\apj] {10.1086/591840}, \href
  {https://ui.adsabs.harvard.edu/abs/2008ApJ...687...59G} {687, 59}

\bibitem[\protect\citeauthoryear{{Genzel} et~al.,}{{Genzel}
  et~al.}{2015}]{genzel2015}
{Genzel} R.,  et~al., 2015, \mn@doi [\apj] {10.1088/0004-637X/800/1/20}, \href
  {https://ui.adsabs.harvard.edu/abs/2015ApJ...800...20G} {800, 20}

\bibitem[\protect\citeauthoryear{{Gieles} \& {Portegies Zwart}}{{Gieles} \&
  {Portegies Zwart}}{2011}]{gieles2011}
{Gieles} M.,  {Portegies Zwart} S.~F.,  2011, \mn@doi [\mnras]
  {10.1111/j.1745-3933.2010.00967.x}, \href
  {https://ui.adsabs.harvard.edu/abs/2011MNRAS.410L...6G} {410, L6}

\bibitem[\protect\citeauthoryear{{Girard}, {Dessauges-Zavadsky}, {Combes},
  {Chisholm}, {Patr{\'\i}cio}, {Richard}  \& {Schaerer}}{{Girard}
  et~al.}{2019}]{girard2019}
{Girard} M.,  {Dessauges-Zavadsky} M.,  {Combes} F.,  {Chisholm} J.,
  {Patr{\'\i}cio} V.,  {Richard} J.,   {Schaerer} D.,  2019, \mn@doi [\aap]
  {10.1051/0004-6361/201935896}, \href
  {https://ui.adsabs.harvard.edu/abs/2019A&A...631A..91G} {631, A91}

\bibitem[\protect\citeauthoryear{{Goldbaum}, {Krumholz}  \&
  {Forbes}}{{Goldbaum} et~al.}{2016}]{goldbaum2016}
{Goldbaum} N.~J.,  {Krumholz} M.~R.,   {Forbes} J.~C.,  2016, \mn@doi [\apj]
  {10.3847/0004-637X/827/1/28}, \href
  {http://adsabs.harvard.edu/abs/2016ApJ...827...28G} {827, 28}

\bibitem[\protect\citeauthoryear{{Green} et~al.,}{{Green}
  et~al.}{2014}]{green2014}
{Green} A.~W.,  et~al., 2014, \mn@doi [\mnras] {10.1093/mnras/stt1882}, \href
  {https://ui.adsabs.harvard.edu/abs/2014MNRAS.437.1070G} {437, 1070}

\bibitem[\protect\citeauthoryear{{Guo}, {Giavalisco}, {Ferguson}, {Cassata}  \&
  {Koekemoer}}{{Guo} et~al.}{2012}]{guo2012}
{Guo} Y.,  {Giavalisco} M.,  {Ferguson} H.~C.,  {Cassata} P.,   {Koekemoer}
  A.~M.,  2012, \mn@doi [\apj] {10.1088/0004-637X/757/2/120}, \href
  {https://ui.adsabs.harvard.edu/abs/2012ApJ...757..120G} {757, 120}

\bibitem[\protect\citeauthoryear{{Herenz}, {Hayes}, {Papaderos}, {Cannon},
  {Bik}, {Melinder}  \& {{\"O}stlin}}{{Herenz} et~al.}{2017}]{herenz2017}
{Herenz} E.~C.,  {Hayes} M.,  {Papaderos} P.,  {Cannon} J.~M.,  {Bik} A.,
  {Melinder} J.,   {{\"O}stlin} G.,  2017, \mn@doi [\aap]
  {10.1051/0004-6361/201731809}, \href
  {https://ui.adsabs.harvard.edu/abs/2017A&A...606L..11H} {606, L11}

\bibitem[\protect\citeauthoryear{{Hoffmann}, {Mack}, {Avila}, {Martlin},
  {Cohen}  \& {Bajaj}}{{Hoffmann} et~al.}{2021}]{hoffmann2021}
{Hoffmann} S.~L.,  {Mack} J.,  {Avila} R.,  {Martlin} C.,  {Cohen} Y.,
  {Bajaj} V.,  2021, in American Astronomical Society Meeting Abstracts. p.
  216.02

\bibitem[\protect\citeauthoryear{{Hopkins}, {Quataert}  \& {Murray}}{{Hopkins}
  et~al.}{2012}]{hopkins2012}
{Hopkins} P.~F.,  {Quataert} E.,   {Murray} N.,  2012, \mn@doi [\mnras]
  {10.1111/j.1365-2966.2012.20578.x}, \href
  {https://ui.adsabs.harvard.edu/#abs/2012MNRAS.421.3488H} {421, 3488}

\bibitem[\protect\citeauthoryear{{Immeli}, {Samland}, {Gerhard}  \&
  {Westera}}{{Immeli} et~al.}{2004a}]{immeli2004b}
{Immeli} A.,  {Samland} M.,  {Gerhard} O.,   {Westera} P.,  2004a, \mn@doi
  [\aap] {10.1051/0004-6361:20034282}, \href
  {https://ui.adsabs.harvard.edu/abs/2004A&A...413..547I} {413, 547}

\bibitem[\protect\citeauthoryear{{Immeli}, {Samland}, {Westera}  \&
  {Gerhard}}{{Immeli} et~al.}{2004b}]{immeli2004a}
{Immeli} A.,  {Samland} M.,  {Westera} P.,   {Gerhard} O.,  2004b, \mn@doi
  [\apj] {10.1086/422179}, \href
  {https://ui.adsabs.harvard.edu/abs/2004ApJ...611...20I} {611, 20}

\bibitem[\protect\citeauthoryear{{Jedrzejewski}}{{Jedrzejewski}}{1987}]{jedrzejewski1987}
{Jedrzejewski} R.~I.,  1987, \mn@doi [\mnras] {10.1093/mnras/226.4.747}, \href
  {https://ui.adsabs.harvard.edu/abs/1987MNRAS.226..747J} {226, 747}

\bibitem[\protect\citeauthoryear{{Johnson} et~al.,}{{Johnson}
  et~al.}{2017}]{johnson2017}
{Johnson} T.~L.,  et~al., 2017, \mn@doi [\apjl] {10.3847/2041-8213/aa7516},
  \href {https://ui.adsabs.harvard.edu/abs/2017ApJ...843L..21J} {843, L21}

\bibitem[\protect\citeauthoryear{{Jones}, {Swinbank}, {Ellis}, {Richard}  \&
  {Stark}}{{Jones} et~al.}{2010}]{jones2010}
{Jones} T.~A.,  {Swinbank} A.~M.,  {Ellis} R.~S.,  {Richard} J.,   {Stark}
  D.~P.,  2010, \mn@doi [\mnras] {10.1111/j.1365-2966.2010.16378.x}, \href
  {https://ui.adsabs.harvard.edu/abs/2010MNRAS.404.1247J} {404, 1247}

\bibitem[\protect\citeauthoryear{{Jullo}, {Kneib}, {Limousin},
  {El{\'\i}asd{\'o}ttir}, {Marshall}  \& {Verdugo}}{{Jullo}
  et~al.}{2007}]{jullo2007}
{Jullo} E.,  {Kneib} J.~P.,  {Limousin} M.,  {El{\'\i}asd{\'o}ttir} {\'A}.,
  {Marshall} P.~J.,   {Verdugo} T.,  2007, \mn@doi [New Journal of Physics]
  {10.1088/1367-2630/9/12/447}, \href
  {https://ui.adsabs.harvard.edu/abs/2007NJPh....9..447J} {9, 447}

\bibitem[\protect\citeauthoryear{{Kennicutt} \& {Evans}}{{Kennicutt} \&
  {Evans}}{2012}]{kennicutt2012}
{Kennicutt} R.~C.,  {Evans} N.~J.,  2012, \mn@doi [Annual Review of Astronomy
  and Astrophysics] {10.1146/annurev-astro-081811-125610}, \href
  {https://ui.adsabs.harvard.edu/#abs/2012ARA&A..50..531K} {50, 531}

\bibitem[\protect\citeauthoryear{{Kennicutt} Robert~C. et~al.,}{{Kennicutt}
  et~al.}{2003}]{kennicutt2003}
{Kennicutt} Robert~C. J.,  et~al., 2003, \mn@doi [Publications of the
  Astronomical Society of the Pacific] {10.1086/376941}, \href
  {https://ui.adsabs.harvard.edu/#abs/2003PASP..115..928K} {115, 928}

\bibitem[\protect\citeauthoryear{{Kewley} \& {Dopita}}{{Kewley} \&
  {Dopita}}{2002}]{kewley2002}
{Kewley} L.~J.,  {Dopita} M.~A.,  2002, \mn@doi [\apjs] {10.1086/341326}, \href
  {https://ui.adsabs.harvard.edu/abs/2002ApJS..142...35K} {142, 35}

\bibitem[\protect\citeauthoryear{{Kroupa}}{{Kroupa}}{2001}]{kroupa2001}
{Kroupa} P.,  2001, \mn@doi [\mnras] {10.1046/j.1365-8711.2001.04022.x}, \href
  {https://ui.adsabs.harvard.edu/abs/2001MNRAS.322..231K} {322, 231}

\bibitem[\protect\citeauthoryear{{Krumholz} \& {Dekel}}{{Krumholz} \&
  {Dekel}}{2010}]{krumholz2010}
{Krumholz} M.~R.,  {Dekel} A.,  2010, \mn@doi [\mnras]
  {10.1111/j.1365-2966.2010.16675.x}, \href
  {https://ui.adsabs.harvard.edu/abs/2010MNRAS.406..112K} {406, 112}

\bibitem[\protect\citeauthoryear{{Krumholz}, {McKee}  \& {Bland
  -Hawthorn}}{{Krumholz} et~al.}{2019}]{krumholz2019}
{Krumholz} M.~R.,  {McKee} C.~F.,   {Bland -Hawthorn} J.,  2019, \mn@doi
  [\araa] {10.1146/annurev-astro-091918-104430}, \href
  {https://ui.adsabs.harvard.edu/abs/2019ARA&A..57..227K} {57, 227}

\bibitem[\protect\citeauthoryear{{Larson} et~al.,}{{Larson}
  et~al.}{2020}]{larson2020}
{Larson} K.~L.,  et~al., 2020, \mn@doi [\apj] {10.3847/1538-4357/ab5dc3}, \href
  {https://ui.adsabs.harvard.edu/abs/2020ApJ...888...92L} {888, 92}

\bibitem[\protect\citeauthoryear{{Leitherer} et~al.,}{{Leitherer}
  et~al.}{1999}]{leitherer1999}
{Leitherer} C.,  et~al., 1999, \mn@doi [\apjs] {10.1086/313233}, \href
  {https://ui.adsabs.harvard.edu/abs/1999ApJS..123....3L} {123, 3}

\bibitem[\protect\citeauthoryear{{Livermore} et~al.,}{{Livermore}
  et~al.}{2012}]{livermore2012}
{Livermore} R.~C.,  et~al., 2012, \mn@doi [\mnras]
  {10.1111/j.1365-2966.2012.21900.x}, \href
  {https://ui.adsabs.harvard.edu/#abs/2012MNRAS.427..688L} {427, 688}

\bibitem[\protect\citeauthoryear{{Livermore} et~al.,}{{Livermore}
  et~al.}{2015}]{livermore2015}
{Livermore} R.~C.,  et~al., 2015, \mn@doi [\mnras] {10.1093/mnras/stv686},
  \href {https://ui.adsabs.harvard.edu/#abs/2015MNRAS.450.1812L} {450, 1812}

\bibitem[\protect\citeauthoryear{{Ma} et~al.,}{{Ma} et~al.}{2018}]{ma2018}
{Ma} X.,  et~al., 2018, \mn@doi [\mnras] {10.1093/mnras/sty684}, \href
  {https://ui.adsabs.harvard.edu/abs/2018MNRAS.477..219M} {477, 219}

\bibitem[\protect\citeauthoryear{{Mandelker}, {Dekel}, {Ceverino}, {Tweed},
  {Moody}  \& {Primack}}{{Mandelker} et~al.}{2014}]{mandelker2014}
{Mandelker} N.,  {Dekel} A.,  {Ceverino} D.,  {Tweed} D.,  {Moody} C.~E.,
  {Primack} J.,  2014, \mn@doi [\mnras] {10.1093/mnras/stu1340}, \href
  {https://ui.adsabs.harvard.edu/abs/2014MNRAS.443.3675M} {443, 3675}

\bibitem[\protect\citeauthoryear{{Mandelker}, {Dekel}, {Ceverino}, {DeGraf},
  {Guo}  \& {Primack}}{{Mandelker} et~al.}{2017}]{mandelker2017}
{Mandelker} N.,  {Dekel} A.,  {Ceverino} D.,  {DeGraf} C.,  {Guo} Y.,
  {Primack} J.,  2017, \mn@doi [\mnras] {10.1093/mnras/stw2358}, \href
  {https://ui.adsabs.harvard.edu/abs/2017MNRAS.464..635M} {464, 635}

\bibitem[\protect\citeauthoryear{{Meng} \& {Gnedin}}{{Meng} \&
  {Gnedin}}{2020}]{meng2020}
{Meng} X.,  {Gnedin} O.~Y.,  2020, \mn@doi [\mnras] {10.1093/mnras/staa776},
  \href {https://ui.adsabs.harvard.edu/abs/2020MNRAS.494.1263M} {494, 1263}

\bibitem[\protect\citeauthoryear{{Messa}, {Adamo}, {{\"O}stlin}, {Melinder},
  {Hayes}, {Bridge}  \& {Cannon}}{{Messa} et~al.}{2019}]{messa2019}
{Messa} M.,  {Adamo} A.,  {{\"O}stlin} G.,  {Melinder} J.,  {Hayes} M.,
  {Bridge} J.~S.,   {Cannon} J.,  2019, \mn@doi [\mnras]
  {10.1093/mnras/stz1337}, \href
  {https://ui.adsabs.harvard.edu/abs/2019MNRAS.487.4238M} {487, 4238}

\bibitem[\protect\citeauthoryear{{Me{\v{s}}tri{\'c}}
  et~al.,}{{Me{\v{s}}tri{\'c}} et~al.}{2022}]{mestric2022}
{Me{\v{s}}tri{\'c}} U.,  et~al., 2022, \mn@doi [arXiv e-prints]
  {10.48550/arXiv.2202.09377}, \href
  {https://ui.adsabs.harvard.edu/abs/2022arXiv220209377M} {p. arXiv:2202.09377}

\bibitem[\protect\citeauthoryear{{Mieda}, {Wright}, {Larkin}, {Armus},
  {Juneau}, {Salim}  \& {Murray}}{{Mieda} et~al.}{2016}]{mieda2016}
{Mieda} E.,  {Wright} S.~A.,  {Larkin} J.~E.,  {Armus} L.,  {Juneau} S.,
  {Salim} S.,   {Murray} N.,  2016, \mn@doi [\apj]
  {10.3847/0004-637X/831/1/78}, \href
  {https://ui.adsabs.harvard.edu/abs/2016ApJ...831...78M} {831, 78}

\bibitem[\protect\citeauthoryear{{Nagy}, {Dessauges-Zavadsky}, {Richard},
  {Schaerer}, {Combes}, {Messa}  \& {Chisholm}}{{Nagy} et~al.}{2021}]{nagy2021}
{Nagy} D.,  {Dessauges-Zavadsky} M.,  {Richard} J.,  {Schaerer} D.,  {Combes}
  F.,  {Messa} M.,   {Chisholm} J.,  2021, arXiv e-prints, \href
  {https://ui.adsabs.harvard.edu/abs/2021arXiv210906206N} {p. arXiv:2109.06206}

\bibitem[\protect\citeauthoryear{Newville et~al.,}{Newville
  et~al.}{2021}]{lmfit}
Newville M.,  et~al., 2021, lmfit/lmfit-py 1.0.2,
  \mn@doi{10.5281/zenodo.4516651}, \url
  {https://doi.org/10.5281/zenodo.4516651}

\bibitem[\protect\citeauthoryear{{Noguchi}}{{Noguchi}}{1999}]{noguchi1999}
{Noguchi} M.,  1999, \mn@doi [\apj] {10.1086/306932}, \href
  {https://ui.adsabs.harvard.edu/abs/1999ApJ...514...77N} {514, 77}

\bibitem[\protect\citeauthoryear{{Oklop{\v{c}}i{\'c}}, {Hopkins}, {Feldmann},
  {Kere{\v{s}}}, {Faucher-Gigu{\`e}re}  \& {Murray}}{{Oklop{\v{c}}i{\'c}}
  et~al.}{2017}]{oklopcic2017}
{Oklop{\v{c}}i{\'c}} A.,  {Hopkins} P.~F.,  {Feldmann} R.,  {Kere{\v{s}}} D.,
  {Faucher-Gigu{\`e}re} C.-A.,   {Murray} N.,  2017, \mn@doi [\mnras]
  {10.1093/mnras/stw2754}, \href
  {https://ui.adsabs.harvard.edu/abs/2017MNRAS.465..952O} {465, 952}

\bibitem[\protect\citeauthoryear{{{\"O}stlin} et~al.,}{{{\"O}stlin}
  et~al.}{2014}]{ostlin2014}
{{\"O}stlin} G.,  et~al., 2014, \mn@doi [\apj] {10.1088/0004-637X/797/1/11},
  \href {https://ui.adsabs.harvard.edu/abs/2014ApJ...797...11O} {797, 11}

\bibitem[\protect\citeauthoryear{{Patr{\'\i}cio} et~al.,}{{Patr{\'\i}cio}
  et~al.}{2018}]{patricio2018}
{Patr{\'\i}cio} V.,  et~al., 2018, \mn@doi [\mnras] {10.1093/mnras/sty555},
  \href {https://ui.adsabs.harvard.edu/abs/2018MNRAS.477...18P} {477, 18}

\bibitem[\protect\citeauthoryear{{Planck Collaboration} et~al.,}{{Planck
  Collaboration} et~al.}{2014}]{planck13_cosmo}
{Planck Collaboration} et~al., 2014, \mn@doi [\aap]
  {10.1051/0004-6361/201321591}, \href
  {https://ui.adsabs.harvard.edu/abs/2014A&A...571A..16P} {571, A16}

\bibitem[\protect\citeauthoryear{{Renaud}, {Romeo}  \& {Agertz}}{{Renaud}
  et~al.}{2021}]{renaud2021}
{Renaud} F.,  {Romeo} A.~B.,   {Agertz} O.,  2021, \mn@doi [\mnras]
  {10.1093/mnras/stab2604}, \href
  {https://ui.adsabs.harvard.edu/abs/2021MNRAS.508..352R} {508, 352}

\bibitem[\protect\citeauthoryear{{Richard} et~al.,}{{Richard}
  et~al.}{2010}]{richard2010}
{Richard} J.,  et~al., 2010, \mn@doi [\mnras]
  {10.1111/j.1365-2966.2009.16274.x}, \href
  {https://ui.adsabs.harvard.edu/abs/2010MNRAS.404..325R} {404, 325}

\bibitem[\protect\citeauthoryear{{Richard} et~al.,}{{Richard}
  et~al.}{2014}]{richard2014}
{Richard} J.,  et~al., 2014, \mn@doi [\mnras] {10.1093/mnras/stu1395}, \href
  {https://ui.adsabs.harvard.edu/abs/2014MNRAS.444..268R} {444, 268}

\bibitem[\protect\citeauthoryear{{Rivera-Thorsen} et~al.,}{{Rivera-Thorsen}
  et~al.}{2019}]{riverathorsen2019}
{Rivera-Thorsen} T.~E.,  et~al., 2019, \mn@doi [Science]
  {10.1126/science.aaw0978}, \href
  {https://ui.adsabs.harvard.edu/abs/2019Sci...366..738R} {366, 738}

\bibitem[\protect\citeauthoryear{{Ryon} et~al.,}{{Ryon}
  et~al.}{2015}]{ryon2015}
{Ryon} J.~E.,  et~al., 2015, \mn@doi [\mnras] {10.1093/mnras/stv1282}, \href
  {https://ui.adsabs.harvard.edu/abs/2015MNRAS.452..525R} {452, 525}

\bibitem[\protect\citeauthoryear{{Ryon} et~al.,}{{Ryon}
  et~al.}{2017}]{ryon2017}
{Ryon} J.~E.,  et~al., 2017, \mn@doi [\apj] {10.3847/1538-4357/aa719e}, \href
  {https://ui.adsabs.harvard.edu/abs/2017ApJ...841...92R} {841, 92}

\bibitem[\protect\citeauthoryear{{Salpeter}}{{Salpeter}}{1955}]{salpeter1955}
{Salpeter} E.~E.,  1955, \mn@doi [\apj] {10.1086/145971}, \href
  {https://ui.adsabs.harvard.edu/abs/1955ApJ...121..161S} {121, 161}

\bibitem[\protect\citeauthoryear{{Shapiro} et~al.,}{{Shapiro}
  et~al.}{2008}]{shapiro2008}
{Shapiro} K.~L.,  et~al., 2008, \mn@doi [\apj] {10.1086/587133}, \href
  {https://ui.adsabs.harvard.edu/abs/2008ApJ...682..231S} {682, 231}

\bibitem[\protect\citeauthoryear{{Soto} et~al.,}{{Soto}
  et~al.}{2017}]{soto2017}
{Soto} E.,  et~al., 2017, \mn@doi [\apj] {10.3847/1538-4357/aa5da3}, \href
  {https://ui.adsabs.harvard.edu/abs/2017ApJ...837....6S} {837, 6}

\bibitem[\protect\citeauthoryear{{Speagle}, {Steinhardt}, {Capak}  \&
  {Silverman}}{{Speagle} et~al.}{2014}]{speagle2014}
{Speagle} J.~S.,  {Steinhardt} C.~L.,  {Capak} P.~L.,   {Silverman} J.~D.,
  2014, \mn@doi [\apjs] {10.1088/0067-0049/214/2/15}, \href
  {https://ui.adsabs.harvard.edu/abs/2014ApJS..214...15S} {214, 15}

\bibitem[\protect\citeauthoryear{{Storey} \& {Hummer}}{{Storey} \&
  {Hummer}}{1995}]{storey1995}
{Storey} P.~J.,  {Hummer} D.~G.,  1995, \mn@doi [\mnras]
  {10.1093/mnras/272.1.41}, \href
  {https://ui.adsabs.harvard.edu/abs/1995MNRAS.272...41S} {272, 41}

\bibitem[\protect\citeauthoryear{{Swinbank}, {Bower}, {Smith}, {Wilman},
  {Smail}, {Ellis}, {Morris}  \& {Kneib}}{{Swinbank}
  et~al.}{2007}]{swinbank2007}
{Swinbank} A.~M.,  {Bower} R.~G.,  {Smith} G.~P.,  {Wilman} R.~J.,  {Smail} I.,
   {Ellis} R.~S.,  {Morris} S.~L.,   {Kneib} J.~P.,  2007, \mn@doi [\mnras]
  {10.1111/j.1365-2966.2007.11454.x}, \href
  {https://ui.adsabs.harvard.edu/abs/2007MNRAS.376..479S} {376, 479}

\bibitem[\protect\citeauthoryear{{Swinbank} et~al.,}{{Swinbank}
  et~al.}{2009}]{swinbank2009}
{Swinbank} A.~M.,  et~al., 2009, \mn@doi [\mnras]
  {10.1111/j.1365-2966.2009.15617.x}, \href
  {https://ui.adsabs.harvard.edu/abs/2009MNRAS.400.1121S} {400, 1121}

\bibitem[\protect\citeauthoryear{{Swinbank}, {Smail}, {Sobral}, {Theuns},
  {Best}  \& {Geach}}{{Swinbank} et~al.}{2012}]{swinbank2012}
{Swinbank} A.~M.,  {Smail} I.,  {Sobral} D.,  {Theuns} T.,  {Best} P.~N.,
  {Geach} J.~E.,  2012, \mn@doi [\apj] {10.1088/0004-637X/760/2/130}, \href
  {https://ui.adsabs.harvard.edu/abs/2012ApJ...760..130S} {760, 130}

\bibitem[\protect\citeauthoryear{{Tacconi} et~al.,}{{Tacconi}
  et~al.}{2010}]{tacconi2010}
{Tacconi} L.~J.,  et~al., 2010, \mn@doi [\nat] {10.1038/nature08773}, \href
  {https://ui.adsabs.harvard.edu/abs/2010Natur.463..781T} {463, 781}

\bibitem[\protect\citeauthoryear{{Tacconi} et~al.,}{{Tacconi}
  et~al.}{2013}]{tacconi2013}
{Tacconi} L.~J.,  et~al., 2013, \mn@doi [\apj] {10.1088/0004-637X/768/1/74},
  \href {https://ui.adsabs.harvard.edu/abs/2013ApJ...768...74T} {768, 74}

\bibitem[\protect\citeauthoryear{{Tamburello}, {Mayer}, {Shen}  \&
  {Wadsley}}{{Tamburello} et~al.}{2015}]{tamburello2015}
{Tamburello} V.,  {Mayer} L.,  {Shen} S.,   {Wadsley} J.,  2015, \mn@doi
  [\mnras] {10.1093/mnras/stv1695}, \href
  {https://ui.adsabs.harvard.edu/abs/2015MNRAS.453.2490T} {453, 2490}

\bibitem[\protect\citeauthoryear{{Tamburello}, {Rahmati}, {Mayer}, {Cava},
  {Dessauges-Zavadsky}  \& {Schaerer}}{{Tamburello}
  et~al.}{2017}]{tamburello2017}
{Tamburello} V.,  {Rahmati} A.,  {Mayer} L.,  {Cava} A.,  {Dessauges-Zavadsky}
  M.,   {Schaerer} D.,  2017, \mn@doi [\mnras] {10.1093/mnras/stx784}, \href
  {https://ui.adsabs.harvard.edu/abs/2017MNRAS.468.4792T} {468, 4792}

\bibitem[\protect\citeauthoryear{{Tremonti} et~al.,}{{Tremonti}
  et~al.}{2004}]{tremonti2004}
{Tremonti} C.~A.,  et~al., 2004, \mn@doi [\apj] {10.1086/423264}, \href
  {https://ui.adsabs.harvard.edu/abs/2004ApJ...613..898T} {613, 898}

\bibitem[\protect\citeauthoryear{{\VAN{Vanden}{van}{van} den Bergh}, {Abraham},
  {Ellis}, {Tanvir}, {Santiago}  \& {Glazebrook}}{{\VAN{Vanden}{van}{van} den
  Bergh} et~al.}{1996}]{vandenbergh1996}
{\VAN{Vanden}{van}{van} den Bergh} S.,  {Abraham} R.~G.,  {Ellis} R.~S.,
  {Tanvir} N.~R.,  {Santiago} B.~X.,   {Glazebrook} K.~G.,  1996, \mn@doi [\aj]
  {10.1086/118020}, \href {http://adsabs.harvard.edu/abs/1996AJ....112..359V}
  {112, 359}

\bibitem[\protect\citeauthoryear{{\VAN{Vandon}{van}{van}}~Donkelaar, {Agertz}
  \& {Renaud}}{{\VAN{Vandon}{van}{van}}~Donkelaar
  et~al.}{2021}]{vandonkelaar2021arxiv}
{\VAN{Vandon}{van}{van}}~Donkelaar F.,  {Agertz} O.,   {Renaud} F.,  2021,
  arXiv e-prints, \href {https://ui.adsabs.harvard.edu/abs/2021arXiv211013165V}
  {p. arXiv:2110.13165}

\bibitem[\protect\citeauthoryear{{Vanzella} et~al.,}{{Vanzella}
  et~al.}{2017a}]{vanzella2017a}
{Vanzella} E.,  et~al., 2017a, \mn@doi [\mnras] {10.1093/mnras/stx351}, \href
  {https://ui.adsabs.harvard.edu/abs/2017MNRAS.467.4304V} {467, 4304}

\bibitem[\protect\citeauthoryear{{Vanzella} et~al.,}{{Vanzella}
  et~al.}{2017b}]{vanzella2017b}
{Vanzella} E.,  et~al., 2017b, \mn@doi [\apj] {10.3847/1538-4357/aa74ae}, \href
  {https://ui.adsabs.harvard.edu/abs/2017ApJ...842...47V} {842, 47}

\bibitem[\protect\citeauthoryear{{Vanzella} et~al.,}{{Vanzella}
  et~al.}{2019}]{vanzella2019}
{Vanzella} E.,  et~al., 2019, \mn@doi [\mnras] {10.1093/mnras/sty3311}, \href
  {https://ui.adsabs.harvard.edu/abs/2019MNRAS.483.3618V} {483, 3618}

\bibitem[\protect\citeauthoryear{{Vanzella} et~al.,}{{Vanzella}
  et~al.}{2021a}]{vanzella2021b}
{Vanzella} E.,  et~al., 2021a, arXiv e-prints, \href
  {https://ui.adsabs.harvard.edu/abs/2021arXiv210610280V} {p. arXiv:2106.10280}

\bibitem[\protect\citeauthoryear{{Vanzella} et~al.,}{{Vanzella}
  et~al.}{2021b}]{vanzella2021}
{Vanzella} E.,  et~al., 2021b, \mn@doi [\aap] {10.1051/0004-6361/202039466},
  \href {https://ui.adsabs.harvard.edu/abs/2021A&A...646A..57V} {646, A57}

\bibitem[\protect\citeauthoryear{{V{\'a}zquez} \& {Leitherer}}{{V{\'a}zquez} \&
  {Leitherer}}{2005}]{vazquez2005}
{V{\'a}zquez} G.~A.,  {Leitherer} C.,  2005, \mn@doi [\apj] {10.1086/427866},
  \href {https://ui.adsabs.harvard.edu/abs/2005ApJ...621..695V} {621, 695}

\bibitem[\protect\citeauthoryear{{Wisnioski}, {Glazebrook}, {Blake}, {Poole},
  {Green}, {Wyder}  \& {Martin}}{{Wisnioski} et~al.}{2012}]{wisnioski2012}
{Wisnioski} E.,  {Glazebrook} K.,  {Blake} C.,  {Poole} G.~B.,  {Green} A.~W.,
  {Wyder} T.,   {Martin} C.,  2012, \mn@doi [\mnras]
  {10.1111/j.1365-2966.2012.20850.x}, \href
  {https://ui.adsabs.harvard.edu/#abs/2012MNRAS.422.3339W} {422, 3339}

\bibitem[\protect\citeauthoryear{{Wisnioski} et~al.,}{{Wisnioski}
  et~al.}{2018}]{wisnioski2018}
{Wisnioski} E.,  et~al., 2018, \mn@doi [\apj] {10.3847/1538-4357/aab097}, \href
  {https://ui.adsabs.harvard.edu/abs/2018ApJ...855...97W} {855, 97}

\bibitem[\protect\citeauthoryear{{Wuyts}, {Rigby}, {Gladders}  \&
  {Sharon}}{{Wuyts} et~al.}{2014}]{wuyts2014}
{Wuyts} E.,  {Rigby} J.~R.,  {Gladders} M.~D.,   {Sharon} K.,  2014, \mn@doi
  [\apj] {10.1088/0004-637X/781/2/61}, \href
  {https://ui.adsabs.harvard.edu/abs/2014ApJ...781...61W} {781, 61}

\bibitem[\protect\citeauthoryear{{Zackrisson}, {Rydberg}, {Schaerer},
  {{\"O}stlin}  \& {Tuli}}{{Zackrisson} et~al.}{2011}]{zackrisson2011}
{Zackrisson} E.,  {Rydberg} C.-E.,  {Schaerer} D.,  {{\"O}stlin} G.,   {Tuli}
  M.,  2011, \mn@doi [\apj] {10.1088/0004-637X/740/1/13}, \href
  {https://ui.adsabs.harvard.edu/abs/2011ApJ...740...13Z} {740, 13}

\bibitem[\protect\citeauthoryear{{Zanella} et~al.,}{{Zanella}
  et~al.}{2015}]{zanella2015}
{Zanella} A.,  et~al., 2015, \mn@doi [\nat] {10.1038/nature14409}, \href
  {https://ui.adsabs.harvard.edu/abs/2015Natur.521...54Z} {521, 54}

\bibitem[\protect\citeauthoryear{{Zanella} et~al.,}{{Zanella}
  et~al.}{2019}]{zanella2019}
{Zanella} A.,  et~al., 2019, \mn@doi [\mnras] {10.1093/mnras/stz2099}, \href
  {https://ui.adsabs.harvard.edu/abs/2019MNRAS.489.2792Z} {489, 2792}

\makeatother
\end{thebibliography}




\appendix

\section{Supplementary photometric table and figures}
\label{sec:app:completetab}
\begin{figure*}
    \centering
        \subfigure{\includegraphics[width=0.9\textwidth]{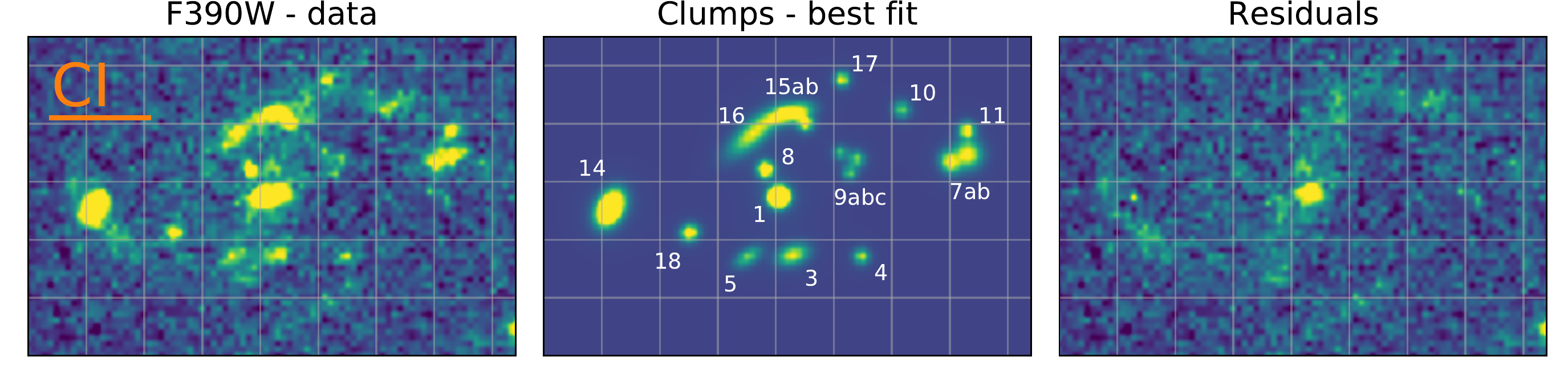}}
    \subfigure{\includegraphics[width=0.9\textwidth]{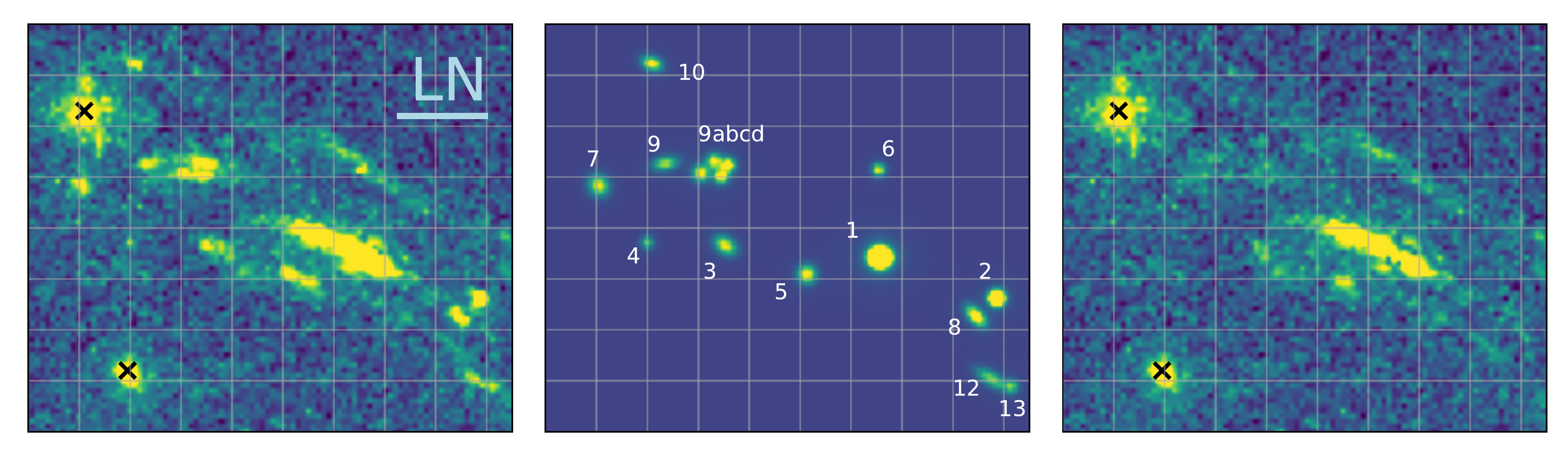}}
    \subfigure{\includegraphics[width=0.9\textwidth]{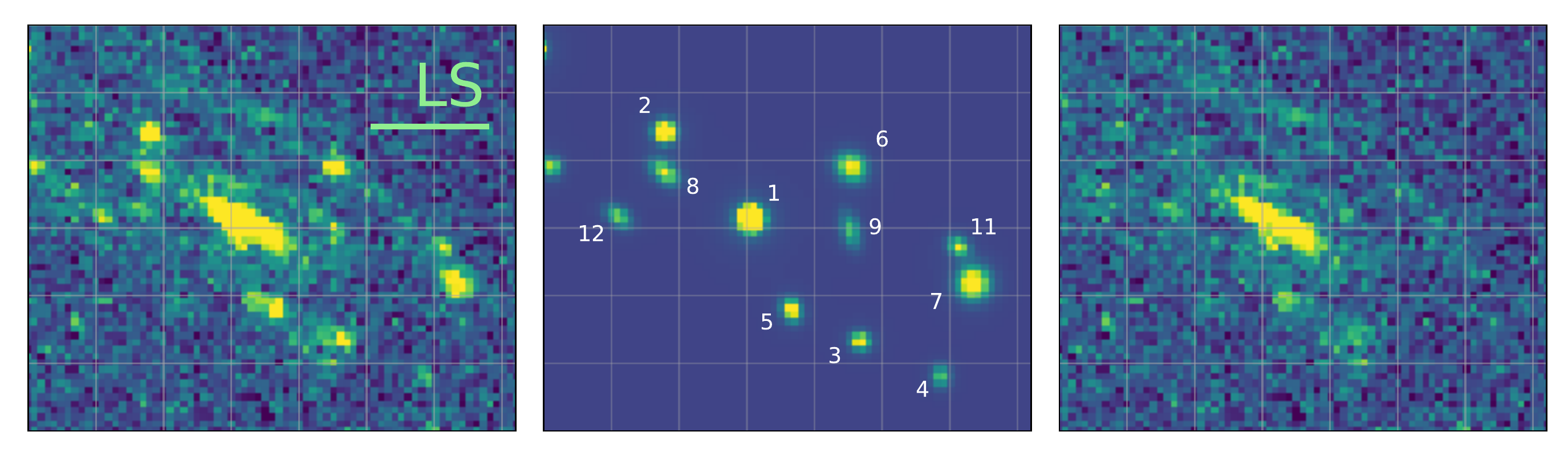}}
    \caption{F390W observations (left column), best-fit F390W models (central column) and residuals (right column), for the three A521-sys1 sub--regions, CI (top row), LN (middle row) and LS (bottom row). In each case a line corresponding to 1 arcsec is given at the bottom of the sub--region name. Foreground galaxies are marked as black crosses in the LN panels. Clumps IDs are reported in the central panels. The bright residual in the inner part of the galaxy corresponds to the `tail' of clump 1; while it is not considered as a source in the reference photometry, it is analyzed when the alternative photometric method is adopted, as discussed in Section~\ref{sec:alternative_results} of the main text. Grids with 0.6 arcsec size are plotted to facilitate the comparison between panels.}
    \label{fig:app:hstdata_390}
\end{figure*}

\begin{figure*}
    \centering
    \includegraphics[width=0.87\textwidth]{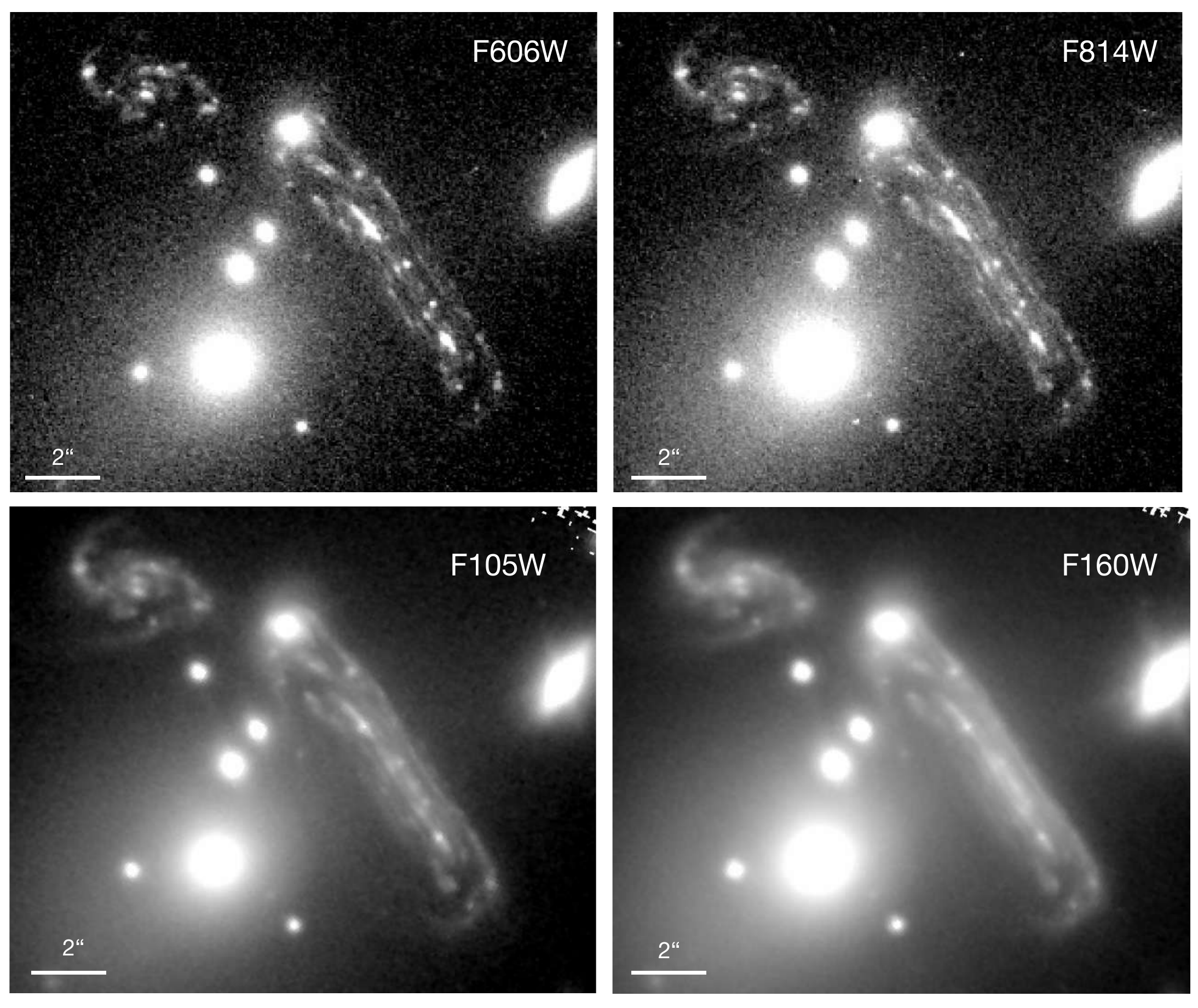}
    \caption{Observations for F606W, F814W, F105W and F160W, corresponding to rest-frame central wavelengths of 2900, 3900, 5200 and 7500 \AA. The complete photometry of the clump sample is presented in Tab.~\ref{tab:photometry}. As discussed in Section~\ref{sec:clumpiness} of the main text, the galaxy appears less clumpy when moving to longer wavelengths.}
    \label{fig:app:hstdata}
\end{figure*}

We report in Tab.~\ref{tab:photometry} the clump photometry in all filters; we provide apparent magnitudes (and uncertainties), corrected for Galactic reddening, but uncorrected for lensing. 
Data, best--fit clump models and fit residuals in F390W are shown in Fig.~\ref{fig:app:hstdata_390}; the observations in all the the other filters are shown in Fig.~\ref{fig:app:hstdata}.

\begin{table*}
    \centering
    \begin{tabular}{lccccc}
        \multicolumn{1}{c}{ID} & $\rm mag_{F390W}$ & $\rm mag_{F606W}$ & $\rm mag_{F814W}$ & $\rm mag_{F105W}$ & $\rm mag_{F160W}$ \\ 
        \multicolumn{1}{c}{(0)} & (1) & (2) & (3) & (4) & (5)  \\ 
\hline
\hline
ci\_1	& $24.45^{\pm0.05}$ & $24.34^{\pm0.05}$ & $24.30^{\pm0.10}$ & $24.12^{\pm0.21}$ & $24.71^{\pm0.05}$  \\
ci\_3	& $25.75^{\pm0.21}$ & $25.56^{\pm0.09}$ & $24.89^{\pm0.11}$ & $24.74^{\pm0.07}$ & $24.42^{\pm0.07}$  \\
ci\_4	& $26.79^{\pm0.21}$ & $25.72^{\pm0.08}$ & $25.17^{\pm0.08}$ & $24.67^{\pm0.07}$ & $24.22^{\pm0.07}$  \\
ci\_5	& $26.39^{\pm0.21}$ & $26.69^{\pm0.21}$ & $26.02^{\pm0.21}$ & $26.04^{\pm0.26}$ & $26.64^{\pm0.64}$  \\
ci\_7a	& $26.02^{\pm0.23}$ & $25.39^{\pm0.14}$ & $24.82^{\pm0.12}$ & $24.36^{\pm0.08}$ & $24.26^{\pm0.10}$  \\
ci\_7b	& $25.43^{\pm0.22}$ & $25.10^{\pm0.13}$ & $24.74^{\pm0.16}$ & $24.50^{\pm0.09}$ & $24.36^{\pm0.11}$  \\
ci\_8	& $26.01^{\pm0.13}$ & $25.89^{\pm0.11}$ & $25.25^{\pm0.10}$ & $24.23^{\pm0.07}$ & $23.33^{\pm0.07}$  \\
ci\_9a	& $27.44^{\pm0.33}$ & $27.63^{\pm0.05}$ & $27.84^{\pm0.05}$ & $26.42^{\pm0.05}$ & $26.00^{\pm0.05}$  \\
ci\_9b	& $26.91^{\pm0.24}$ & $27.30^{\pm0.05}$ & $26.98^{\pm0.05}$ & $26.63^{\pm0.05}$ & $---$  \\
ci\_9c	& $27.26^{\pm0.32}$ & $27.28^{\pm0.05}$ & $26.33^{\pm0.05}$ & $26.05^{\pm0.05}$ & $26.88^{\pm0.05}$  \\
ci\_10	& $26.83^{\pm0.38}$ & $26.37^{\pm0.18}$ & $25.81^{\pm0.18}$ & $25.15^{\pm0.15}$ & $24.76^{\pm0.12}$  \\
ci\_11	& $26.38^{\pm0.15}$ & $25.96^{\pm0.12}$ & $25.82^{\pm0.19}$ & $25.60^{\pm0.05}$ & $25.36^{\pm0.25}$  \\
ci\_14	& $24.30^{\pm0.10}$ & $24.06^{\pm0.07}$ & $23.58^{\pm0.08}$ & $23.30^{\pm0.07}$ & $23.09^{\pm0.08}$  \\
ci\_15a	& $25.06^{\pm0.05}$ & $24.66^{\pm0.08}$ & $24.32^{\pm0.11}$ & $23.91^{\pm0.05}$ & $24.21^{\pm0.05}$  \\
ci\_15b	& $26.43^{\pm0.05}$ & $26.13^{\pm0.16}$ & $25.46^{\pm0.13}$ & $25.26^{\pm0.05}$ & $24.40^{\pm0.05}$  \\
ci\_16	& $24.88^{\pm0.23}$ & $24.64^{\pm0.09}$ & $24.07^{\pm0.14}$ & $23.93^{\pm0.08}$ & $24.29^{\pm0.19}$  \\
ci\_17	& $26.68^{\pm0.23}$ & $25.96^{\pm0.09}$ & $26.18^{\pm0.13}$ & $25.93^{\pm0.11}$ & $26.63^{\pm0.36}$  \\
ci\_18	& $26.18^{\pm0.14}$ & $25.49^{\pm0.08}$ & $25.29^{\pm0.10}$ & $24.98^{\pm0.11}$ & $24.52^{\pm0.12}$  \\
ln\_1	& $23.34^{\pm0.05}$ & $23.14^{\pm0.05}$ & $23.32^{\pm0.06}$ & $23.78^{\pm0.05}$ & $23.51^{\pm0.08}$  \\
ln\_2	& $25.00^{\pm0.12}$ & $24.51^{\pm0.06}$ & $24.68^{\pm0.07}$ & $24.58^{\pm0.07}$ & $24.47^{\pm0.05}$  \\
ln\_3	& $25.86^{\pm0.23}$ & $25.23^{\pm0.10}$ & $24.68^{\pm0.09}$ & $24.36^{\pm0.08}$ & $24.22^{\pm0.08}$  \\
ln\_4	& $27.19^{\pm0.27}$ & $26.18^{\pm0.11}$ & $25.66^{\pm0.11}$ & $25.09^{\pm0.09}$ & $24.73^{\pm0.08}$  \\
ln\_5	& $25.95^{\pm0.17}$ & $25.90^{\pm0.13}$ & $25.68^{\pm0.19}$ & $25.37^{\pm0.11}$ & $25.39^{\pm0.10}$  \\
ln\_6	& $26.72^{\pm0.33}$ & $26.01^{\pm0.15}$ & $25.86^{\pm0.17}$ & $25.47^{\pm0.18}$ & $26.08^{\pm0.26}$  \\
ln\_7	& $25.84^{\pm0.16}$ & $25.41^{\pm0.10}$ & $24.81^{\pm0.10}$ & $24.62^{\pm0.11}$ & $24.74^{\pm0.18}$  \\
ln\_8	& $25.71^{\pm0.16}$ & $25.20^{\pm0.08}$ & $24.77^{\pm0.11}$ & $24.35^{\pm0.08}$ & $23.88^{\pm0.05}$  \\
ln\_9	& $26.13^{\pm0.15}$ & $25.95^{\pm0.13}$ & $25.34^{\pm0.13}$ & $25.07^{\pm0.12}$ & $25.52^{\pm0.41}$  \\
ln\_9a	& $26.25^{\pm0.18}$ & $25.98^{\pm0.05}$ & $25.53^{\pm0.18}$ & $25.17^{\pm0.05}$ & $25.19^{\pm0.05}$  \\
ln\_9b	& $26.29^{\pm0.18}$ & $25.63^{\pm0.05}$ & $25.21^{\pm0.14}$ & $24.94^{\pm0.05}$ & $24.84^{\pm0.05}$  \\
ln\_9c	& $26.19^{\pm0.19}$ & $25.53^{\pm0.05}$ & $24.77^{\pm0.10}$ & $24.27^{\pm0.05}$ & $24.06^{\pm0.05}$  \\
ln\_9d	& $26.35^{\pm0.20}$ & $26.07^{\pm0.05}$ & $25.75^{\pm0.23}$ & $25.67^{\pm0.05}$ & $26.73^{\pm0.05}$  \\
ln\_10	& $26.10^{\pm0.13}$ & $25.79^{\pm0.10}$ & $25.31^{\pm0.14}$ & $25.29^{\pm0.19}$ & $25.98^{\pm0.85}$  \\
ln\_12	& $25.96^{\pm0.21}$ & $25.43^{\pm0.09}$ & $25.18^{\pm0.05}$ & $25.33^{\pm0.09}$ & $---$  \\
ln\_13	& $27.02^{\pm0.37}$ & $28.44^{\pm0.72}$ & $28.77^{\pm0.05}$ & $---$ & $---$  \\
ls\_1	& $24.00^{\pm0.05}$ & $23.81^{\pm0.06}$ & $23.93^{\pm0.07}$ & $23.91^{\pm0.05}$ & $24.31^{\pm0.11}$  \\
ls\_2	& $25.33^{\pm0.08}$ & $24.74^{\pm0.07}$ & $24.51^{\pm0.07}$ & $24.75^{\pm0.07}$ & $24.73^{\pm0.05}$  \\
ls\_3	& $26.42^{\pm0.19}$ & $25.67^{\pm0.07}$ & $25.46^{\pm0.09}$ & $24.94^{\pm0.08}$ & $24.68^{\pm0.05}$  \\
ls\_4	& $27.00^{\pm0.23}$ & $26.00^{\pm0.08}$ & $25.29^{\pm0.09}$ & $24.87^{\pm0.07}$ & $24.50^{\pm0.07}$  \\
ls\_5	& $26.25^{\pm0.18}$ & $26.17^{\pm0.12}$ & $26.02^{\pm0.18}$ & $25.62^{\pm0.18}$ & $25.67^{\pm0.22}$  \\
ls\_6	& $25.67^{\pm0.17}$ & $25.39^{\pm0.11}$ & $24.89^{\pm0.12}$ & $24.72^{\pm0.11}$ & $24.56^{\pm0.05}$  \\
ls\_7	& $25.20^{\pm0.15}$ & $24.88^{\pm0.08}$ & $24.22^{\pm0.09}$ & $23.96^{\pm0.08}$ & $23.85^{\pm0.09}$  \\
ls\_8	& $26.10^{\pm0.15}$ & $25.15^{\pm0.08}$ & $24.86^{\pm0.10}$ & $24.32^{\pm0.09}$ & $23.92^{\pm0.09}$  \\
ls\_9	& $26.55^{\pm0.25}$ & $26.03^{\pm0.11}$ & $25.90^{\pm0.13}$ & $25.36^{\pm0.11}$ & $25.75^{\pm0.05}$  \\
ls\_11	& $26.46^{\pm0.13}$ & $26.03^{\pm0.09}$ & $25.98^{\pm0.16}$ & $25.30^{\pm0.09}$ & $25.20^{\pm0.05}$  \\
ls\_12	& $26.57^{\pm0.22}$ & $26.17^{\pm0.16}$ & $26.11^{\pm0.25}$ & $26.37^{\pm0.26}$ & $27.33^{\pm0.88}$  \\
\hline
    \end{tabular}
    \caption{Apparent AB magnitudes (and relative uncertainties), corrected for Galactic reddening. Empty entries indicate a non-detection in the corresponding filter.}
   \label{tab:photometry}
\end{table*}

\section{Update on lensing model}
\label{sec:app:lensmodel}
The starting point of our lens model is the LoCuSS cluster mass model presented in \citet{richard2010}, which was based on a limited number of star-forming clumps in the giant arc at $z=1$. The cluster RXCJ0454 has the smallest Einstein radius ($3.6"$) among the 20 LoCuSS clusters analysed in \citet{richard2010}, making it more similar to a group-like lens dominated by the brightest cluster galaxy (BCG). We have followed here the same approach in the parametrisation but improved the model to include new constraints from HST images and cluster members identified in the MUSE observations, and summarise here the elements of the modelling. The mass distribution of the cluster is parametrised as the sum of double Pseudo Isothermal Elliptical (dPIE) potentials: 1 cluster-scale component and multiple galaxy-scale components. These potentials are characterised by the center, ellipticity and position angle, velocity dispersion $\sigma$ and two characteristic radii $r_{\rm core}$ and $r_{\rm cut}$.

\begin{figure*}
    \centering
    \includegraphics[width=\textwidth]{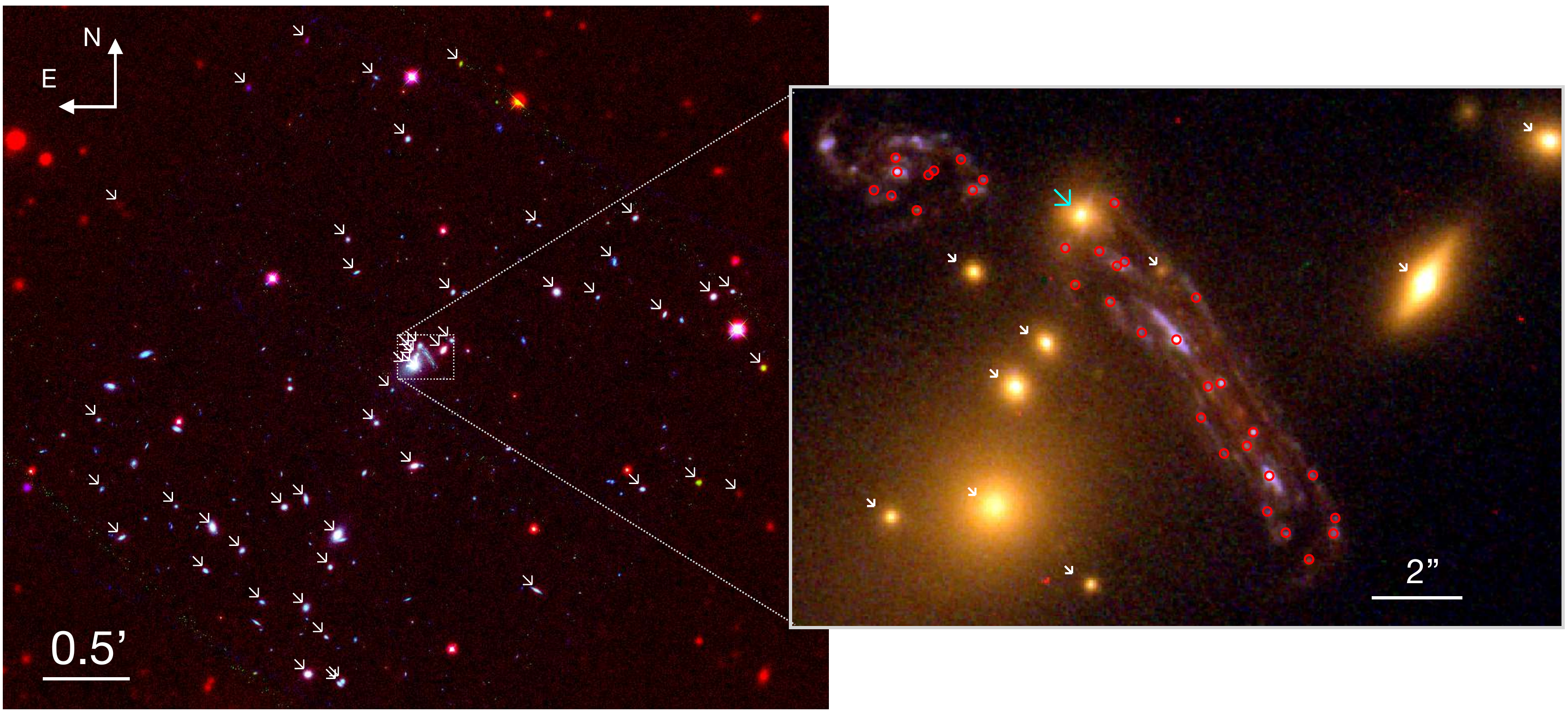}
    \caption{A521 cluster members (white arrows) and multiple images used to constrain the lens model (red circles). The inset show a zoom-in to the central region of the cluster, containing the BCG, the galaxy GAL1 (large cyan arrow) and the images of A521-sys1.}
    \label{fig:app:members}
\end{figure*}

We have selected color-selected cluster members from \citet{richard2010}, complemented by spectroscopically-confirmed cluster members from MUSE, leading to a total of 52 galaxy-scale cluster members (indicated with white arrows in Fig.~\ref{fig:app:members}). To reduce the number of free parameters in the model we have assumed as in previous works (e.g. \citealt{richard2014}) a mass-traces-light approach for these galaxy-scale components, where the geometry (center, ellipticity and position angle) follow the light distribution and the other dPIE parameters are scaled with respect to the values of an L$^*$ galaxy ($\sigma^*$, $r_{\rm core}$ and $r_{\rm cut}$). The two exceptions are the BCG and the brightest galaxy located in the arc, whose $\sigma$ and $r_{\rm cut}$ parameters are fit independently. Regarding the cluster-scale component, we only assumed $r_{\rm cut}=1000$ kpc as it is unconstrained. In total our model is comprised of 12 free parameters.

Regarding the constraints, we have complemented the constraints used in \citet{richard2010} and reach 13 multiple systems of matched clumps in the giant arc, forming a total of 33 multiple images; all of them are included at their spectroscopic redshift. Unfortunately the Einstein radius is too small and the MUSE data is not deep enough to provide us with additional spectroscopic redshifts for multiple images. Accounting for the image multiplicity and the unknown source location, these clump locations 
give us 40 constraints, which gives us a well-constrained model with regard to the 12 free parameters. The 33 multiple images of the clumps used to constrain the lens model are shown in Fig.~\ref{fig:app:members} as red circles.

The best fit parameters of the \textsc{lenstool} mass model are presented in Tab~\ref{tab:lenstool_params}. This model gives us an rms of $0.08"$ between the observed and the predicted location of all constraints, which is close to the precision of the HST locations. The velocity dispersion of the main dark matter halo (cluster-scale) component is $\sim$ 600 km/s, again confirming that the lens is somewhere in between a massive group and a low-mass cluster.
\begin{table*}
    \begin{tabular}{c|c|c|c|c|c|c|c}
\hline\hline  
Potential & $\Delta\alpha$ & $\Delta\delta$ & $e$ & $\theta$ & r$_{\rm core}$ & r$_{\rm cut}$ & $\sigma$ \\
   & [arcsec] & [arcsec] & & [deg] & kpc & kpc & km$\,$s$^{-1}$ \\
\hline 
DM1 & $ -0.7^{+  0.2}_{ -0.2}$ & $ -0.5^{+  0.2}_{ -0.3}$ & $ 0.65^{+ 0.02}_{-0.03}$ & $ 53.2^{+  0.2}_{ -0.2}$ & $23^{+1}_{-1}$ & $[1000]$ & $610^{+4}_{-5}$ \\
BCG & $[  0.0]$ & $[ -0.0]$ & $[0.24]$ & $[ 47.6]$ & $[0]$ & $81^{+47}_{-18}$ & $215^{+33}_{-12}$ \\
GAL1 & $[  2.1]$ & $[  6.8]$ & $[0.13]$ & $[ 58.0]$ & $[0]$ & $5^{+1}_{-1}$ & $27^{+29}_{-50}$ \\
L$^{*}$ galaxy &  & & & & $[0.15]$ & $10^{+3}_{1}$ & $180^{+4}_{-13}$\\
\hline
\end{tabular}
\caption{Best fit parameters of the \textsc{lenstool} mass model.}
\label{tab:lenstool_params}
\end{table*}

\section{Minimum resolvable size}
\label{sec:app:reffmin}
In order to test what is the minimum clump size measurable with our method, we simulate synthetic sources with asymmetric Gaussian profiles and we fit them in the same way as the real clumps. In more details, we produce 3 sets of synthetic sources, with axis ratios uniformly distributed in the ranges $[1.0; 1.5]$, $[1.5; 2.0]$ and $[2.0; 4.0]$, respectively. We add a fourth set of sources with axis ratio fixed at $axr=1.0$, i.e. with a fixed circular symmetric Gaussian profile. 
For each set we simulate 500 sources with sizes uniformly distributed in the range $\rm \log(\sigma_{x,in}/[px])=[-2; 0.6]$, fluxes uniformly distributed in the range $\rm \log(flux_{in}/[e/s])=[0.0; 0.5]$ and random angle $\theta$. These ranges are chosen to cover the range of properties of the A521-sys1 clump catalog. The sources are introduced at a random position in the region of the observations covered by the images of the A521-sys1 galaxy and then fitted one at a time, in order to avoid the manually-introduced crowding we would have by adding all the 500 sources together. 

\begin{figure*}
\centering
\subfigure{\includegraphics[width=0.49\textwidth]{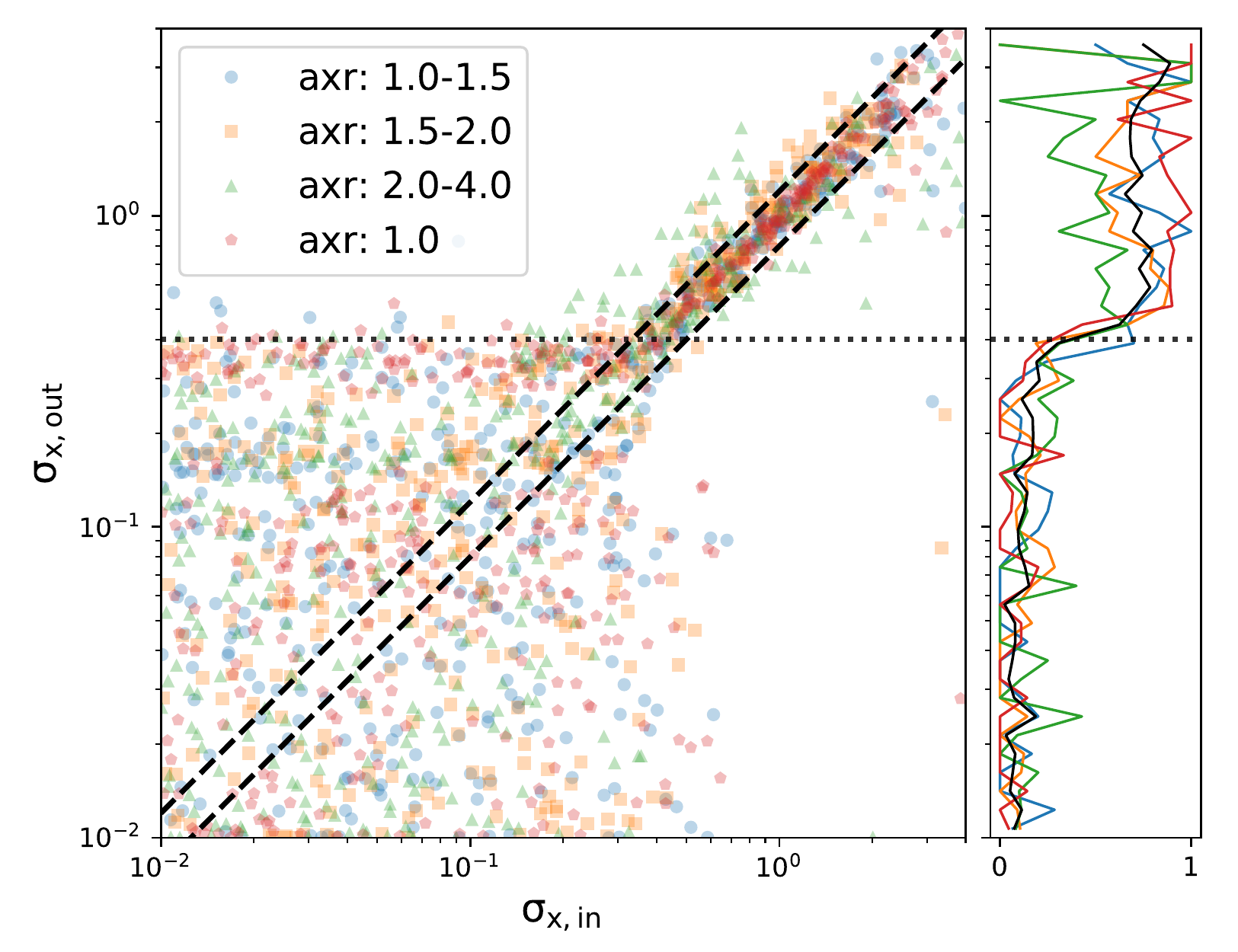}}
\subfigure{\includegraphics[width=0.49\textwidth]{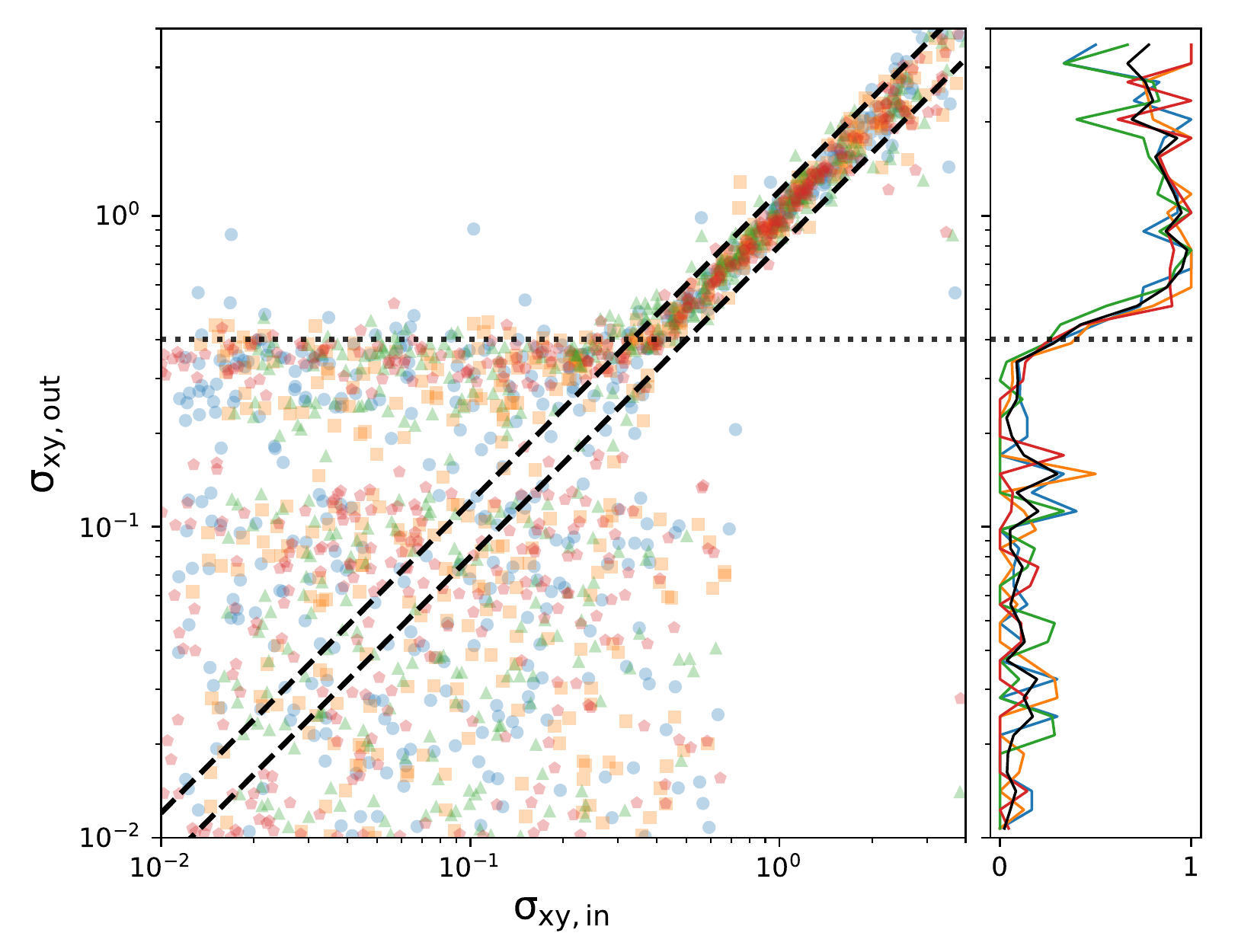}}
    \caption{Standard deviation resulting from the fit of synthetic sources in function of their input values. Left panel shows $\rm \sigma$ for the minor axis ($\rm \sigma_x$), right panel shows the geometrical mean $\rm \sigma_{xy}\equiv\sqrt{\sigma_x\cdot\sigma_y}$. Different colors and symbols refer to sources with different axis ratios, as reported in the legend. The dashed lines enclose the good fits, i.e. where the relative error on the retrieved size is less than $20\%$. The horizontal dotted line mark the $\rm \sigma_{x,min}=0.4$ px value chosen as the minimum resolvable size. 
    For each panel, the sub--panel on the side shows the fraction of sources with good fit in function of the output standard deviation. }
    \label{fig:test_reffmin}
\end{figure*}

We define the Gaussian standard deviations derived from the fit as $\rm \sigma_{x,out}$, in contrast to the intrinsic ones, used as input for the simulated clusters, $\rm \sigma_{x,in}$. We consider good fits the ones where the relative difference  $\rm \sigma_{x,rel}\equiv|\sigma_{x,out}-\sigma_{x,in}|/\sigma_{x,out}$ is less than 0.2, i.e. the relative error on the retrieved size is less than $20\%$. We show the results of the test in Fig.~\ref{fig:test_reffmin}. In the left panel it can be observed how the fraction of good fits steeply increases for $\rm \sigma_{x,out}>0.4$ px. Above this value, the fraction of good fits stabilizes above $\sim50\%$, with a clear dependence on the axis ratio, as for more circular sources better fits are returned, on average. If, instead of $\rm \sigma_x$, we consider the geometrical mean of the minor and major axes of the gaussian $\rm \sigma_{xy}\equiv\sqrt{\sigma_{y}\cdot\sigma_{y}}=\sigma_x\sqrt{axr}$, as done for estimating the effective radius of the real clumps, we see that the fraction of good fits with $
\rm \sigma_{xy}>0.4$ flattens to a value $\sim80\%$, indicating that, for large $axr$, the derived $\rm \sigma_{xy}$ is more robust than $\rm \sigma_{x}$ and $\rm \sigma_{y}$ considered alone. We observe a small decline of the fraction of good fits for large sizes, possibly driven by their lower average surface brightness. We deal in detail with the completeness in surface brightness in Appendix~\ref{sec:app:completeness}. We consider $\rm \sigma_{x}=0.4$ px as the lowest size recognizable by our routine, as below such value the derived sizes seem to be totally uncorrelated to the input values. 
We use $\rm \sigma_{x,out}$ instead of $\rm \sigma_{x,in}$ as reference as this is the quantity we derive for the real clumps. 

\section{Completeness test}
\label{sec:app:completeness}
We test the luminosity completeness of our observation in a similar way as described in Appendix~\ref{sec:app:reffmin}, i.e. by introducing synthetic sources in the field of view of the galaxy and fitting them in the same way as for the real clumps.
We use the map of the galaxy after having subtracted the flux of the real clumps. Despite the fact that most of the observed clumps have profiles consistent with the instrumental PSF, we simulate sources with different sizes, in order to derive a surface brightness limit. In more details, we simulate 3 sets of clumps, with $\rm \sigma_x=0.4$, $1.0$ and $2.0$ px ($0.024"$, $0.06"$ and $0.12"$ respectively); sources with larger sizes are not measured in this galaxy and therefore are not necessary to be simulated. For all sets we simulate circularly symmetrical sources, i.e. we set $axr\equiv1$. For each set we simulate 500 sources with fluxes randomly drawn from a uniform distribution in the range $\rm \log(flux_{in}/[e/s])=[-2.0; 1.0]$ for sources with $\sigma_x=0.4$ and $1.0$ px, and in the range $\rm \log(flux_{in}/[e/s])=[-1.0; 2.0]$ for sources with $\sigma_x=2.0$ px. 

Some of the synthetic clumps have recovered fluxes $\rm flux_{out}$ consistent with zero ($<10^{-4}$ e/s, i.e. more than two orders of magnitude lower than the input values), meaning that the fitting process do not recognize the source and consider the cutout as only filled by background emission. Those are 27 sources with $\rm \sigma_{x,in}=0.4$ px and $\rm flux_{in}<0.07$ e/s ($\rm 28.3\ mag$), 31 with $\rm \sigma_{x,in}=1.0$ px and $\rm flux_{in}<0.12$ e/s ($\rm 27.7\ mag$), and 11 with $\rm \sigma_{in}=2.0$ px and $\rm flux_{in}<0.39$ e/s ($\rm 26.4\ mag$). We call these $\rm flux_{in}$ values detectability limits, $\rm lim_{det}$. 
We observe that some of the sources with fluxes higher that the detectability limits are still not well-fitted and we therefore investigate the precision in recovering the input properties.

\begin{figure*}
\centering
\subfigure{\includegraphics[width=0.49\textwidth]{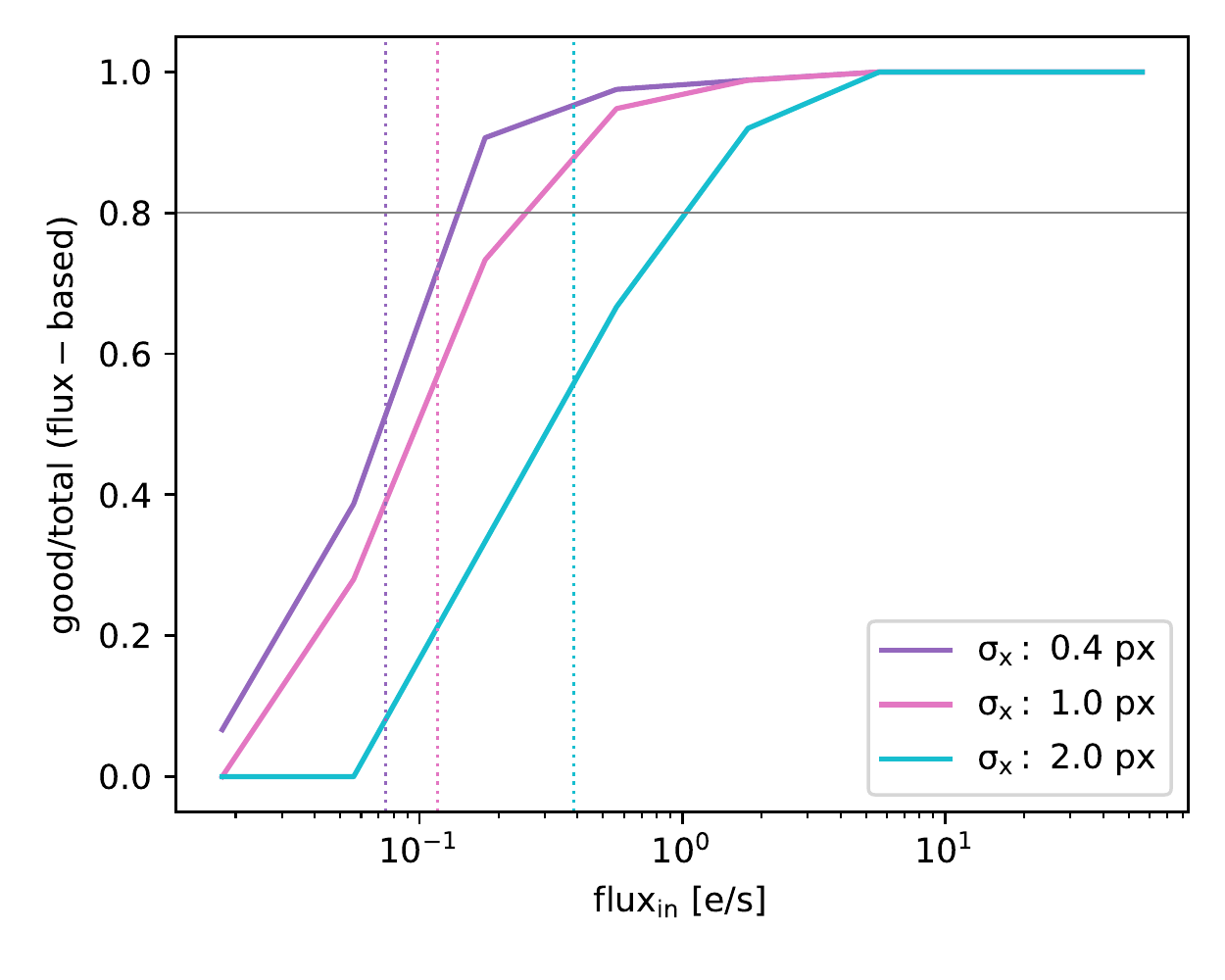}}
\subfigure{\includegraphics[width=0.49\textwidth]{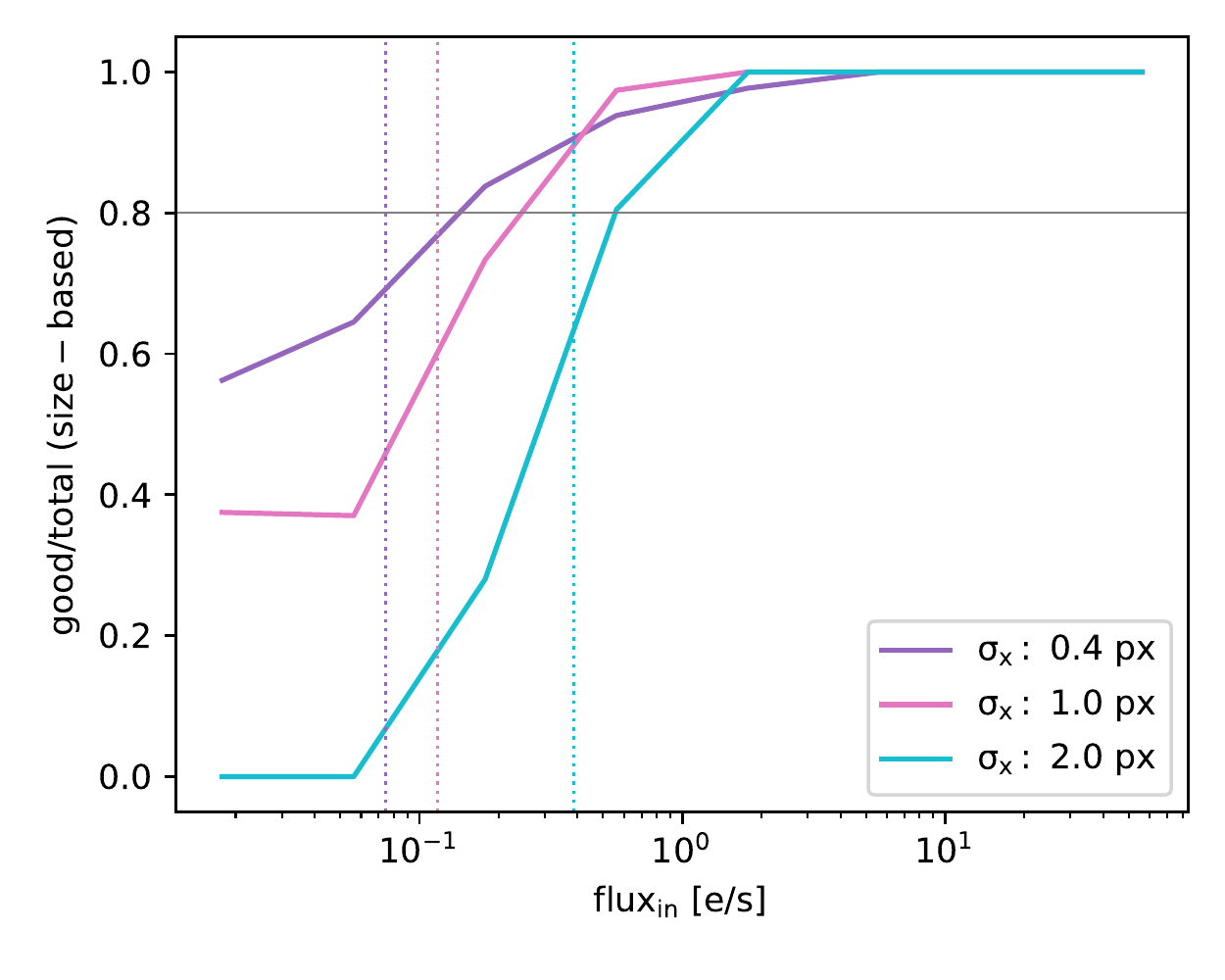}}
    \caption{Fraction of good fits over the total number of simulated sources, in function of the input flux of the sources, $\rm flux_{in}$. Good fits are defined as the ones whose relative flux (left panel) or size relative error (right panel) is below $50\%$. Each `completeness' curve refer to a different input size (0.4, 1.0 and 2.0 px for purple, pink and cyan curve, respectively). The horizontal line indicate $80\%$ completeness, used to derive the completeness limits ($\rm lim_{com}$, defined as the flux values where the curves reaches the $80\%$ completeness). The dotted vertical lines refer to the detection limits $\rm lim_{det}$ described in the text.}
    \label{fig:app:completeness}
\end{figure*}

We calculate for each of the synthetic sources the relative error on the recovered flux, i.e. $\rm flux_{rel}=|flux_{in}-flux_{out}|/flux_{in}$. The values of $\rm flux_{rel}$ are clustered around zero for bright sources, but they start deviating to larger values (suggesting larger uncertainties in fitting the source) when considering dimmer sources. 
The cases where the relative error on the recovered flux is above $50\%$, i.e. $\rm flux_{rel}\ge0.5$, can be considered unreliable fits. We plot the fraction of acceptable fits, satisfying $\rm flux_{rel}<0.5$, in function of $\rm flux_{in}$ in the left panel of Fig.~\ref{fig:app:completeness}. We name the flux values at which the fraction goes above $80\%$ completeness limits, $\rm lim_{com}$; these are more conservative values compared to the detectability limits described above. The completeness limits for the three sets of sources are $\rm lim_{com,0.4}=0.15$ e/s ($\rm 27.5\ mag$), $\rm lim_{com,1.0}=0.30$ e/s ($\rm 26.7\ mag$) and $\rm lim_{com,2.0}=1.20$ e/s ($\rm 25.2\ mag$). 
We repeat this process by calculating the relative error on the recovered size, i.e. $\rm \sigma_{rel}=|\sigma_{in}-\sigma_{out}|/\sigma_{in}$, and plotting the fraction of acceptable fits with $\rm \sigma_{rel}<0.5$ in the right panel of Fig.~\ref{fig:app:completeness}. The $\rm flux_{in}$ values corresponding to fractions above $80\%$ are the same or smaller than $\rm lim_{com}$ discussed above therefore we kept the latter as more conservative values. 
In Section~\ref{sec:sizelum} of the main text we compare $\rm lim_{com}$ values found with this analysis to the magnitudes of the observed clumps. 
As final remark, we tested an average completeness over the entire area covered by the 3 images of A521-sys1; keeping separated the 3 regions defined in Section~\ref{sec:data_hst} would not affect very much the values recovered. 

\section{Extinction map from MUSE}\label{sec:app:extinction}
\begin{figure}
\centering
\includegraphics[width=0.49\textwidth]{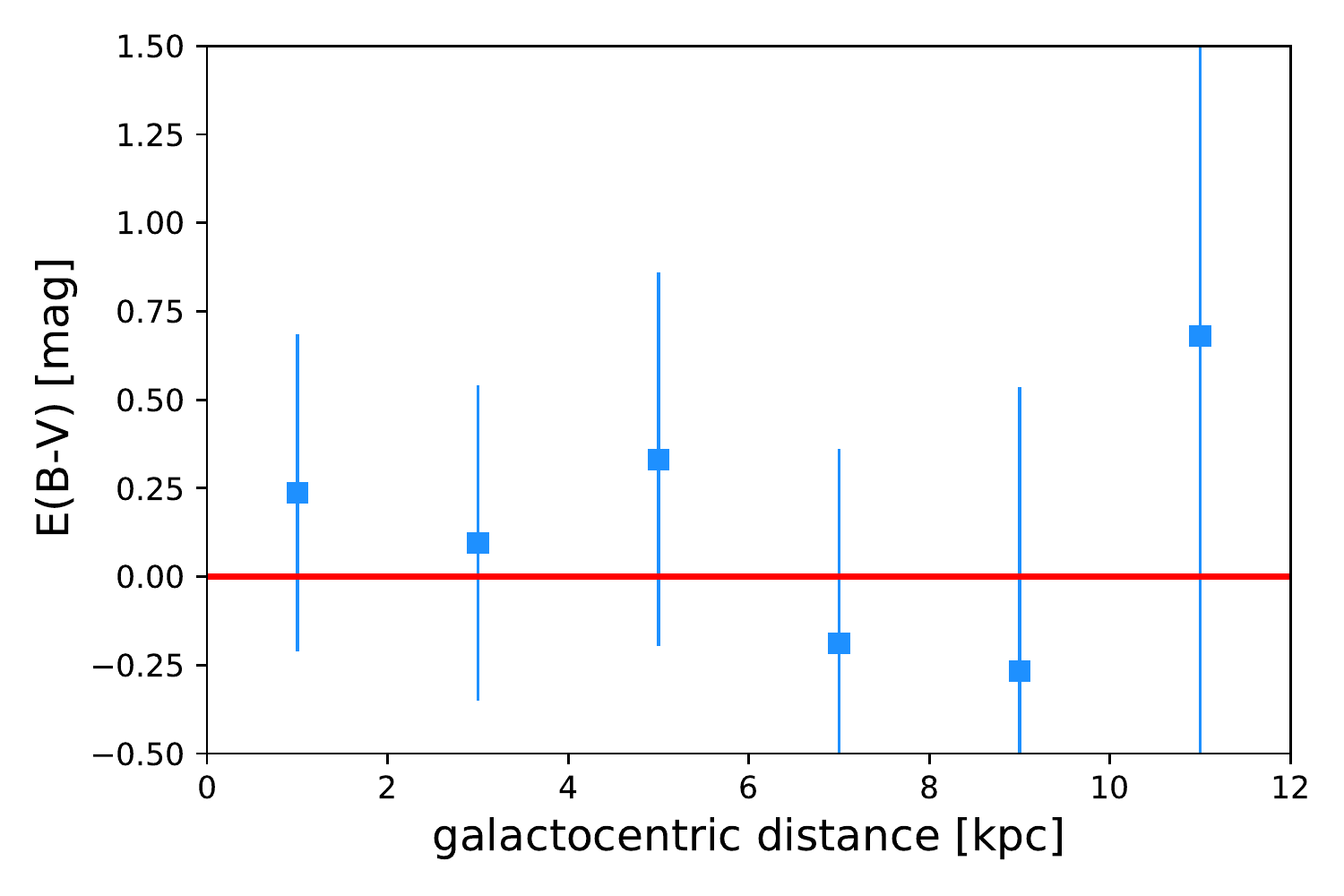}
\caption{Color excess values, E(B-V), derived from the MUSE data in 6 concentric annular sub--regions of A521-sys1, of 2 kpc radius. The values are derived assuming an electron temperature $\rm T_e=10^4\ K$. The unphysical uncertainty on last bin is due to the low signal in the outskirts of the galaxy. We consider the (non-physical) negative values as consistent with no extinction.}
\label{fig:app:ebv}
\end{figure}
We leverage the VLT-MUSE observations of A521 to estimate the nebular extinction of the galaxy. The spectrum at the redshift of A521-sys1 covers the wavelengths of two Balmer lines, namely $\rm H\gamma$ and $\rm H\delta$. At fixed gas density and temperature these lines have a fixed ratio i.e. $\rm R_{\gamma\delta,intr}\equiv L_{H\gamma}/L_{H_\delta}=1.81$ for electron density $\rm n_e=10^2\ cm^{-3}$ and electron temperature $\rm T_e=10000\ K$. The ratio change only by $\pm0.01$ if $\rm T_e$ varies in the range $5000-20000$ K (values from \citealp{dopita2003}, based on \citealp{storey1995}). 
A non-zero extinction changes the value of the ratio by a factor proportional to the magnitude of the extinction itself. We can use the observed line ratio $\rm R_{\gamma\delta,obs}$ to derive the color excess $\rm E(B-V)$ from:
\begin{equation}\label{eq:ebv_muse}
    \rm R_{\gamma\delta,obs} = R_{\gamma\delta,intr}\cdot10^{0.4\cdot E(B-V) [k(H\gamma)-k(H\delta)]}
\end{equation}
where $\rm k(H\gamma)$ and $\rm k(H\delta)$ are set by the extinction curve considered, in this case the Milky Way one \citep{cardelli1989}.
We divide the galaxy in 6 concentric annular regions with radii of 2 kpc, using the source--plane image to define the annuli and transposing them to the CI, LN and LS images using the lensing model, as described in \citet{nagy2021}. This division assumes that the largest extinction differences would appear studying the galaxy radially. 

In each of the 6 bins, we use the \texttt{pPXF} tool \citep{cappellari2017} to fit and subtract the spectral continuum (including self-absorption of the lines) and the \textsc{Pyplatefit} tool to perform the line fit of the $\rm H\gamma$ and $\rm H\delta$ lines\footnote{\textsc{Pyplatefit} is a tool developed for the MUSE deep fields and is a simplified python version of the \textsc{Platefit IDL} routines developed by \citet{tremonti2004} and \citet{brinchmann2004} for the SDSS project.}.  
Before deriving $\rm R_{\gamma\delta,obs}$ we de--redden the line flux for the Milky Way extinction ($\rm A_{V,MW}=0.21$ mag), using the \citet{cardelli1989} extinction function. We consider the same extinction function to derive $\rm E(B-V)$ in Eq.~\ref{eq:ebv_muse}. 

The derived $\rm E(B-V)$ values are shown in Fig.~\ref{fig:app:ebv}, along with the uncertainties coming from the line and continuum fitting. Due to the large uncertainties, all values are consistent within $1\sigma$ with zero extinction. However, we notice that extinction in the 3 internal bins is consistently higher than in the external bins (where the face values goes to unphysical negative values). The outermost bin has lower S/N compared to the other ones, translating into very large uncertainties that makes it unreliable.
If differential extinction is considered, as in \citet{calzetti2000}, the nebular extinction we derived should be rescaled, $\rm E(B-V)_{star} = 0.44\cdot E(B-V)_{gas}$; in this case, the stellar extinction within the galaxy would be even lower.

Despite not being able to put hard constraint on the extinction values, this analysis suggests the presence of only low average extinction in A521-sys1, ranging up to $E(B-V)\approx0.5$ mag in the internal regions and close to $E(B-V)\approx0.0$ mag in the outskirts. These overall values are consistent with the extinction values of the individual clumps, mainly distributed in the range $E(B-V)$ range $0.0-0.5$ mag (Section~\ref{sec:results_sed}).

\section{Comparison between fiducial and alternative extraction and photometry}
\label{sec:app:alternative}
To test the reliability of our results we implement an alternative method for extracting and analyzing the clumps. 
We measure the properties of the galactic diffuse background (median value and standard deviation, $\rm \sigma$) in a region within the galaxy devoid of clumps. We use contours at $3\sigma$ level (using a smoothing of 3 pixels) above the median value of the background to extract clumps and define their extent. The sizes of clumps are calculated using ellipses that better trace the $3\sigma$ contours. We used $6\sigma$ contours to separate multiple peaks within the same $3\sigma$ contours, considering them as separate clumps. 
When two $6\sigma$ peaks are in the same $3\sigma$ contour, two ellipses are considered, trying to cover the entire region within the contour without intersecting them. We consider the geometric mean of the major and minor axis of each ellipses, $R_3 = \sqrt{ab}$, where  the subset $3$ is used to indicate that this radius refer to the extent of the $3\sigma$ contours. In order to convert $R_3$ into an effective radius we assume that clumps have Gaussian profiles and we first derive an \textit{observed} effective radius: 
\begin{equation}
\rm    R_{eff,obs}=R_3\sqrt{\frac{\ln{(2)}}{\ln{(r_{peak}/3)}}},
\end{equation}
where $\rm r_{peak}$ is the ratio of the peak of each region over the RMS value. Then we find the intrinsic effective radius by subtracting, in quadrature, the HWHM of the instrumental PSF, which, for F390W, is $0.8$ px, 
\begin{equation}\label{eq:app:reff}
\rm R_{eff} = \sqrt{R_{eff,obs}^2-0.8^2}.
\end{equation}
Where $\rm R_{eff,obs}$ is smaller than the HWHM of the PSF we set manually the intrinsic $\rm R_{eff}$ to the minimum value detectable, $\rm R_{eff,min} = \sigma_{x,min}\sqrt{2\ln{2}} \approx 1.8\sigma_{x,min} = 0.47$ px, described in Section~\ref{sec:minreff}. 

Photometry is performed using aperture photometry in the ellipses defined above, and subtracting the background estimated as the median value of the sky evaluated in a annular region around the aperture. Aperture correction is needed to correct the flux for losses due to finite apertures. We simulate sources with the sizes found using Eq.~\ref{eq:app:reff}, we perform aperture photometry using the same apertures used on the real data and then we calculate what is the fraction of flux we are missing. The missing flux is then converted into an aperture correction; we therefore have a specific aperture correction for each source. The values hence found may constitute, in some cases, overestimates; some of the clumps present a bright peak and then some more diffuse light filling the $3\sigma$ contour and the assumption of a 2D--Gaussian profile may not be accurate in these cases. 
The conversion of sizes from pixels to parsecs and of flux into observed and absolute magnitudes is done in the same way as for the reference sample, as well as the de-lensing\footnote{with the only exception that we use different apertures, i.e. the ones also used for photometry, to estimate the median amplification factors and their uncertainties.} and the SED fitting (see Sections~\ref{sec:convert_intrinsic} and \ref{sec:BBSED}).


%
%

\bsp	
\label{lastpage}
\end{document}